\documentclass[iop,preprint2,numberedappendix]{emulateapj}

\usepackage{graphicx}
\usepackage{longtable}
\usepackage{booktabs}
\usepackage{amssymb}
\usepackage{empheq}
\usepackage[colorlinks=true,linkcolor=blue,anchorcolor=blue,citecolor=blue,urlcolor=blue]{hyperref}  
\hypersetup{pdfauthor={Mauricio Carrasco}, pdftitle={VLT/Magellan spectroscopy of 29 strong lensing selected galaxy clusters}}

\usepackage{pict2e}
\usepackage{xcolor}
\usepackage{appendix}

\newcommand\etal{et al.\ }

\newcommand\rmag{\ifmmode r_{625}\else$r_{625}$\fi}
\newcommand\imag{\ifmmode i_{775}\else$i_{775}$\fi}
\newcommand\zmag{\ifmmode z_{850}\else$z_{850}$\fi}
\newcommand\Vmag{\ifmmode V_{606}\else$V_{606}$\fi}
\newcommand\Imag{\ifmmode I_{814}\else$I_{814}$\fi}

\newcommand\rtwoh{\ifmmode {\rm r}_{200}\else r$_{200}$\fi}
\newcommand\ls{\mathrel{\hbox{\rlap{\hbox{\lower4pt\hbox{$\sim$}}}\hbox{$<$}}}}
\newcommand\gs{\mathrel{\hbox{\rlap{\hbox{\lower4pt\hbox{$\sim$}}}\hbox{$>$}}}}

\renewcommand\deg{^\circ}

\def\simgreat{\ifmmode{\mathrel{\mathpalette\@versim>}}
    \else{$\mathrel{\mathpalette\@versim>}$}\fi}
\def\simless{\ifmmode{\mathrel{\mathpalette\@versim<}}
    \else{$\mathrel{\mathpalette\@versim<}$}\fi}

\newcounter{thefigs}

\newcounter{thetabs}

\newcommand{\hGpc}{\ifmmode{h^{-1}{\rm Gpc}}\;\else${h^{-1}}${\rm Gpc}\fi}
\newcommand{\hkpc}{\ifmmode{h^{-1}_{70}\ {\rm kpc}}\;\else${h^{-1}_{70}}$ {\rm kpc}\fi}
\newcommand{\hMpc}{\ifmmode{h^{-1}_{70}\ {\rm Mpc}}\;\else${h^{-1}_{70}}$ {\rm Mpc}\fi}
\newcommand{\LCDM}{$\Lambda$CDM}

\def\simless{\mathbin{\lower 3pt\hbox
	{$\,\rlap{\raise 5pt\hbox{$\char'074$}}\mathchar"7218\,$}}} 
\def\simgreat{\mathbin{\lower 3pt\hbox
	{$\,\rlap{\raise 5pt\hbox{$\char'076$}}\mathchar"7218\,$}}} 

\shortauthors{Carrasco \etal }
\slugcomment{Submitted to the Astrophysical Journal, \today}

\begin{document}

\title{VLT/Magellan spectroscopy of 29 strong lensing selected galaxy clusters}

\shorttitle{VLT/Magellan spectroscopy of 29 strong lensing selected galaxy clusters}

\author{Mauricio Carrasco\altaffilmark{1,2}}
\author{L. Felipe Barrientos\altaffilmark{2,3}}  
\author{Timo Anguita\altaffilmark{4,3}}  
\author{Cristina Garc\'ia-Vergara\altaffilmark{2,5}}  
\author{Matthew Bayliss\altaffilmark{6,7}} 
\author{Michael Gladders\altaffilmark{8}}
\author{David Gilbank\altaffilmark{9}}
\author{H.K.C. Yee\altaffilmark{10}}
\author{Michael West\altaffilmark{11}}

\email{carrasco@uni-heidelberg.de}
\altaffiltext{1}{Zentrum f\"ur Astronomie, Institut f\"ur Theoretische Astrophysik, Philosophenweg 12, 69120 Heidelberg, Germany}
\altaffiltext{2}{Instituto de Astrof\'isica, Pontificia Universidad Cat\'olica de Chile. Avda Vicu\~na Mackenna 4860, Santiago, Chile}
\altaffiltext{3}{Millennium Institute of Astrophysics, Chile}
\altaffiltext{4}{Departamento de Ciencias Fisicas, Universidad Andres Bello, Fernandez Concha 700, Las Condes, Santiago, Chile}
\altaffiltext{5}{Max-Planck-Institut f\"ur Astronomie, K\"onigstuhl 17, 69117 Heidelberg, Germany}
\altaffiltext{6}{Department of Physics \& Astronomy, Colby College, 5800 Mayflower Hill, Waterville, Maine 04901}
\altaffiltext{7}{Department of Physics, Harvard University, 17 Oxford Street, Cambridge, MA 02138, USA}
\altaffiltext{8}{Department of Astronomy and Astrophysics, and the Kavli Institute for Cosmological Physics, University of Chicago, 
5640 South Ellis Avenue, Chicago, IL 60637, USA}
\altaffiltext{9}{South African Astronomical Observatory, PO Box 9, 7935 Observatory, South Africa}
\altaffiltext{10}{Department of Astronomy and Astrophysics, University of Toronto, 50 St. George Street, Toronto, Ontario, M5S 3H4, Canada}
\altaffiltext{11}{Maria Mitchell Observatory, 4 Vestal Street, Nantucket, MA 02554, USA}



\begin{abstract}
We present an extensive spectroscopic follow-up campaign of 29 strong 
lensing (SL) selected galaxy clusters discovered primarily in the 
Second Red-Sequence Cluster Survey (RCS-2). Our spectroscopic analysis
yields redshifts for 52 gravitational arcs present in the core of our 
galaxy clusters, which correspond to 35 distinct background sources 
that are clearly distorted by the gravitational potential of these 
clusters. These lensed galaxies span a wide redshift range of 
$0.8 \le z \le 2.9$, with a median redshift of $z_s = 1.8 \pm 0.1 $.
We also measure reliable redshifts for 1004 cluster members, allowing us
to obtain robust velocity dispersion measurements for 23 of these clusters, 
which we then use to determine their dynamical masses by using 
a simulation-based $\sigma_{DM} - M_{200}$  scaling relation.
The redshift and mass ranges covered by our SL sample are $0.22 \le z \le 1.01$
and $5  \times10^{13} \le M_{200}/h^{-1}_{70}M_{\odot} \le 1.9\times10^{15}$, 
respectively. We analyze and quantify some possible effects that might
bias our mass estimates,  such as the presence of substructure, the region
where cluster members are selected for spectroscopic follow-up, the 
final number of confirmed members, and line-of-sight effects. 
We find that 10 clusters of our sample with $N_{\textnormal{mem}} \gtrsim 20$ 
show signs of dynamical substructure.  
However, the velocity data of only one system is inconsistent with 
a uni-modal distribution. We therefore assume that the substructures
are only marginal and not of comparable size to the clusters themselves.
Consequently, our velocity dispersion and mass estimates can be used 
as \textit{priors} for SL mass reconstruction studies and
also represent an important step toward a better understanding 
of the properties of the SL galaxy cluster population.
\end{abstract}


\section{Introduction} \label{sec:intro}

Galaxy clusters have long been considered one of the most important cosmological probes due to their 
privileged position in the hierarchical formation scenario as the most massive and virialized 
objects. As such, galaxy clusters can inform important tests of the standard concordance
 cosmological model, and provide unique laboratories for the understanding of dark matter (DM) properties (see \citealt{Voit2005}, for a review). 
For example, the number density of galaxy clusters is sensitive to the amplitude of the primordial 
density fluctuations, and can therefore be used as a powerful tool for  
estimating cosmological parameters, so long as the masses and redshifts of clusters are measured
with good precision \citep[][]{Evrard02, Vikhlinin09, Rozo10}.

In recent years extensive efforts have been made to determine the mass and redshift distribution of samples 
of galaxy clusters \citep[][]{Bohringer04,GladdersYee05,Burenin07,Bleem2015, Gilbank11}. Most of these 
mass measurements are based on baryonic signatures --- e.g. optical light, X-ray emission, and the 
Sunyaev Zel'dovich effect (SZe). However, there are pernicious 
systematic uncertainties that affect these baryonic signatures, which trace only a small fraction of 
the total matter content of galaxy clusters. The most robust and direct way to map the total (baryonic 
and dark) matter distribution in galaxy clusters is through the analysis of gravitational lensing effects, 
whereby background galaxy profiles are modified by the gravitational potential of foreground clusters, 
producing shape distortions and apparent flux magnification 
\citep[see reviews by][]{Bartelmann2010reviewB, KneibNatarajan11}. 

In the outer regions of galaxy clusters the gravitational potential produces only tiny distortions that 
do not constrain the cluster profile individually, and therefore it is necessary to perform statistical 
measurements of the distortions affecting large ensembles of background galaxies. 
This is the so-called weak lensing (WL) regime, and the application of WL measurements has 
become a powerful tool in the recent years \citep{van_Uitert_2015,Dahle06, HoekstraJain08,Okabe10b, Umetsu14,High2012}. 
In the innermost region of clusters --- the strong lensing (SL) regime --- the mass density can be 
high enough to produce dramatic distortions in the form of giant elongated arcs or multiple images 
of background sources, which can inform high-precision mass reconstruction studies of the cluster 
core \citep[e.g.][]{Zitrin09a,Zitrin2012CLASH1206,Zitrin2012CLASH0329,Limousin2012_M0717}. 
Statistical analysis of samples of galaxy clusters that exhibit the hallmarks --- giant arcs and 
multiply-imaged sources --- of strong lensing can also be used in tests of the $\Lambda$CDM 
cosmological model. For example, several different studies comparing observations and predictions 
for the abundance of giant arcs (``giant arc statistics'') find that  simulations under-predict the number 
of giant arcs on the sky by perhaps as much as an order of magnitude \citep{Bartelmann98, Gladders03, Li06}. 

In addition to their cosmological applications, the large magnification effects that typically act on 
strongly lensed background galaxies allows the study of high-redshift galaxies which would 
otherwise be too faint to be observed. However, cases of bright, high-magnification strong lensing remain 
relatively rare phenomena, occurring in a only small fraction of galaxy clusters \citep{Horesh2010}. It 
is therefore always valuable to identify and follow-up new samples of strong lensing galaxy clusters. 
Identifying new strong lensing samples serves to 
effectively increase the volume of the early universe that is available for observations 
\citep[e.g.][]{Bradley08, Bouwens09, Zheng09, Zheng12, Wuyts10, Wuyts14, Bayliss11b,Bayliss2010,Coe13}.

In this paper we present the results of a spectroscopic follow-up campaign of 29 SL-selected galaxy 
clusters with $6.5 - 8.2$m class telescopes,  discovered primarily in the Second Red-Sequence Cluster 
Survey  \citep[RCS-2;][]{Gilbank11}. The data presented in this work reveal the nature of the giant 
arc and multiple-image candidates, increasing the number of spectroscopically confirmed 
lensed galaxies at high-redshift. From these data we also identify redshifts for more than one thousand 
cluster member galaxies, which are used to estimate dynamical masses of these clusters. The 
combination of these results will allow us to derive robust reconstructions of the matter distribution 
in these galaxy clusters.

This paper is organized as follows: In \S \ref{sec:observations} we describe the cluster sample
and summarize the spectroscopic follow-up campaign. In \S\ref{sec:results} we present the redshift results
for the lensed galaxies and cluster members, as well as the final redshift and velocity dispersion of
the clusters. In \S\ref{sec:mass_clusters} we show the dynamical mass results, while in
\S \ref{sec:discussion} we analyze the possible systematics that might affect our measurements.
Lastly, we summarize the main results and present the final conclusions in \S \ref{sec:conclusion}.
Throughout the paper we assume a flat \LCDM\ cosmology with  $\Omega_m = 0.27$, 
$\Omega_\Lambda = 0.73$, and  $H_0 = 70$ $h_{70}$ km s$^{-1}$ Mpc$^{-1}$.

\section{Observations} \label{sec:observations}

\subsection{The SL-selected cluster sample}

The cluster sample presented here is a subset of a larger sample of more than one hundred 
SL-selected clusters that have been identified primarily in RCS-2 imaging data. The median 
seeing of the RCS-2 survey is $\sim 0.7\arcsec$, making it ideal for the detection and 
classification of giant arcs. Furthermore, the RCS-2 survey was designed to detect galaxy 
clusters out to $z\sim1.1$ using the cluster red-sequence technique \citep{GladdersYee00} in 
deep wide-field images. We have searched the resulting galaxy cluster catalog for SL systems, 
where the SL selection criterion is the presence of one or more giant arcs around a galaxy 
cluster's core. The SL-selected clusters identified in the RCS-2 comprise a new sample of giant 
arcs, the Red-Sequence Cluster Survey Giant Arc \citep[RCSGA;][Gladders et al. in prep.]{Bayliss12} 
catalog. Giant arc candidates are flagged by multiple members of our team performing independent 
visual inspections of the entire RCS-2 red-sequence catalog of galaxy clusters, and systems that 
are flagged by most or all inspectors constitute the SL cluster sample, as described in more detail by 
\citet{Bayliss12}. We supplement the RCS-2 SL-selected cluster sample with a few systems 
chosen to fill RA gaps that were similarly selected from the Sloan Digital Sky Survey 
\citep[SDSS;][]{York00}.

We performed a comprehensive spectroscopic follow-up of 29 SL-selected galaxy clusters primarily from the 
RCSGA; 12 of these clusters are previously unpublished. 
The cluster sample is presented in Table \ref{table:Summary_spec_obs_RCSGAsample},
where  clusters formerly reported in other studies are marked with their corresponding references.
We have adopted the naming convention described in \cite{Bayliss11b} for giant arcs discovered in 
RCSGA, given
by RCSGA $-$ Jhhmmss$+$ddmmss \citep[e.g.][]{Wuyts10, Bayliss12}.

\begin{deluxetable*}{l c c c c c}
\tabletypesize{\scriptsize}
\tablecaption{Summary of VLT/Magellan spectroscopic observations \label{table:Summary_spec_obs_RCSGAsample}}
\tablewidth{0pt}
\tablehead{
{} & {} & {} & \multicolumn{2}{c}{Total exposure time} \vspace{0.05cm}  & {}\\
{Target} & 
{R.A.}$^a$ & 
{Dec}$^a$ &
{FORS2/VLT} &  
{IMACS/Magellan} \vspace{0.05cm} & {Prev. Rep.}$^c$\\
{} & 
{(J2000)} & 
{(J2000)} & 
\multicolumn{2}{c}{(Hrs / N$^{\circ}$ masks)$^b$}  & {}
}
\startdata
SDSS J0004$-$0103      & 00 04 52.001  &  $-$01 03 16.58  &  2.00 / 2  &  -- --          &   \cite{Rigby2014}  \\
RCS2 J0034+0225$^d$    & 00 34 28.134  &    +02 25 22.34  &  4.33 / 2  &  -- --          &   \cite{Voges99}    \\ 
RCS2 J0038+0215        & 00 38 55.898  &    +02 15 52.35  &  4.90 / 2  &  -- --          &    -- -- \\
RCS2 J0047+0508        & 00 47 50.787  &    +05 08 20.02  &  2.00 / 1  &  1.67 / 2       &    -- -- \\ 
RCS2 J0052+0433        & 00 52 10.352  &    +04 33 33.31  &  1.67 / 1  &  3.33 / 2       &    -- -- \\ 
RCS2 J0057+0209        & 00 57 27.869  &    +02 09 33.98  &  3.48 / 2  &  -- --          &    -- -- \\
RCS2 J0252$-$1459      & 02 52 41.474  &  $-$14 59 30.38  &  1.33 / 1  &  0.67 / 1       &    \cite{Horesh2010} \\
RCS2 J0309$-$1437      & 03 09 44.096  &  $-$14 37 34.38  &  2.15 / 2  &  1.83 / 1       &    \cite{Bayliss12} \\
RCS2 J0327$-$1326      & 03 27 27.174  &  $-$13 26 22.90  &  2.00 / 1  &  2.00 / 2       &    \cite{Wuyts10}  \\ 
RCS2 J0859$-$0345      & 08 59 14.486  &  $-$03 45 14.63  &  3.33 / 2  &  -- --          &    \cite{Cabanac2007_SL2S}\\
RCS2 J1055$-$0459      & 10 55 35.647  &  $-$04 59 41.60  &  5.17 / 4  &  1.00 / 1       &    \cite{Wittman06}  \\
RCS2 J1101$-$0602      & 11 01 54.093  &  $-$06 02 32.02  &  1.00 / 1  &  -- --          &    \cite{Anguita2012}    \\
RCS2 J1108$-$0456      & 11 08 16.835  &  $-$04 56 37.62  &  2.00 / 2  &  1.17 / 1       &    -- -- \\
SDSS J1111+1408        & 11 11 24.483  &    +14 08 50.82  &  2.17 / 2  &  -- --          &    \cite{Wuyts12_SGAS}  \\
RCS2 J1119$-$0728      & 11 19 11.925  &  $-$07 28 17.51  &  2.17 / 2  &  -- --          &    -- -- \\
RCS2 J1125$-$0628      & 11 25 28.940  &  $-$06 28 39.04  &  2.67 / 2  &  1.33 / 1       &    -- --  \\
RCS2 J1250+0244        & 12 50 41.890  &    +02 44 26.57  &  1.67 / 2  &  1.67 / 1       &    \cite{White97}  \\
RCS2 J1511+0630        & 15 11 44.681  &    +06 30 31.79  &  2.67 / 2  &  -- --          &    -- --  \\
SDSS J1517+1003        & 15 17 02.587  &    +10 03 29.27  &  2.00 / 2  &  2.33 / 1       &    \cite{Szabo11} \\
SDSS J1519+0840        & 15 19 31.213  &    +08 40 01.43  &  4.00 / 3  &  1.33 / 1       &    \cite{Hao10_SDSS_DR7} \\
RCS2 J1526+0432        & 15 26 14.914  &    +04 32 48.01  &  2.00 / 2  &  -- --          &    -- --  \\
SDSS J2111$-$0114      & 21 11 19.307  &  $-$01 14 23.95  &  2.17 / 2  &  2.67 / 1       &    \cite{Bayliss11b} \\
SDSS J2135$-$0102      & 21 35 12.040  &  $-$01 02 58.27  &  6.11 / 4  &  2.00 / 1       &    \cite{Szabo11} \\
RCS2 J2147$-$0102      & 21 47 37.172  &  $-$01 02 51.93  &  3.00 / 2  &  1.50 / 1       &    -- --  \\
RCS2 J2151$-$0138$^d$  & 21 51 25.950  &  $-$01 38 50.14  &  1.50 / 2  &  1.00 / 1       &    \cite{Voges99} \\
SDSS J2313$-$0104      & 23 13 54.514  &  $-$01 04 48.46  &  1.50 / 2  &  1.17 / 1       &    \cite{Geach11_SDSS_GC} \\
RCS2 J2329$-$1317      & 23 29 09.528  &  $-$13 17 49.26  &  4.08 / 4  &  1.33 / 1       &    -- --  \\ 
RCS2 J2329$-$0120      & 23 29 47.782  &  $-$01 20 46.89  &  2.00 / 1  &  2.00 / 1       &    \cite{White97}   \\
RCS2 J2336$-$0608      & 23 36 20.838  &  $-$06 08 35.81  &  3.00 / 2  &  -- --          &    -- -- \\
\enddata
\tablenotetext{a}{Coordinates correspond to the  BCG centroids in sexagesimal degrees (J2000).}
\tablenotetext{b}{Total integration time for lensed galaxy candidates 
(for instrument/telescope), distributed across the number of masks listed in the spectroscopic follow-up. 
Slits targeting to cluster members varied between each mask; therefore, 
the total integration time shown in this table does not necessarily correspond to the total integration time of these objects.}
\tablenotetext{c}{Clusters formerly identified and/or described  in previous studies.}
\tablenotetext{d}{Clusters reported in the ROSAT all-sky bright source catalog \citep{Voges99} as RX J0034.4+0225 and
RX J2151.4$-$0139, respectively.}
\end{deluxetable*}

\subsection{Imaging}
\label{sec:observations-imaging}

Most of the imaging data presented here have been obtained from the RCS-2 survey. They were collected in 
queue-scheduled mode with MegaCam at the $3.6$m Canada-France-Hawaii Telescope (CFHT), between the 
semesters 2003A and 2007B inclusive. The RCS-2 data consist in single exposures (without dithering) 
of 4, 8, and 6 minutes, for the $g',r',$ and $z'$ bands, yielding 5-$\sigma$ limiting magnitudes of
24.3, 24.4, and 22.8, respectively \citep{Gilbank11}. 
We used the $r'$ and $z'$ bands to identify the red-sequence cluster member candidates, which have 
been used as targets for the spectroscopic follow-up. 
Color images from all three bands were used to identify strongly lensed arcs. 
We have also obtained pre-imaging of our clusters in $B, R,$ and $I$ bands, with the Focal Reducer 
and low dispersion Spectrograph 2 \citep[FORS2;][]{Appenzeller98} at the  ESO\footnote{ESO: the 
European Southern Observatory; http://www.eso.org/} 8.2m Very Large Telescope (VLT),  in queue mode. 
These pre-imaging data have been mainly used to design the spectroscopic masks but they 
 also help in the search of multiple-image systems.

\subsection{Spectroscopy and data reduction}

\subsubsection{FORS2/VLT}
\label{sec:observations-spec-VLT}
The FORS2/VLT observations were carried out between October 2006
and March 2010 using the Multi-object spectroscopy with exchangeable masks (MXU) mode.
The MXU mode allows one both to increase the  density of the slits  and to freely manipulate the width,
length, and orientation of the slits, making it ideal for spectroscopy in dense regions like the 
cluster cores. The masks were strategically positioned in the center of each cluster in order to 
prioritize giant arcs and lensed galaxy candidates.  
The central slit widths were set to $1''$, while their length was varied depending on the arc candidate size, 
typically between $12''$ to $25''$. Subsequently, the masks were filled in with slits placed on 
red-sequence selected cluster members, prioritizing brighter galaxies ahead of fainter ones.
The length of these slits was set between $6''$ and $12''$,  
depending on each member candidate size. 
Two masks were usually constructed for each cluster, although some of them had up to four. In order 
to increase the flux of our lensed galaxy candidates, we fixed the slits located at the arc 
positions. On the other hand, with the purpose of increasing the number of spectroscopically 
confirmed cluster members, we varied the slits located on member candidates between each 
mask. We used a low spectral resolution  instrument setup by 
combining the GRIS$\_$150I+27 grism and GG435$+$81 filter, and adopted a 2$\times$2 binning in order 
to improve the signal-to-noise ratio of the spectra.
This configuration results in a final dispersion of 6.9 {\AA} per image pixel, and covers a spectral 
range of $\Delta \lambda \sim 4300 - 10500$ {\AA}, with our highest sensitivity  in the 
interval $\Delta\lambda \sim 4500 - 9000$ {\AA}, due to the transmission efficiency
of the filter used, as well as to the detector's quantum efficiency. 
Depending on the intrinsic features of each target, we varied the exposure times between $1200 - 
3000$ seconds. The total integration time for each galaxy cluster is reported in Table  
\ref{table:Summary_spec_obs_RCSGAsample}.

The basic data reduction steps consisted of bias subtraction, flat-fielding,  wavelength 
calibration, and sky subtraction. 
These steps were carried out by using the ESO Recipe Execution Tool 
(EsoRex\footnote{http://www.eso.org/sci/software/cpl/esorex.html})
and the Common Pipeline Library (CPL\footnote{http://www.eso.org/sci/software/cpl}). 
The wavelength calibration was done by comparison to the standard He+Ne+Ar lamp observations. 
Due to the nature of giant arcs, long and elongated objects, the sky subtraction was a complicated 
task and was carried out independently.
The other advanced steps consisted of the removal of cosmic rays, the 1D spectra extraction and the
average of multiple spectra for each source. These steps were performed using our own IDL 
routines
, inspired by the optimal extraction 
algorithm by \cite{Horne86}.

\begin{figure*}[t!]
\begin{center}
\begin{tabular}{c}
\includegraphics[width=\textwidth, trim= 8mm 3mm 2mm 1mm,clip]{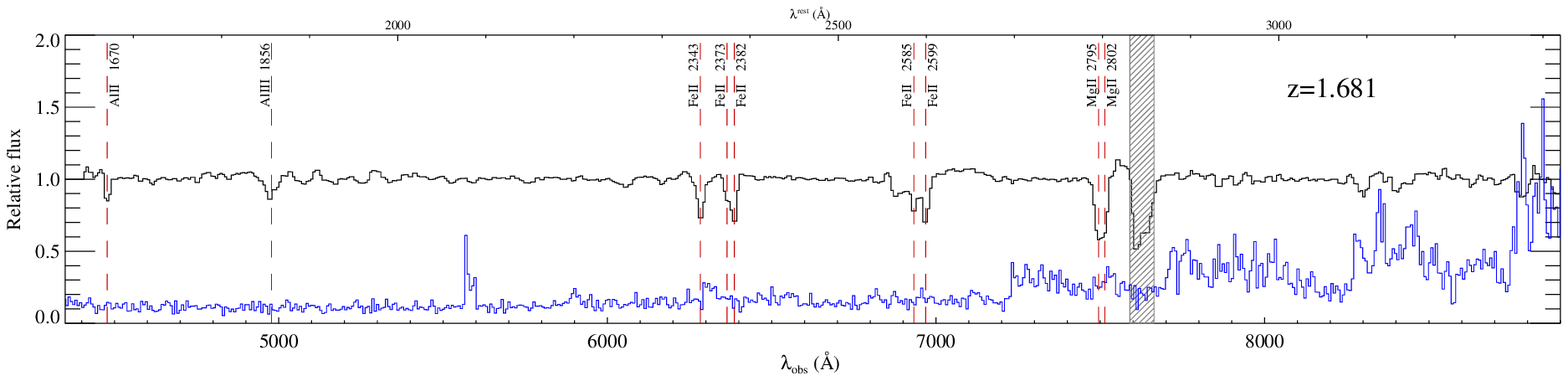} \\
\includegraphics[width=\textwidth, trim= 8mm 3mm 2mm 1mm,clip]{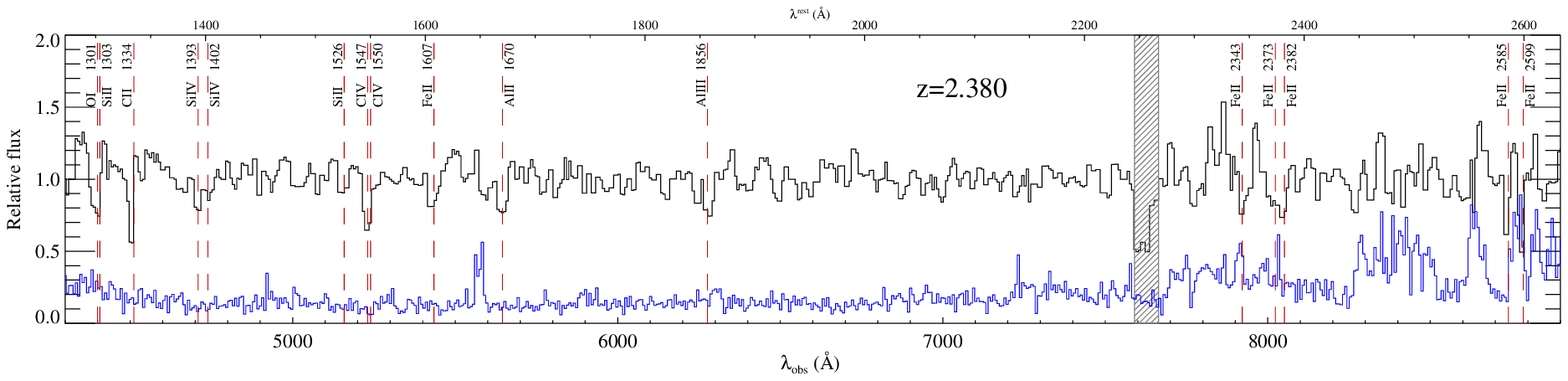} \\
\includegraphics[width=\textwidth, trim= 8mm 3mm 2mm 1mm,clip]{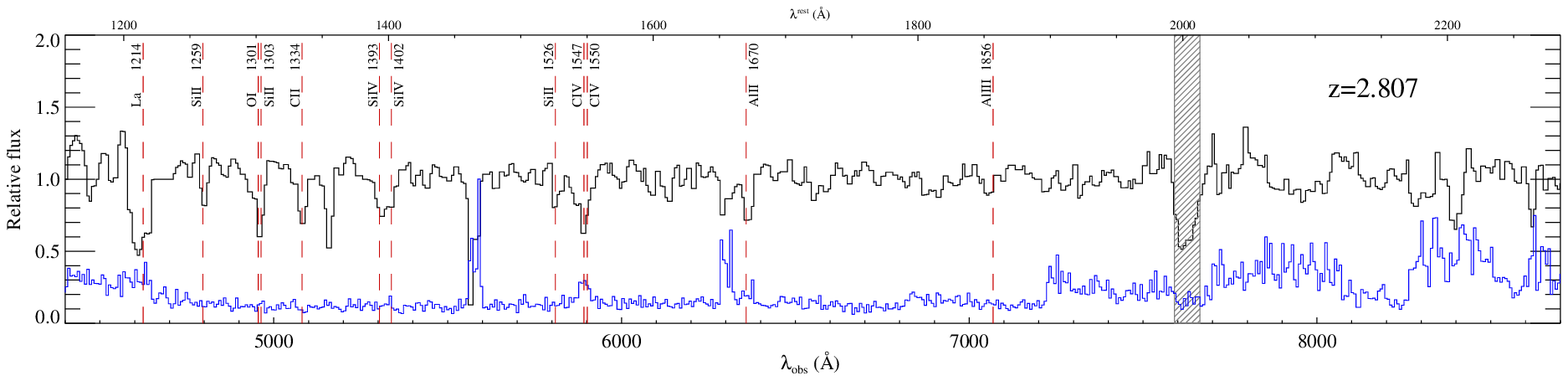} \\
\includegraphics[width=\textwidth, trim= 8mm 3mm 2mm 1mm,clip]{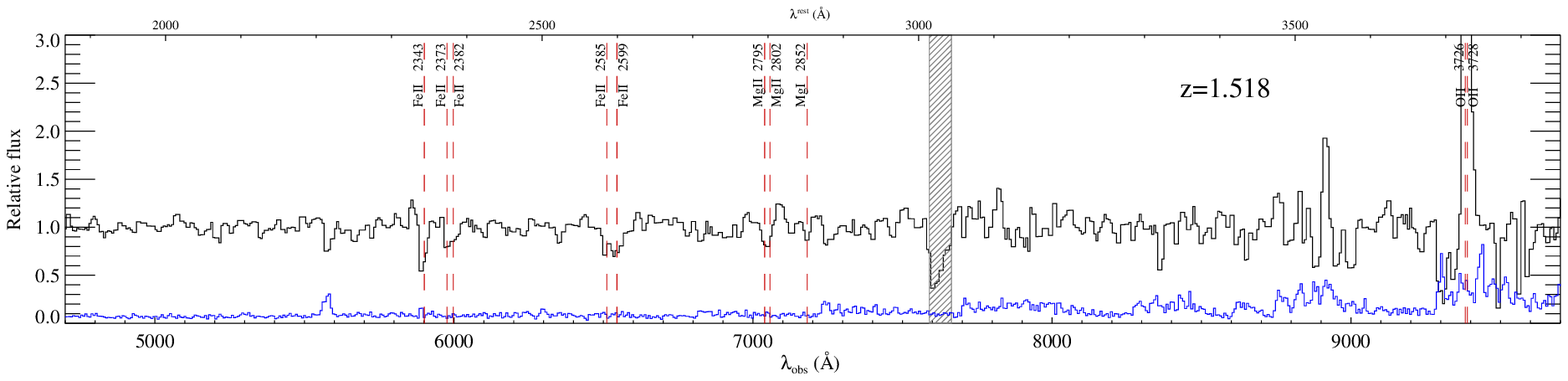} \\
\includegraphics[width=\textwidth, trim= 8mm 3mm 2mm 1mm,clip]{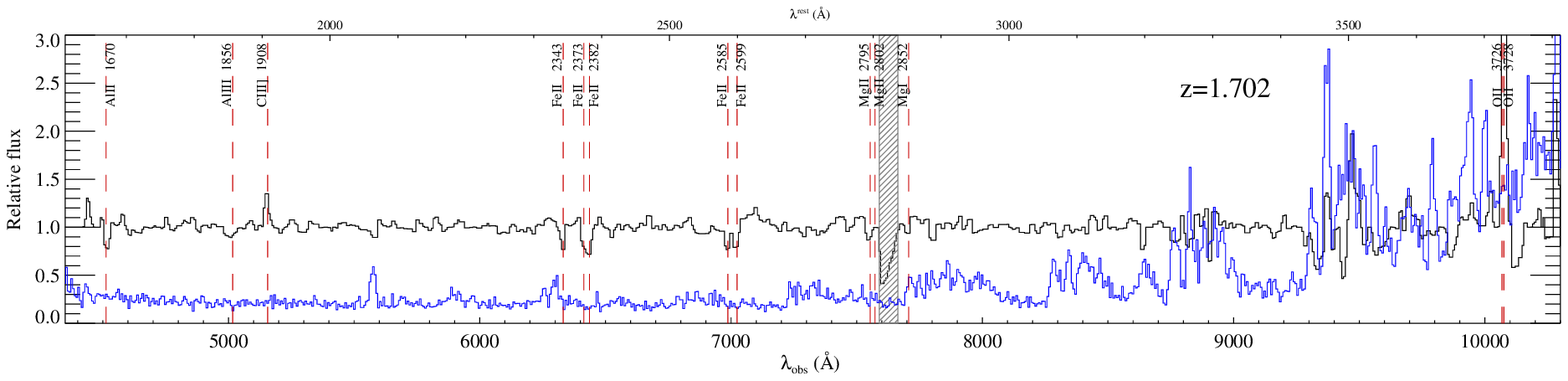}
\end{tabular}
\put(-60, 295){\bf \color{black} a)}
\put(-60, 170){\bf \color{black} b)}
\put(-60, 45){\bf \color{black} c)}
\put(-60, -85){\bf \color{black} d)}
\put(-60, -210){\bf \color{black} e)}
\caption{\label{fig:class3}
FORS2/VLT spectra for five lensed galaxies with highly reliable redshift measurements, labeled as class 3. Spectra are displayed in the
observer-frame and smoothed to match the spectral resolution of the data. The blue histograms correspond to the error array
for the spectra, and the locations of spectral lines are identified by red dashed lines and labeled with their corresponding
ion name and rest-frame wavelength. The telluric A Band absorption feature is indicated by a vertical shaded region. 
From top to bottom the spectra in each panel correspond to the following sources in Table \ref{table:lensed_galaxy} -- 
 a) SDSS J0004$-$0103, S1;  b) RCS2 J0034$+$0225, S1; c) RCS2 J1055$-$0459, S1; d) RCS2 J0309$-$1437, S1; e) RCS2 J0327$+$1326 S1.
}
\end{center}
\end{figure*}

\subsubsection{IMACS/Magellan}

We have also performed spectroscopic observations with the Inamori Magellan Areal Camera and Spectrograph \citep[IMACS;][]{Dressler06, Dressler11} on the 6.5m Magellan (Baade) telescope at LCO\footnote{LCO: Las Campanas 
Observatory; http://www.lco.cl/}.
The IMACS/Magellan observations were collected during 6 different runs between June 2008 and March 2011. 
We used the Gladders Image­-Slicing Multislit Option 
(GISMO\footnote{http://www.lco.cl/telescopes-information/magellan/instru- ments/imacs/gismo}) 
at IMACS, which allows high-efficiency spectroscopy for galaxy clusters, obtaining spectra from $\sim 20-60$ members 
in a single spectroscopic mask. 
The mask design was performed following the same strategy  described above,
 although the slit length was considerably shorter than the VLT data. 
The GISMO-IMACS masks had slits that were consistently $\sim 5.5''-6.5''$ long.
The IMACS f/4 camera and the grating of 300 lines/mm were used to obtain $2\times2$ binning spectra, 
covering an effective spectral range of $\Delta \lambda \sim 4800 - 8000$ {\AA}, with final dispersion 
of $\sim 1.5$ {\AA} per image pixel. 
The exposure time of the GISMO-IMACS observations varied between $2400 - 9600$ seconds.
The total integration time for each cluster is reported in Table \ref{table:Summary_spec_obs_RCSGAsample}. 

The GISMO-IMACS data were reduced using the Carnegie Observatories System for MultiObject Spectroscopy
(COSMOS\footnote{http://code.obs.carnegiescience.edu/cosmos}) pipeline. 
The COSMOS reduction process consisted of bias subtraction, flat-fielding 
and a wavelength calibration using comparison arcs. 
The wavelength solutions have typical rms $\lesssim 0.5$ {\AA}. 
The sky subtraction was carried out by constructing a cosmic ray-cleaned median sky spectrum for each slit, 
which was then subtracted from the slit image. 
Finally, the 1D spectra extraction and the combination of different exposures
were performed using our own IDL routines.

\begin{figure*}[t!]
\begin{center}
\begin{tabular}{c}
\includegraphics[width=\textwidth, trim= 8mm 3mm 2mm 1mm,clip]{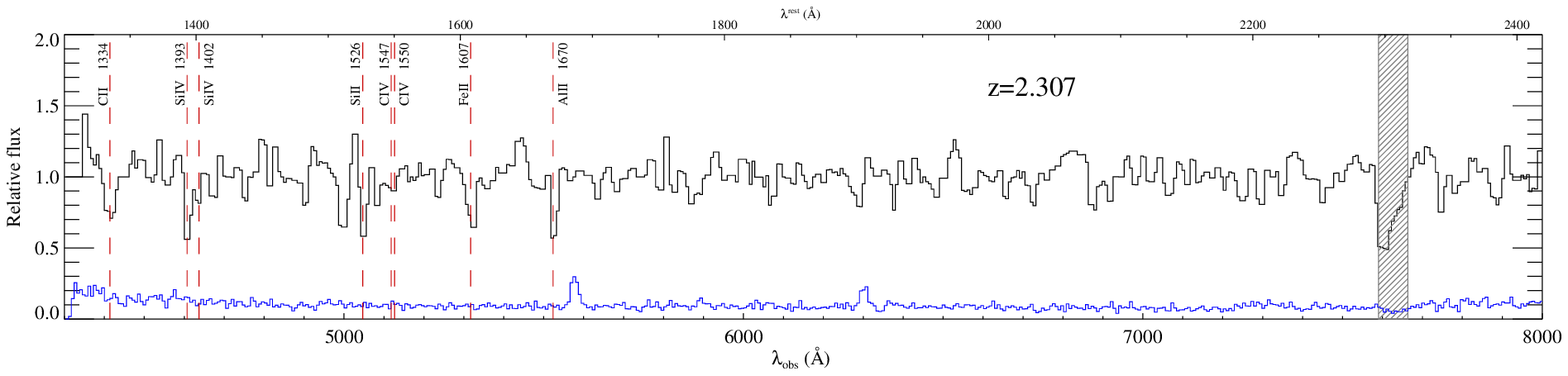} \\
\includegraphics[width=\textwidth, trim= 8mm 3mm 2mm 1mm,clip]{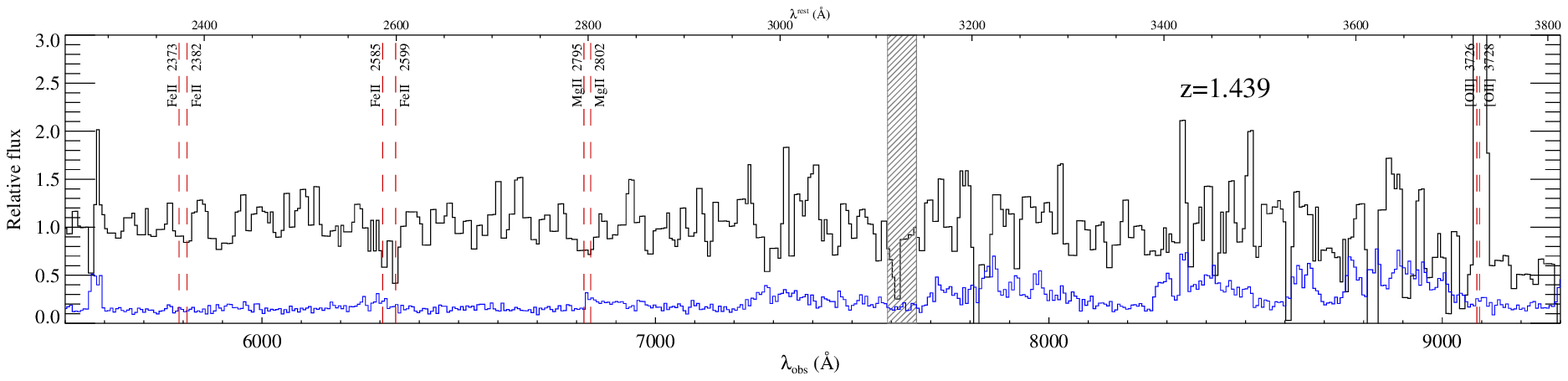} \\
\end{tabular}
\put(-55, 105){\bf \color{black} a)}
\put(-55, -22){\bf \color{black} b)}
\caption{\label{fig:class2}
FORS2/VLT spectra for two sources with medium confidence level, labeled as class 2. The spectra are shown in the same manner as in 
Figure \ref{fig:class3}.
From top to bottom the spectra in each panel correspond to the following sources in Table \ref{table:lensed_galaxy} -- 
a) RCS2 J1250$+$0244, S1; b) RCS2 J2329$-$1317, S1.
} 
\end{center}
\end{figure*}

\begin{figure*}[t!]
\begin{center}
\begin{tabular}{c}
\includegraphics[width=\textwidth, trim= 8mm 3mm 2mm 1mm,clip]{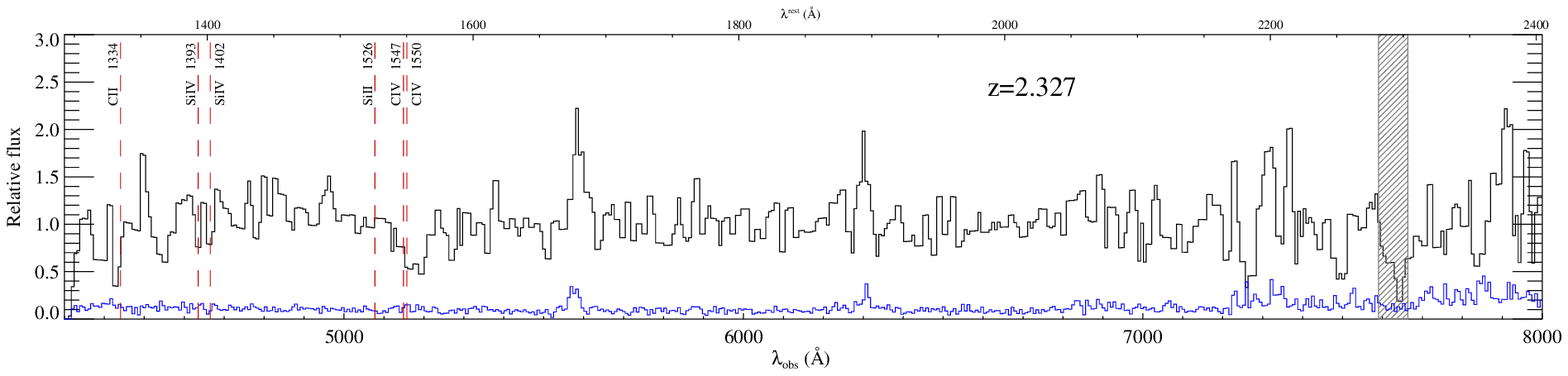} \\
\includegraphics[width=\textwidth, trim= 8mm 3mm 2mm 1mm,clip]{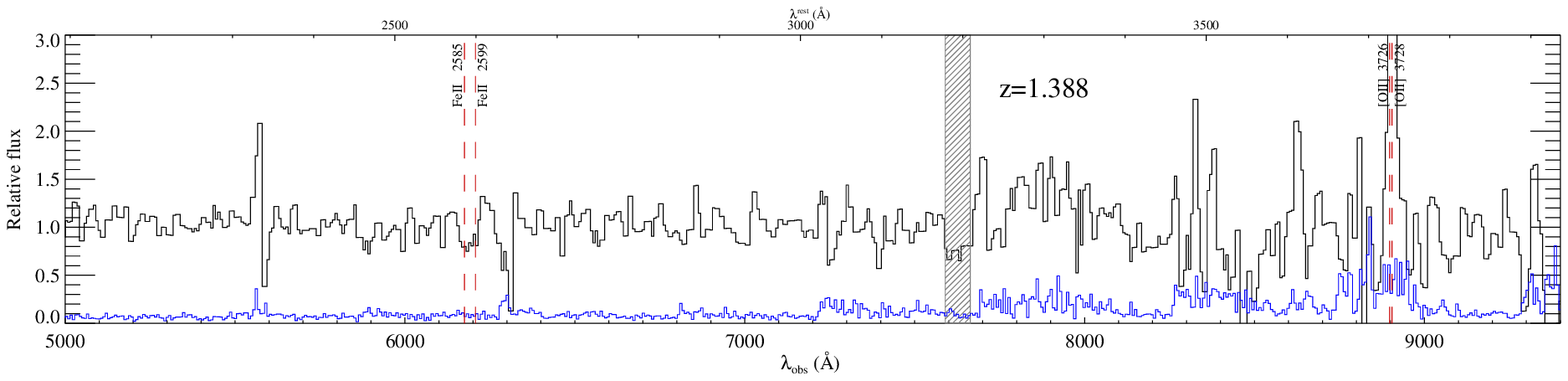} \\
\end{tabular}
\put(-55, 105){\bf \color{black} a)}
\put(-55, -22){\bf \color{black} b)}
\caption{\label{fig:class1}
FORS2/VLT spectra for two sources with low confidence level, labeled as class 1. The spectra are displayed in the same 
fashion as in Figure \ref{fig:class3}.
From top to bottom the spectra in each panel correspond to the following sources in Table \ref{table:lensed_galaxy} -- 
a) RCS2 J2147$+$0102, S1; b) RCS2 J1108$-$0456, S1.
The spectrum in the top panel is mainly identified at $z = 2.327$ by two weak UV metal absorption lines: 
SiIV $\lambda$1393.76\AA{} and SiIV $\lambda$1402.77\AA. The spectrum in the bottom panel is identified at $z = 1.388$ by assuming
the lone robust emission line feature corresponds to [OII] $\lambda$3727.09, 3729.88\AA. There are also 
two weak UV metal absorption lines matching at this redshift: FeII $\lambda$2586.65\AA{} and FeII $\lambda$2600.17\AA.
} 
\end{center}
\end{figure*}

\section{Results} \label{sec:results}

\subsection{Redshift measurements} \label{sec:results_3.1}

The spectroscopic redshifts were determined with two independent methods 
for Magellan and VLT data. In the first procedure, 
galaxy redshifts were  measured by cross-correlating the  
spectra with galaxy spectral templates of the SDSS Data Release 7 
\citep[DR7;][]{Abazajian09} using the RVSAO/XCSAO package for IRAF\footnote{IRAF is distributed by 
the National Optical Astronomy Observatories, which are operated by the Association of Universities 
for Research in Astronomy, Inc., under cooperative agreement with the National Science Foundation.} 
\citep{KurtzMink98}. 
The spectral features in each spectrum were confirmed by visual inspection with the 2D spectra.
This technique yielded accurate redshift measurements for high/medium signal-to-noise spectra. However, most 
of the lensed galaxies have a low signal-to-noise ratio, and 
thus, the redshift results of these cross-correlations have low reliability. In order to determine 
reliable spectroscopic redshift for all lensed galaxies, we used a second method; we  assigned 
redshifts to individual spectra by identifying a set of lines at a common redshift, fitting a 
Gaussian profile to each line in order to determine their central wavelength, and taking the mean 
redshift of the entire set of lines. 
The sets of emission and absorption lines used in the redshift measurements of the lensed galaxies 
varied significantly among the different source spectra \citep{Bayliss11b}.
Most emission lines observed in the giant arc spectra coincided with CIII] $\lambda$1908.73\AA, 
[OII] $\lambda$3727.09, 3729.88\AA, H$\beta$ $\lambda$4862.68\AA, [OIII] $\lambda$4960.30, and 
5008.24\AA, 
which usually correspond to emission lines that 
come from star forming regions, matching the expectations for high-redshift blue galaxies.
Due to the spectral range covered by our instrument setups, these emission lines are observed only 
in lensed galaxies at $z\lesssim1.7$.
For background sources at higher redshifts, we had to rely on the rest-frame UV features to 
determine their redshifts. 
The most common UV metal absorption lines observed in the lensed galaxy spectra were MgII 
$\lambda$2796.35, 2803.53\AA, MgI $\lambda$2852.96\AA, FeII $\lambda$2344.21, 2374.46, 2382.76, 
2586.65, 2600.17\AA, CIV $\lambda$1548.20, 1550.78\AA, SiII $\lambda$1260.42, 1304.37, 1526.71\AA, 
and SiIV $\lambda$1393.76, 1402.77\AA. 
For completeness, this second method was also applied to  rest of the spectra in our sample. In this case, 
the redshift measurements of cluster member galaxies were derived from at least three lines; the 
most used ones correspond to the characteristic lines in older stellar populations (e.g. CaII 
H$\&$K $\lambda$3934.77, 3969.59\AA, g-band $\lambda$4305.61\AA, MgI $\lambda$5168.74, 5174.14, 
5185.04\AA, and NaI $\lambda$5891.61, 5894.13, 5897.57\AA), although some emission lines were
also observed several times (e.g. [OII] $\lambda$3727.09, 3729.88\AA).
The spectral lines used in this analysis were taken from \cite{Shapley03},  \cite{Erb12}, and also 
from the SDSS spectral line tables\footnote{http://www.sdss2.org/dr2/algorithms/linestable.html}. It 
should be noted that all vacuum wavelengths were converted to air wavelengths by applying the IAU 
standard conversion \citep{Morton91}, in order to measure redshifts in the air frame.

Redshift errors are mainly due to the combination of the uncertainty in our wavelength calibrations 
and the statistical uncertainty in the identification of line centers. The median RMS in the 
wavelength calibration is $\sim 1.4$ \AA, which at a central wavelength of $\sim7000$ \AA, results 
in redshift errors of $\sim \pm0.0002$. 
The redshift errors for the high signal-to-noise spectra are distributed around $\sim \pm 0.0005$. 
These errors are in agreement with the expected ones, since we have to add the uncertainties in the 
line center identifications.
However, the errors increase for low signal-to-noise spectra, assuming values of the order of $\sim\pm 
0.001$.

Following the work done by \cite{Bayliss11b}, we classified our redshift measurements into four 
classes, which describe the confidence level of the redshift measurements.
Class 3 redshifts are the highest confidence measurements, typically obtained from more than 4 
absorption and/or emission features. These redshift measurements are secure.
Given to the high signal-to-noise ratio of the cluster member spectra, 
virtually all their redshift measurements  fall into this category. 
Approximately  $55\%$ of the redshift estimates of the lensed galaxies are 
also classified as class 3 spectra.
Five examples of class 3 lensed galaxy spectra are shown in Figure \ref{fig:class3}. Class 2 redshifts
are medium-confidence measurements. These estimates are based on at least 
two prominent lines and/or a larger number of low-significance features. The redshifts with this 
classification are very likely to be the real redshifts of the corresponding spectra, but there is a 
small chance that some of them could have been misidentified. 
Of the order of $21\%$ of the lensed galaxy spectra fall into this classification.
Two examples of class 2 lensed galaxy spectra are 
shown in Figure \ref{fig:class2}. 
Class 1 redshifts are low-confidence measurements, which are based on few low-significance spectral 
features and represent the ``best-guess'' redshift using the available spectral data. Figure 
\ref{fig:class1} shows two examples of class 1 lensed galaxy spectra.
Redshift estimates falling in this category correspond to $< 23\%$ of all lensed galaxy spectra.
Finally, we have the class 0 redshift for those cases where the spectral  analysis shows no evidence 
of spectral features. 
There is only one lensed galaxy spectrum in our sample where the redshift measurement completely failed.

\subsection{Lensed galaxies}
\begin{figure}[h!]
\begin{center}
\includegraphics[width=0.48\textwidth, trim= 8mm 0mm 0mm 0mm,clip]{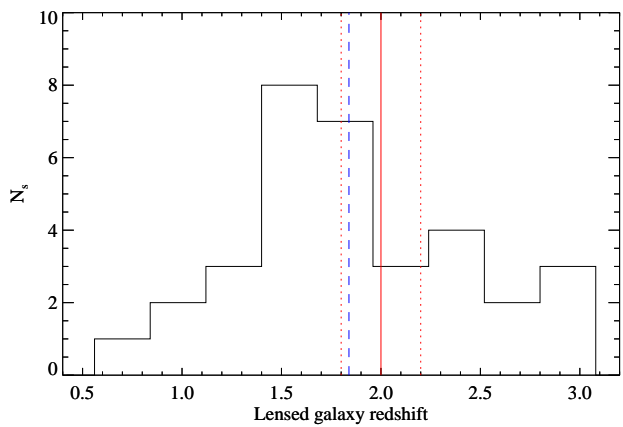} 
\caption{\label{fig:z_dist_arcs}
The redshift distribution of the lensed galaxy sample presented in this work. 
The blue dashed line corresponds to the median redshift of our sample, $z_s = 1.8 \pm 0.1$, while
the red solid line correspond to the median redshift found 
by the spectroscopic and color analysis of high-redshift lensed galaxies - primarily the larger full RCSGA sample - by \cite{Bayliss12}.
} 
\end{center}
\end{figure}

Our spectroscopic analysis has revealed the nature of 52 gravitational arcs present in the core of 
our galaxy clusters, which correspond to 35 background sources at high-redshift that are clearly 
distorted by the gravitational potential of these clusters. 
These lensed galaxies are distributed in a wide redshift range from $0.8 \le z \le 2.9$, 
with a median redshift of ${z}_{s} = 1.8 \pm 0.1$, which is consistent
with the spectroscopic and color analysis of high-redshift lensed galaxies performed by \cite{Bayliss11a} and \cite{Bayliss12}
for hundreds of giant arcs identified in the RCS-2 and SDSS surveys.
It should be noted that $> 75\%$ of the spectra have a confidence level 
in their redshift measurements $\ge 2$.
The redshift distribution of our lensed galaxy sample and the median redshift found by these previous works 
are shown in Figure \ref{fig:z_dist_arcs}.
The redshift measurements of all confirmed lensed galaxies are reported in Table \ref{table:lensed_galaxy},
with labels that correspond to the label markers in Figures 
\ref{fig:SLclusters_fov_01} -- \ref{fig:SLclusters_fov_08}, in Appendix \ref{appendix:SL_clusters}.

Furthermore, this dataset extends the number of galaxy clusters with spectroscopic confirmation 
of their SL features available to perform lensing reconstructions of their mass distribution,
especially at $z \gtrsim 0.2$.

\begin{figure}[h!]
\begin{center}
\includegraphics[width=0.48\textwidth, trim= 8mm 0mm 0mm 5mm,clip]{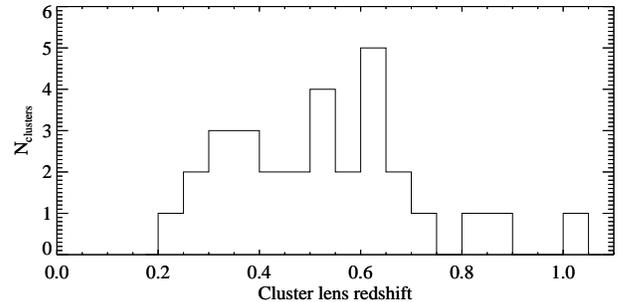} 
\caption{\label{fig:zspec_zphot}
The redshift distribution of our cluster sample, which spans a wide
redshift range $0.22 \le z \le 1.01$.
} 
\end{center}
\end{figure}

\subsection{Cluster redshifts and velocity dispersions}

The correct determination of cluster members is crucial to avoid biases in the velocity dispersion 
and mass measurements \citep{Beers90,Ruel2014}.   
Our selection method of cluster members is an iterative process that starts by applying a cut in 
the (rest-frame) velocity space of $4000$ km s$^{-1}$, centered at the median redshift of all 
candidates. Then, the $3\sigma$ clipping method is applied to remove the interlopers and the median 
redshift is recomputed.  
This process is iterated until the number of members is stable, which usually occurs after 
the second or third iteration. 
We have checked possible systematic effects of this procedure by applying the shifting 
gapper method \citep{Fadda96} to those clusters with a large number of galaxies falling into the 
$\pm4000$ km s$^{-1}$. The results in both methods are fully consistent. 

From this analysis, we have recovered a total of $1004$ spectroscopically confirmed cluster, 
that results in an average of $ \left<N_{\textnormal{mem}}\right> \sim$ 35 member galaxies per cluster. 
The spectroscopic redshift information of all cluster members reported in this work is available as 
supplementary material in the machine-readable format at the Astrophysical Journal.
A portion is shown in Table \ref{table:all_z_cluster_members} for guidance regarding its form and content.
These data have been used to compute robust measurements of redshift and velocity dispersion of our SL-selected galaxy 
clusters by applying the bi-weight estimator for robust statistics \citep{Beers90}.
The errors on redshift and velocity dispersion of each cluster were estimated through many 
bootstrapped realizations, identifying the upper and lower 68$\%$ confidence intervals. 
It should be noted that the errors of individual galaxy redshifts were not 
considered in the velocity dispersion estimates, because the bias introduced by this exclusion 
is $< 0.1\%$  for massive clusters \citep{Danese80} as in our sample.

The redshift distribution of our cluster sample  (Figure \ref{fig:zspec_zphot}) spans a wide redshift range, from $0.22 \le z 
\le 1.01$, making it ideal for studies of evolution of cluster properties. 
The cluster redshift, velocity dispersion, and other properties of each cluster are 
summarized in Table \ref{table:vel_disp_m200_summary}, while the velocity histograms for the 
full sample are shown in the Figure \ref{fig:vel_hist_1}, Figure \ref{fig:vel_hist_2}, and Figure
\ref{fig:vel_hist_3}, in Appendix \ref{appendix:vel_hist}.


\begin{deluxetable*}{l c c c c c}
\tabletypesize{\scriptsize}
\tablecaption{Spectroscopic redshift of individual lensed galaxies \label{table:lensed_galaxy}}
\tablewidth{0pt}
\tablehead{
{Cluster Lens} & 
{Lensed galaxy}$^a$ &
{R.A.}$^b$ & 
{Dec}$^b$ & 
{z}$^c$ & 
{Classification}$^d$ \\ 
{} & 
{} & 
{(J2000)} & 
{(J2000)} & 
{} &
{}
}
\startdata
 SDSS J0004$-$0103  & S1.1  &\hspace{0.5cm}  00 04 51.59\hspace{0.5cm} &\hspace{0.5cm} -01 03 19.8\hspace{0.5cm} &  1.681 &  3  \\
 RCS2 J0034+0225  & S1.1  &\hspace{0.5cm}  00 34 27.35\hspace{0.5cm} &\hspace{0.5cm} +02 25 14.1\hspace{0.5cm} &  2.379 &  3  \\
              --  & S1.2  &\hspace{0.5cm}  00 34 27.39\hspace{0.5cm} &\hspace{0.5cm} +02 25 22.1\hspace{0.5cm} &  2.379 &  3  \\
 RCS2 J0038+0215  & S1.1  &\hspace{0.5cm}  00 38 55.92\hspace{0.5cm} &\hspace{0.5cm} +02 15 48.9\hspace{0.5cm} &  2.817 &  3  \\
              --  & S1.2  &\hspace{0.5cm}  00 38 55.90\hspace{0.5cm} &\hspace{0.5cm} +02 15 56.8\hspace{0.5cm} &  2.817 &  3  \\
 RCS2 J0047+0508  & S1.1  &\hspace{0.5cm}  00 47 51.12\hspace{0.5cm} &\hspace{0.5cm} +05 08 27.7\hspace{0.5cm} &  1.629 &  2  \\
              --  & S2.1  &\hspace{0.5cm}  00 47 51.47\hspace{0.5cm} &\hspace{0.5cm} +05 07 52.1\hspace{0.5cm} &  0.728 &  3  \\
 RCS2 J0052+0433  & S1.1  &\hspace{0.5cm}  00 52 07.73\hspace{0.5cm} &\hspace{0.5cm} +04 33 34.5\hspace{0.5cm} &  1.853 &  2  \\
              --  & S2.1  &\hspace{0.5cm}  00 52 10.94\hspace{0.5cm} &\hspace{0.5cm} +04 33 28.8\hspace{0.5cm} &  1.732 &  2  \\
              --  & S2.2  &\hspace{0.5cm}  00 52 10.57\hspace{0.5cm} &\hspace{0.5cm} +04 33 25.3\hspace{0.5cm} &  1.732 &  2  \\
 RCS2 J0057+0209  & S1.1  &\hspace{0.5cm}  00 57 27.98\hspace{0.5cm} &\hspace{0.5cm} +02 09 26.6\hspace{0.5cm} &  0.775 &  1  \\
              --  & S1.2  &\hspace{0.5cm}  00 57 27.66\hspace{0.5cm} &\hspace{0.5cm} +02 09 27.5\hspace{0.5cm} &  0.775 &  1  \\
 RCS2 J0252$-$1459  & S1.1  &\hspace{0.5cm}  02 52 41.74\hspace{0.5cm} &\hspace{0.5cm} -14 59 33.2\hspace{0.5cm} &  1.096 &  3  \\
 RCS2 J0309$-$1437  & S1.1  &\hspace{0.5cm}  03 09 44.99\hspace{0.5cm} &\hspace{0.5cm} -14 37 16.1\hspace{0.5cm} &  1.519 &  3  \\
              --  & S2.1  &\hspace{0.5cm}  03 09 45.33\hspace{0.5cm} &\hspace{0.5cm} -14 37 16.2\hspace{0.5cm} &  1.413 &  2  \\
 RCS2 J0327$-$1326  & S1.1  &\hspace{0.5cm}  03 27 26.63\hspace{0.5cm} &\hspace{0.5cm} -13 26 15.4\hspace{0.5cm} &  1.701 &  3  \\
              --  & S1.2  &\hspace{0.5cm}  03 27 27.19\hspace{0.5cm} &\hspace{0.5cm} -13 26 54.3\hspace{0.5cm} &  1.701 &  3  \\
              --  & S1.3  &\hspace{0.5cm}  03 27 28.36\hspace{0.5cm} &\hspace{0.5cm} -13 26 15.6\hspace{0.5cm} &  1.701 &  3  \\
 RCS2 J0859$-$0345  & S1.1  &\hspace{0.5cm}  08 59 14.30\hspace{0.5cm} &\hspace{0.5cm} -03 45 12.5\hspace{0.5cm} &  -- --  &  0  \\
 RCS2 J1055$-$0459  & S1.1  &\hspace{0.5cm}  10 55 36.28\hspace{0.5cm} &\hspace{0.5cm} -04 59 41.7\hspace{0.5cm} &  2.804 &  3  \\
              --  & S1.2  &\hspace{0.5cm}  10 55 35.91\hspace{0.5cm} &\hspace{0.5cm} -04 59 38.4\hspace{0.5cm} &  2.804 &  3  \\
 RCS2 J1101$-$0602  & S1.1  &\hspace{0.5cm}  11 01 53.95\hspace{0.5cm} &\hspace{0.5cm} -06 02 31.3\hspace{0.5cm} &  1.674  &  3  \\
 RCS2 J1108$-$0456  & S1.1  &\hspace{0.5cm}  11 08 16.24\hspace{0.5cm} &\hspace{0.5cm} -04 56 23.3\hspace{0.5cm} &  1.521 &  1  \\
              --  & S2.1  &\hspace{0.5cm}  11 08 17.31\hspace{0.5cm} &\hspace{0.5cm} -04 56 17.0\hspace{0.5cm} &  1.388 &  1  \\
 SDSS J1111+1408  & S1.1  &\hspace{0.5cm}  11 11 24.26\hspace{0.5cm} &\hspace{0.5cm} +14 09 01.6\hspace{0.5cm} &  2.141 &  3  \\
 RCS2 J1119$-$0728  & S1.1  &\hspace{0.5cm}  11 19 12.26\hspace{0.5cm} &\hspace{0.5cm} -07 28 14.0\hspace{0.5cm} &  2.062 &  1  \\
 RCS2 J1125$-$0628  & S1.1  &\hspace{0.5cm}  11 25 29.21\hspace{0.5cm} &\hspace{0.5cm} -06 28 48.9\hspace{0.5cm} &  1.518 &  1  \\
              --  & S1.2  &\hspace{0.5cm}  11 25 28.96\hspace{0.5cm} &\hspace{0.5cm} -06 28 49.8\hspace{0.5cm} &  1.518 &  1  \\
 RCS2 J1250+0244  & S1.1  &\hspace{0.5cm}  12 50 42.20\hspace{0.5cm} &\hspace{0.5cm} +02 44 31.2\hspace{0.5cm} &  2.307 &  2  \\
 RCS2 J1511+0630  & S1.1  &\hspace{0.5cm}  15 11 44.39\hspace{0.5cm} &\hspace{0.5cm} +06 30 31.2\hspace{0.5cm} &  1.295 &  1  \\
 SDSS J1517+1003  & S1.1  &\hspace{0.5cm}  15 17 03.75\hspace{0.5cm} &\hspace{0.5cm} +10 03 32.9\hspace{0.5cm} &  2.239 &  3  \\
              --  & S1.2  &\hspace{0.5cm}  15 17 03.95\hspace{0.5cm} &\hspace{0.5cm} +10 03 25.9\hspace{0.5cm} &  2.239 &  3  \\
 SDSS J1519+0840  & S1.1  &\hspace{0.5cm}  15 19 30.04\hspace{0.5cm} &\hspace{0.5cm} +08 40 05.4\hspace{0.5cm} &  2.371 &  3  \\
              --  & S1.2  &\hspace{0.5cm}  15 19 30.07\hspace{0.5cm} &\hspace{0.5cm} +08 39 53.2\hspace{0.5cm} &  2.371 &  3  \\
 RCS2 J1526+0432  & S1.1  &\hspace{0.5cm}  15 26 13.94\hspace{0.5cm} &\hspace{0.5cm} +04 33 02.0\hspace{0.5cm} &  1.443 &  3  \\
              --  & S2.1  &\hspace{0.5cm}  15 26 15.73\hspace{0.5cm} &\hspace{0.5cm} +04 32 42.7\hspace{0.5cm} &  2.636 &  2  \\
 SDSS J2111$-$0114  & S1.1  &\hspace{0.5cm}  21 11 18.91\hspace{0.5cm} &\hspace{0.5cm} -01 14 31.9\hspace{0.5cm} &  2.856 &  3  \\
              --  & S1.2  &\hspace{0.5cm}  21 11 20.22\hspace{0.5cm} &\hspace{0.5cm} -01 14 33.0\hspace{0.5cm} &  2.856 &  3  \\
              --  & S1.3  &\hspace{0.5cm}  21 11 20.40\hspace{0.5cm} &\hspace{0.5cm} -01 14 30.8\hspace{0.5cm} &  2.856 &  3  \\
 SDSS J2135$-$0102  & S1.1  &\hspace{0.5cm}  21 35 12.08\hspace{0.5cm} &\hspace{0.5cm} -01 03 36.7\hspace{0.5cm} &  2.319 &  3  \\
              --  & S1.2  &\hspace{0.5cm}  21 35 11.46\hspace{0.5cm} &\hspace{0.5cm} -01 03 34.3\hspace{0.5cm} &  2.319 &  3  \\
              --  & S1.3  &\hspace{0.5cm}  21 35 09.94\hspace{0.5cm} &\hspace{0.5cm} -01 03 18.0\hspace{0.5cm} &  2.319 &  3  \\
              --  & S2.1  &\hspace{0.5cm}  21 35 10.17\hspace{0.5cm} &\hspace{0.5cm} -01 03 33.8\hspace{0.5cm} &  0.903 &  3  \\
 RCS2 J2147$-$0102  & S1.1  &\hspace{0.5cm}  21 47 37.12\hspace{0.5cm} &\hspace{0.5cm} -01 02 56.3\hspace{0.5cm} &  2.327 &  1  \\
 RCS2 J2151$-$0138  & S1.1  &\hspace{0.5cm}  21 51 26.87\hspace{0.5cm} &\hspace{0.5cm} -01 38 41.1\hspace{0.5cm} &  0.835 &  2  \\
 SDSS J2313$-$0104  & S1.1  &\hspace{0.5cm}  23 13 54.54\hspace{0.5cm} &\hspace{0.5cm} -01 04 56.6\hspace{0.5cm} &  1.845 &  1  \\
 RCS2 J2329$-$0120  & S1.1  &\hspace{0.5cm}  23 29 47.18\hspace{0.5cm} &\hspace{0.5cm} -01 20 45.4\hspace{0.5cm} &  1.790 &  2  \\
              --  & S2.1  &\hspace{0.5cm}  23 29 47.87\hspace{0.5cm} &\hspace{0.5cm} -01 20 53.6\hspace{0.5cm} &  1.570 &  3  \\
              --  & S2.2  &\hspace{0.5cm}  23 29 47.74\hspace{0.5cm} &\hspace{0.5cm} -01 20 53.6\hspace{0.5cm} &  1.570 &  3  \\
 RCS2 J2329$-$1317  & S1.1  &\hspace{0.5cm}  23 29 10.33\hspace{0.5cm} &\hspace{0.5cm} -13 17 44.3\hspace{0.5cm} &  1.441 &  2  \\
              --  & S1.2  &\hspace{0.5cm}  23 29 10.03\hspace{0.5cm} &\hspace{0.5cm} -13 17 41.7\hspace{0.5cm} &  1.441 &  2  \\
 RCS2 J2336$-$0608  & S1.1  &\hspace{0.5cm}  23 36 20.54\hspace{0.5cm} &\hspace{0.5cm} -06 08 38.4\hspace{0.5cm} &  1.295 &  1  \\
              --  & S1.2  &\hspace{0.5cm}  23 36 20.38\hspace{0.5cm} &\hspace{0.5cm} -06 08 33.8\hspace{0.5cm} &  1.295 &  1  \\
\enddata
\tablenotetext{a}{Lensed galaxy labels that correspond to the label markers in Figures 
\ref{fig:SLclusters_fov_01} -- \ref{fig:SLclusters_fov_08}, in Appendix \ref{appendix:SL_clusters}.
S1.X spectra correspond to the "primary" arcs of each cluster, following the notation in \cite{Bayliss11b}.}
\tablenotetext{b}{Coordinates of the lensed galaxies in sexagesimal degrees (J2000).}
\tablenotetext{c}{Spectroscopic redshift.}
\tablenotetext{d}{Classification of the redshift measurements, as discussed in Section \ref{sec:results_3.1}.}
\end{deluxetable*}

\begin{deluxetable}{l c c c}
\tabletypesize{\scriptsize}
\tablecaption{The VLT/Magellan spectroscopic redshift catalog of the cluster members reported in this work \label{table:all_z_cluster_members} }
\tablewidth{0pt}
\tablehead{
{Cluster name}$^a$ & {R.A.}$^b$ &  {Dec.}$^b$ & {$z$}$^c$ \\
{} & {(J2000)} & {(J2000)} & {} }
\startdata
RCS2 J0004$-$0103  &   1.16403723   &   -1.03105652   &  0.5175    \\
RCS2 J0004$-$0103  &   1.18741393   &   -1.03259635   &  0.5172    \\
RCS2 J0004$-$0103  &   1.21820974   &   -1.07451379   &  0.5107    \\
RCS2 J0004$-$0103  &   1.25292516   &   -1.05631912   &  0.5148    \\
RCS2 J0004$-$0103  &   1.26846313   &   -1.05855823   &  0.5103    \\
RCS2 J0004$-$0103  &   1.26440382   &   -1.06821537   &  0.5157    \\
RCS2 J0004$-$0103  &   1.26328409   &   -1.07717276   &  0.5162    \\
RCS2 J0004$-$0103  &   1.17243600   &   -1.03644502   &  0.5140    \\
RCS2 J0004$-$0103  &   1.17733526   &   -1.03910434   &  0.5127    \\
RCS2 J0034+0225    &   8.60755825   &    2.47035551   &  0.3845    \\
\enddata
\tablecomments{Table \ref{table:all_z_cluster_members} is published in its entirety in the machine-readable format.
A portion is shown here for guidance regarding its form and content. The full table contains 4 columns and 1004 redshifts.}
\tablenotetext{a}{Cluster identifier.}
\tablenotetext{b}{Coordinates of the cluster member galaxies in decimal degrees (J2000).}
\tablenotetext{c}{Spectroscopic redshift.}
\end{deluxetable}

\begin{deluxetable*}{l c c c c c c c c c}
\tabletypesize{\scriptsize}
\tablecaption{Summary of dynamical properties of the SL-selected galaxy clusters \label{table:vel_disp_m200_summary}}
\tablewidth{0pt}
\tablehead{
{Name} & {R.A.$^a$} & {Dec$^a$} & {$z$} & {N$_{mem}$} & {$\sigma_{c}${}$^b$} & {$M_{200}$}  & {$r_{200}$} & {\textit{P}-v (DS)$^c$} & {\textit{P}-v (KS)$^d$}\\ 
{} &  {(J2000)} &  {(J2000)} & {} & {} &{[km s$^{-1}]$} & {$[h^{-1}_{70}10^{14}M_{\odot}]$} & {$[h^{-1}_{70}Mpc]$} & {}  & {}
}
\startdata                                                                                                                                                           
SDSS J0004$-$0103$^e$      & 00 04 52.001  &  $-$01 03 16.58  &  0.5144   &   9  &  502 $\pm$ 144  &   1.10  $\pm$  0.71   &   0.83  $\pm$  0.23  & -- --                   &     -- --                \\   
RCS2 J0034+0225            & 00 34 28.134  &    +02 25 22.34  &  0.3842   &  26  &  713 $\pm$ 179  &   3.37  $\pm$  2.19   &   1.26  $\pm$  0.31  & $0.10^{+0.06}_{-0.03}$  &   $0.25^{+0.08}_{-0.10}$  \\  
RCS2 J0038+0215$^g$        & 00 38 55.898  &    +02 15 52.35  &  0.6959   &  20  &  778 $\pm$ 105  &   3.63  $\pm$  1.47   &   1.14  $\pm$  0.16  & $0.04^{+0.10}_{-0.02}$  &   $0.52^{+0.14}_{-0.23}$  \\  
RCS2 J0047+0508            & 00 47 50.787  &    +05 08 20.02  &  0.4286   &  50  &  738 $\pm$  94  &   3.64  $\pm$  1.40   &   1.27  $\pm$  0.16  & $0.24^{+0.06}_{-0.06}$  &   $0.79^{+0.05}_{-0.15}$  \\  
RCS2 J0052+0433            & 00 52 10.352  &    +04 33 33.31  &  0.7248   &  24  &  618 $\pm$ 166  &   1.80  $\pm$  1.17   &   0.89  $\pm$  0.24  & $0.12^{+0.06}_{-0.04}$  &   $0.31^{+0.26}_{-0.18}$  \\  
RCS2 J0057+0209$^g$        & 00 57 27.869  &    +02 09 33.98  &  0.2928   &  19  &  940 $\pm$ 160  &   8.07  $\pm$  4.11   &   1.74  $\pm$  0.30  & $0.09^{+0.08}_{-0.05}$  &   $0.31^{+0.17}_{-0.20}$  \\  
RCS2 J0252$-$1459$^g$      & 02 52 41.474  &  $-$14 59 30.38  &  0.2646   &  36  &  931 $\pm$ 121  &   8.00  $\pm$  3.10   &   1.76  $\pm$  0.23  & $0.05^{+0.07}_{-0.03}$  &   $0.54^{+0.11}_{-0.16}$  \\  
RCS2 J0309$-$1437$^{f,g}$  & 03 09 44.096  &  $-$14 37 34.38  &  0.8079   &  27  &  620 $\pm$  56  &   1.74  $\pm$  0.47   &   0.86  $\pm$  0.07  & $0.02^{+0.06}_{-0.01}$  &   $0.04^{+0.12}_{-0.02}$  \\  
RCS2 J0327$-$1326$^g$      & 03 27 27.174  &  $-$13 26 22.90  &  0.5633   &  44  & 1029 $\pm$ 122  &   9.04  $\pm$  3.20   &   1.63  $\pm$  0.19  & $0.05^{+0.05}_{-0.02}$  &   $0.49^{+0.09}_{-0.14}$  \\  
RCS2 J0859$-$0345$^e$      & 08 59 14.486  &  $-$03 45 14.63  &  0.6483   &   8  &  912 $\pm$ 555  &   6.01  $\pm$  3.90   &   1.37  $\pm$  0.83  & -- --                   &    -- --                 \\  
RCS2 J1055$-$0459          & 10 55 35.647  &  $-$04 59 41.60  &  0.6076   &  40  &  720 $\pm$  68  &   3.04  $\pm$  0.86   &   1.11  $\pm$  0.10  & $0.21^{+0.05}_{-0.08}$  &   $0.89^{+0.04}_{-0.32}$  \\  
RCS2 J1101$-$0602$^e$      & 11 01 54.093  &  $-$06 02 32.02  &  0.4861   &   8  &  573 $\pm$ 201  &   1.66  $\pm$  1.09   &   0.96  $\pm$  0.33  & -- --                   &    -- --                  \\  
RCS2 J1108$-$0456          & 11 08 16.835  &  $-$04 56 37.62  &  0.4088   &  51  &  643 $\pm$  52  &   2.46  $\pm$  0.60   &   1.11  $\pm$  0.09  & $0.12^{+0.06}_{-0.04}$  &   $0.89^{+0.07}_{-0.19}$  \\  
SDSS J1111+1408$^g$        & 11 11 24.483  &    +14 08 50.82  &  0.2204   &  20  &  908 $\pm$ 121  &   7.59  $\pm$  3.01   &   1.74  $\pm$  0.23  & $0.08^{+0.10}_{-0.04}$  &   $0.74^{+0.10}_{-0.29}$  \\  
RCS2 J1119$-$0728$^e$      & 11 19 11.925  &  $-$07 28 17.51  &  1.0128   &   4  &  670 $\pm$ 338  &   1.93  $\pm$  1.26   &   0.81  $\pm$  0.41  & -- --                   &   -- --                   \\  
RCS2 J1125$-$0628          & 11 25 28.940  &  $-$06 28 39.04  &  0.4746   &  33  &  738 $\pm$ 127  &   3.56  $\pm$  1.81   &   1.23  $\pm$  0.21  & $0.16^{+0.03}_{-0.04}$  &   $0.15^{+0.09}_{-0.08}$  \\  
RCS2 J1250+0244            & 12 50 41.890  &    +02 44 26.57  &  0.6902   &  33  &  636 $\pm$  96  &   2.00  $\pm$  0.90   &   0.93  $\pm$  0.14  & $0.15^{+0.06}_{-0.06}$  &   $0.31^{+0.05}_{-0.11}$  \\  
RCS2 J1511+0630$^e$        & 15 11 44.681  &    +06 30 31.79  &  0.5513   &  15  &  444 $\pm$ 210  &   0.74  $\pm$  0.49   &   0.71  $\pm$  0.33  & -- --                   &   -- --                   \\  
SDSS J1517+1003            & 15 17 02.587  &    +10 03 29.27  &  0.6435   &  57  &  789 $\pm$  88  &   3.91  $\pm$  1.30   &   1.19  $\pm$  0.13  & $0.29^{+0.10}_{-0.04}$  &   $0.49^{+0.12}_{-0.09}$  \\  
SDSS J1519+0840$^g$        & 15 19 31.213  &    +08 40 01.43  &  0.3177   &  49  & 1177 $\pm$  91  &  15.57  $\pm$  3.61   &   2.14  $\pm$  0.17  & $0.04^{+0.08}_{-0.02}$  &   $0.35^{+0.07}_{-0.10}$  \\  
RCS2 J1526+0432            & 15 26 14.914  &    +04 32 48.01  &  0.6341   &  30  &  735 $\pm$  89  &   3.19  $\pm$  1.16   &   1.11  $\pm$  0.13  & $0.18^{+0.07}_{-0.09}$  &   $0.43^{+0.08}_{-0.12}$  \\  
SDSS J2111$-$0114          & 21 11 19.307  &  $-$01 14 23.95  &  0.6361   &  46  & 1014 $\pm$  88  &   8.29  $\pm$  2.14   &   1.53  $\pm$  0.13  & $0.21^{+0.05}_{-0.11}$  &   $0.88^{+0.06}_{-0.16}$  \\  
SDSS J2135$-$0102$^g$      & 21 35 12.040  &  $-$01 02 58.27  &  0.3272   &  93  & 1234 $\pm$  83  &  17.86  $\pm$  3.60   &   2.23  $\pm$  0.16  & $0.03^{+0.06}_{-0.02}$  &   $0.31^{+0.19}_{-0.14}$  \\  
RCS2 J2147$-$0102          & 21 47 37.172  &  $-$01 02 51.93  &  0.8801   &  26  &  386 $\pm$  79  &   0.40  $\pm$  0.24   &   0.51  $\pm$  0.10  & $0.11^{+0.05}_{-0.04}$  &   $0.16^{+0.08}_{-0.10}$  \\  
RCS2 J2151$-$0138          & 21 51 25.950  &  $-$01 38 50.14  &  0.3160   &  53  & 1063 $\pm$  99  &  11.57  $\pm$  3.20   &   1.94  $\pm$  0.19  & $0.14^{+0.08}_{-0.06}$  &   $0.91^{+0.05}_{-0.19}$  \\  
SDSS J2313$-$0104          & 23 13 54.514  &  $-$01 04 48.46  &  0.5271   &  26  &  681 $\pm$ 112  &   2.70  $\pm$  1.33   &   1.10  $\pm$  0.19  & $0.10^{+0.07}_{-0.04}$  &   $0.25^{+0.07}_{-0.11}$  \\  
RCS2 J2329$-$1317$^g$      & 23 29 09.528  &  $-$13 17 49.26  &  0.3910   &  80  & 1019 $\pm$ 102  &   9.73  $\pm$  2.91   &   1.79  $\pm$  0.17  & $0.03^{+0.06}_{-0.02}$  &   $0.31^{+0.05}_{-0.09}$  \\  
RCS2 J2329$-$0120          & 23 29 47.782  &  $-$01 20 46.89  &  0.5311   &  40  & 1009 $\pm$ 109  &   8.70  $\pm$  2.80   &   1.63  $\pm$  0.17  & $0.15^{+0.04}_{-0.05}$  &   $0.89^{+0.03}_{-0.15}$  \\  
RCS2 J2336$-$0608$^g$      & 23 36 20.838  &  $-$06 08 35.81  &  0.3924   &  36  & 1041 $\pm$ 113  &  10.43  $\pm$  3.37   &   1.81  $\pm$  0.20  & $0.06^{+0.05}_{-0.03}$  &   $0.88^{+0.04}_{-0.24}$  \\  
\enddata   
\tablenotetext{a}{Coordinates are BCG centroids in sexagesimal degrees (J2000).}
\tablenotetext{b}{The rest-frame velocity dispersion of the clusters.}
\tablenotetext{c}{\textit{P}-values of the DS test. The uncertainties are given by the upper and lower $68\%$ confidence intervals. The threshold for substructure detection is set to a s.l. of 0.05 within uncertainties. }
\tablenotetext{d}{\textit{P}-values of the KS test. The uncertainties are given by the upper and lower $68\%$ confidence intervals. The threshold for rejecting the null hypothesis is set to a s.l. of 0.05 within uncertainties.}
\tablenotetext{e}{Systems with few spectroscopically confirmed cluster members, $N_{\textnormal{mem}} \lesssim 20$. Their velocity dispersion and mass measurements represent only a first guess and should not be considered as the final estimates.}
\tablenotetext{f}{The null hypothesis of the KS test is rejected, suggesting that the velocity data of this system is not consistent with a uni-modal distribution. Its velocity dispersion and mass measurements represent only a first guess and should not be considered as the final estimates.}
\tablenotetext{g}{Systems showing signs of dynamical substructure based on DS test results.}
\end{deluxetable*}

\section{Dynamical masses} \label{sec:mass_clusters}
Dynamical information of galaxy clusters offers an unique possibility for estimating the
virial mass of these systems through the relationship between the velocity dispersion
of galaxy members and the cluster mass.
Mass estimates based on simple variations of the virial theorem are biased high by a 
factor of $10-20\%$ compared with masses obtained from the Jeans analysis \citep{Carlberg97} and the 
caustic technique \citep{Diaferio_Geller97}. 
In order to account for this bias and obtain an universal virial scaling relation for massive 
DM halos, \cite{Evrard08} studied an ensemble of cold DM simulations in a variety of cosmologies. 
They concluded that the large majority ($\sim90\%$) of massive halos 
($M_{200} \ge 10^{14} M_{\odot}$) are, on average and in all cosmologies, consistent with 
a virialized state and obey a power-law relation between one-dimensional DM particle velocity 
dispersion, $\sigma_{DM}$, and halo mass. 
Accordingly, the mass enclosed within the virial radius, $r_{200}$, scales as

\begin{equation}\label{eq:Mass_Evrard08}
M_{200} = \frac{10^{15}}{h(z)} \left(\frac{\sigma_{DM}}{\sigma_{15}} \right)^{1/\alpha}  M_{\odot},
\end{equation}

\noindent  where $\sigma_{15} = 1082.9 \pm 4.0$ km s$^{-1}$ is the normalization for a halo mass of 
$10^{15}h^{-1} M_{\odot}$, $\alpha = 0.3361 \pm 0.0026$ is the logarithmic slope,  
and $h(z) = H(z) / 100 $ km s$^{-1}$ Mpc$^{-1}$ is the normalized Hubble parameter at redshift $z$ for a flat universe.

Computing dynamical mass from Eq. \ref{eq:Mass_Evrard08} requires a good understanding of the relationship
between the DM particle velocity dispersion of cluster halos and the velocity dispersion of cluster members, $\sigma_c$,
often parameterized by the velocity bias $b_v = \sigma_c / \sigma_{DM}$.
Previous studies based on simulation have shown that this velocity bias assumes values in the range $b_v \sim 1.0 -1.3$
\citep{Diemand04, Colin00, Ghigna00}, although more recent simulations have indicated 
that the way in which subhalos are tracked and defined affects the resulting velocity bias predictions
\citep{Evrard08,White10}. In particular, subhalos that are treated in this way show no evidence of a possible velocity bias, i.e.
$b_v \sim 1.0$. For consistency with these latest simulation results, we assume that galaxies are unbiased tracers of the 
total cluster mass (adopting $b_v = 1$), and derive dynamical masses of all clusters in our sample by applying 
Eq. \ref{eq:Mass_Evrard08}. 
The mass range of our SL sample goes from $5  \times10^{13} h^{-1}_{70}M_{\odot}$ to $1.9\times10^{15} h^{-1}_{70}M_{\odot}$.
This wide mass range puts our sample in an excellent position to study relationships between
the halo mass of galaxy clusters and their properties, such as the concentration, ellipticity, triaxiality, etc.

The uncertainties in the mass estimates include both the systematic and statistical errors introduced by
the $\sigma_{DM} - M_{200}$ scaling relation, redshift measurements, and velocity dispersion estimates.
The potential biases due to the lack of cluster members and line-of-sight projection together with other
effects that might affect our measurements will be analyzed in the next section. 
The mass estimates and their respective errors are reported in the Table \ref{table:vel_disp_m200_summary}.

\section{Discussion} \label{sec:discussion}

Before we discuss the implications of these results we explore some effects that may affect our measurements.
The most important factors that might  bias the results are the presence of substructure in our clusters,
the region where cluster members were selected for the spectroscopic follow-up, and the final number of confirmed members 
used in the velocity dispersion estimates.
The other biases are the result of assumptions made about the isotropy of galaxy orbits and
the nature of the lensing cluster population itself. 

\subsection{Substructure}

Numerical and observational studies have shown that a significant fraction of galaxy clusters
contain substructure, which is frequently attributed to the active merging histories 
of massive halos \citep{White10,Battaglia12}. 
However, this does not affect the velocity dispersion measurements
when the clusters under consideration have a uni-modal velocity distribution \citep{Girardi97}, 
i.e. when the substructure is only marginal and not of comparable size to the cluster itself 
(having $<10\%$ of the cluster members in an average extension of $\sim 0.2$ Mpc h$^{-1}$).

In order to search for substructure in our sample we apply the DS test \citep{Dressler_Schectman88}
to those clusters with high enough number of member galaxies ($N_{\textnormal{mem}} \gtrsim 20$). 
The DS test has been proven as one of the most sensitive tests for dynamical substructure \citep{Pinkney96} 
and widely used in the literature \citep{Cen97,Knebe_Muller00,White10,Hou09, Hou12,Sifon12}.
The test is based on the detection of localized subgroups of galaxies that deviate from the global distribution
of velocities by using the substructure estimator $\Delta = \sum_i\delta_i$, with

\begin{equation}\label{eq:DS_test}
\delta_i^2 = \left(\frac{N_{\textnormal{local}} +1}{\sigma_c^2}\right) \left[(\bar{v}_i - \bar{v})^2 + (\sigma_i - \sigma_c)^2\right],
\end{equation}

\noindent where $\bar{v}_i$ and $\sigma_i$ correspond to the mean and standard deviation of the velocity
distribution of the
$N_{\textnormal{local}}$ members closest to the $i$th member (included), while $\bar{v}$ and $\sigma_c$ 
correspond to the mean velocity and
velocity dispersion using all available cluster members, respectively. 
The null hypothesis of the DS test is no correlation between position and velocity, i.e. the mean velocity 
and dispersion should
be the same locally as globally (within counting statistic).
The \textit{P}-values for the DS test are calculated as in \cite{Hou12}; by comparing the observed
substructure
estimator $\Delta_{\textnormal{obs}}$ to the shuffled values 
$\Delta_{\textnormal{shuffled}}$, which are computed by randomly shuffling the observed velocities 
and reassigning these values to
the member positions via 1000 Monte Carlo (MC) simulations, and by taking 
$N_{\textnormal{local}} = \sqrt{N_{\textnormal{mem}}}$. 
The \textit{P}-values are given by

\begin{equation}\label{eq:p_values_DS}
P =  \sum  \left( \Delta_{\textnormal{shuffled}} - \Delta_{\textnormal{obs}} \right)/ n_{\textnormal{shuffle}},
\end{equation}

\noindent where  $n_{\textnormal{shuffle}}$ is the number of MC simulations used to compute the probability. 
The statistic of the DS test is performed by 100 realizations of the process above, with the central value
given by the mean of the \textit{P}-values distribution and the uncertainties given by the upper and lower $68\%$ confidence intervals.
 One can see from Eq. \ref{eq:p_values_DS} that clusters with significant substructure will have low \textit{P}-values, 
since it is unlikely to obtain $\Delta_{\textnormal{obs}}$ randomly. We have therefore set the threshold for substructure detection
to a significance level (s.l.) of 0.05 within uncertainties, where false detections are not expected given the size of our sample \citep{Pinkney96,Sifon12}.
 The results are listed in Table \ref{table:vel_disp_m200_summary}.
There are 10 out of 24 clusters ($N_{\textnormal{mem}} \gtrsim 20$) that have rejected the null hypothesis, indicating signs of dynamical substructure
which is consistent with previous optical and X-ray studies
of local clusters \citep{Girardi97, Schuecker01}.
However, high values of $\Delta$ might also be obtained when the velocities are shuffled,
leading to a higher probability-to-exceed or a lower significance detection of substructure,
resulting in erroneous assumptions that are consistent with no substructure \citep{White10}.
Furthermore, the DS test (and almost all substructure indicators) is highly viewing-angle
dependent which complicates the inferences about the dynamical state of the clusters.

Since substructure does not affect the velocity dispersion measurements when 
the clusters under consideration have a uni-modal velocity distribution \citep{Girardi97, Evrard08, Sifon12}, 
we apply the Kolmogorov-Smirnov test \citep[KS;][ and references therein]{Hou_etal2009_KS}
to those clusters in our sample with $N_{\textnormal{mem}} \gtrsim 20$, to corroborate whether their velocity distributions 
are consistent with a  Gaussian distribution.
The KS test is a non-parametric hypothesis test 
based on the \textit{supremum} statistics that measures the distance between the
empirical distribution function (EDF) of a sample and the cumulative distribution function (CDF) of a chosen reference distribution.  
We set the mean and variance of the reference distribution to the mean velocity and
velocity dispersion squared of each cluster, respectively.
The statistic of the KS test is calculated by 5000 bootstrapped realizations of the velocity data of each cluster, with the final probability
given by the mean of the \textit{P}-values distribution and the uncertainties given by the upper and lower $68\%$ confidence intervals.
The threshold for rejecting the null hypothesis, that the sample is drawn from the reference distribution,
is set to a s.l. of 0.05 within uncertainties.
The velocity distributions of 23 out of 24 clusters (with $N_{\textnormal{mem}} \gtrsim 20$) show high \textit{P}-values, 
suggesting that the deviation between the velocity data of these clusters and a Gaussian distribution 
is smaller than one would expect to arise from two different distributions. 
We therefore assume that the velocity distributions of these 23 SL clusters are uni-modal; 
consequently, we can conclude that their velocity dispersion measurements are robust 
and that their the mass estimates are representative of their total masses, 
i.e. Eq. \ref{eq:Mass_Evrard08}  is valid for this subset.
The results are listed in Table \ref{table:vel_disp_m200_summary}.
For completeness, we also apply a Gaussian fit to the binned data of the velocity histograms of these clusters.  
In all cases, the best-fitting Gaussian parameters are consistent 
with the mean velocity and velocity dispersion squared within $1\sigma$, although the fit usually depends on the choice of the bin width. 
Their velocity histograms together with their respective Gaussian fits are shown 
in Figure \ref{fig:vel_hist_1} and Figure \ref{fig:vel_hist_2}, in Appendix \ref{appendix:vel_hist}.

RCS2 J0309$-$1437 is the only system in our sample (with $N_{\textnormal{mem}} \gtrsim 20$) where the null hypothesis of the KS test is rejected, 
suggesting that its velocity data is not consistent with a uni-modal distribution. 
Nonetheless, we list its velocity dispersion and virial mass in table \ref{table:vel_disp_m200_summary}, 
along with the estimates for five systems with relatively few spectroscopically confirmed cluster members ($\left<N_{\textnormal{mem}}\right> \sim10$).
It should be noted that the measurements for these six systems represent only a first guess and should not be considered as the final estimates.
The velocity histograms for these clusters are shown in Figure \ref{fig:vel_hist_3}, in Appendix \ref{appendix:vel_hist}.

\begin{figure}[h!]
\begin{center}
\includegraphics[width=0.48\textwidth, trim= 8mm 0mm 0mm 0mm,clip]{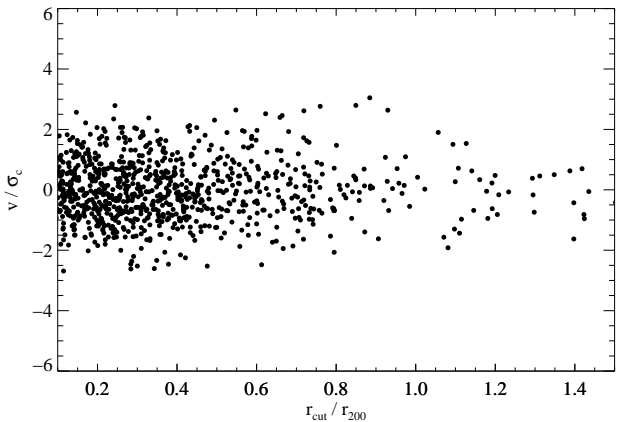} 
\caption{\label{fig:velocities_vs_r_mem}
The stacked analysis of our 29 galaxy clusters. The peculiar velocity of each galaxy  
normalized by the velocity dispersion $\sigma_c$ of the corresponding cluster versus 
the cluster-centric distance normalized by $r_{200}$ for each cluster is shown. 
} 
\end{center}
\end{figure}

\subsection{Region of cluster members}

Observational and theoretical studies have shown that velocity dispersion profiles are flat
from $\sim 0.6 - 0.8$ $r_{200}$ outward \citep{Fadda96, Biviano_Girardi03,Faltenbacher_Diemand06},
and highly biased in the innermost regions of the clusters
($\sim 0.2$ $r_{200}$). Hence, estimating the effect of this potential sampling bias requires the knowledge of 
the virial radius for each cluster and also its projected angular size on the sky. 
We derive the virial radius of all clusters in our sample by assuming  spherical 
symmetry\footnote{The virial radii of our cluster sample are listed in Table \ref{table:vel_disp_m200_summary} } 
and using the previous results of $M_{200}$ (i.e. $M_{200} = 200 \rho_c \times 4\pi r^3_{200} / 3$),
resulting in a median virial radius of $r_{200} = 0.88$ Mpc h$^{-1}$ that corresponds to an
angular size on the sky of $\theta_{r_{200}} = 3.5'$. 
Since our field of view has an average angular radius of $\sim 3.2'$ from the cluster
centers, we are sampling cluster galaxies till approximately $0.9$ $r_{200}$ of the clusters. 
Therefore, the field restriction should not bias our results because the flatness 
of the velocity dispersion profiles begins to smaller radii. 
Furthermore, since we are mainly sampling cluster members on average within
$\sim0.9$ $r_{200}$ (Figure \ref{fig:velocities_vs_r_mem}), the probability to 
include interlopers is approximately $30\%$ lower than for sampling till twice $r_{200}$ \citep{White10, Saro13}. 

\begin{figure}[t!]
\begin{center}
\includegraphics[width=0.48\textwidth, trim= 8mm 13.2mm 0mm 0mm,clip]{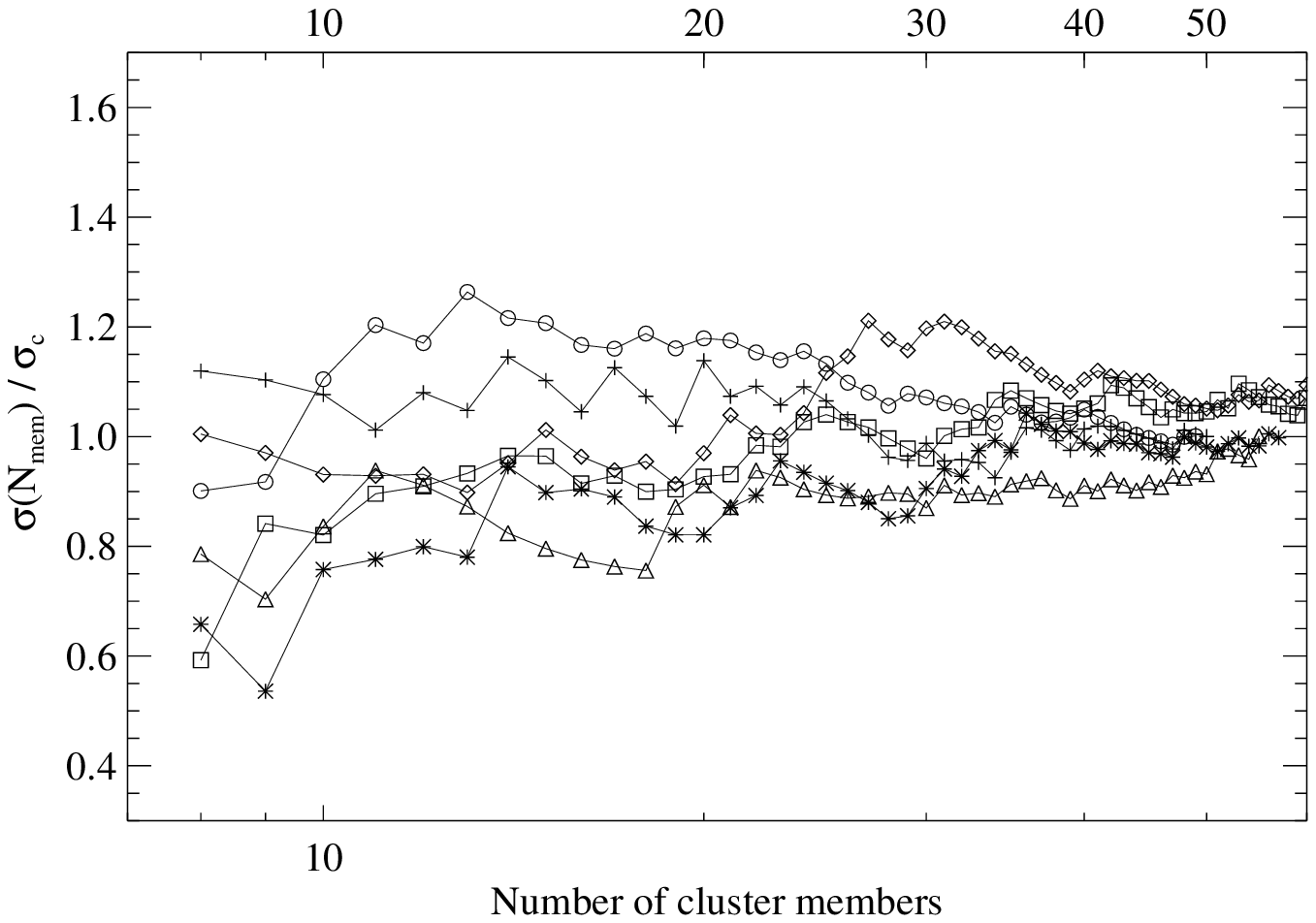}\\
\includegraphics[width=0.48\textwidth, trim= 8mm 0mm 0mm 7mm,clip]{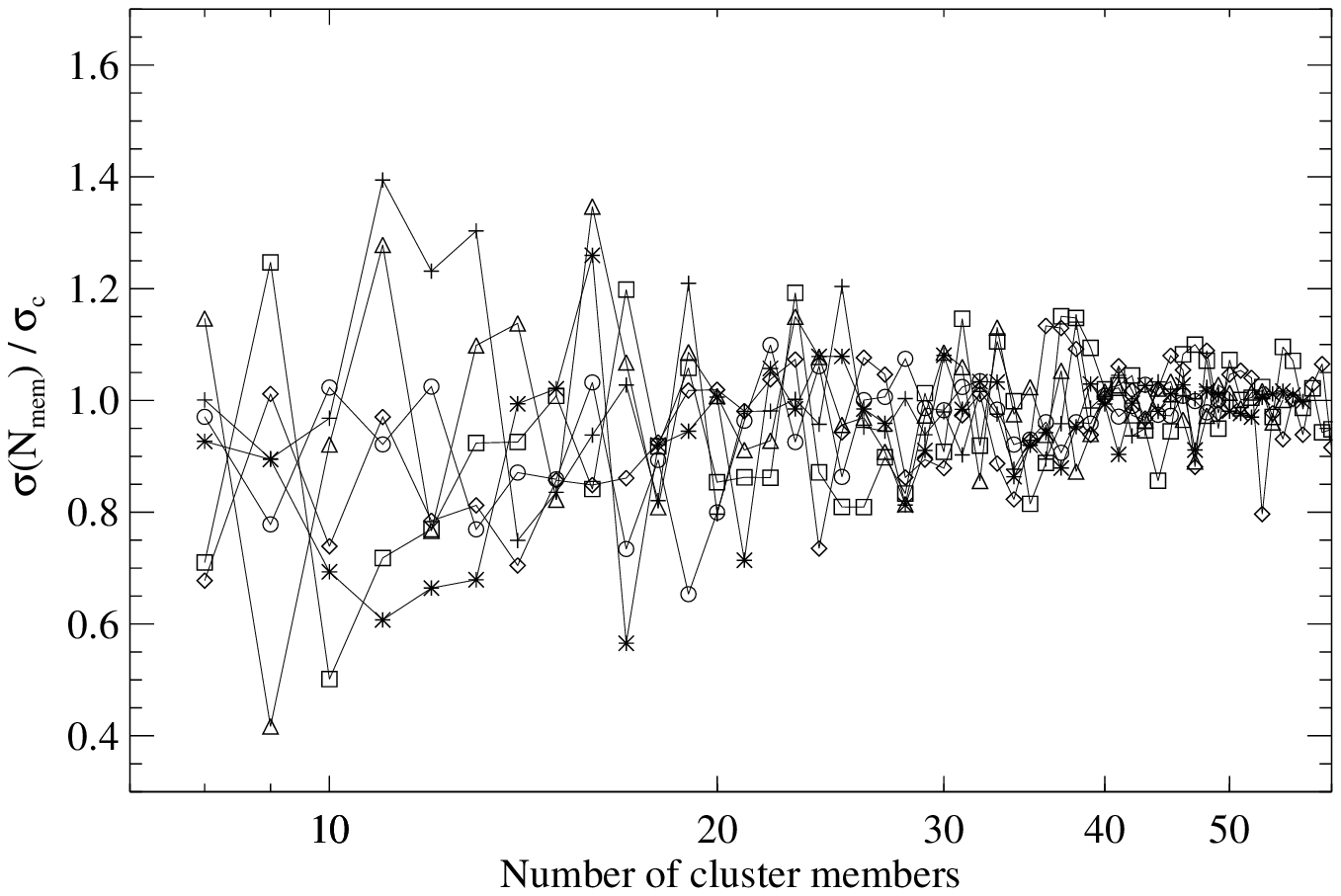} 
\caption{\label{fig:veldisp_vs_Nmem_clusters}
The top panel shows the line-of-sight velocity dispersion versus the number of cluster members ordered by 
luminosity (from highest to lowest) for 6 clusters from our SL lensing sample
with more than 49 member galaxies. 
The bottom panel shows the line-of-sight velocity dispersion but with the inclusion of 
randomly selected $N_{\textnormal{mem}}$ member galaxies.
In both panels the resultant line-of-sight velocity dispersion has been normalized 
by the total dispersion of the corresponding cluster, $\sigma_c$. 
In both cases the scatter is symmetric about the ration of 1 when $N_{\textnormal{mem}} \gtrsim 20$, 
indicating that the estimates are mainly affected by the lack of galaxy tracers
rather than by dynamical friction.
The cluster RCS2 J0047$+$0507 is represented by ``+'';
RCS2 J1517$+$1003 by ``$\ast$''; RCS2 J1519$+$0840 by ``$\circ$'';
SDSS J2135$-$0102 by ``$\diamond$''; RCS2 J2151$-$0138 by ``$\triangle$'';
and the cluster RCS2 J2329$-$1317 by ``$\square$''.
} 
\end{center}
\end{figure}

\subsection{Number of cluster members}

The number of cluster members plays a decisive role in the analysis of dynamical cluster
properties and velocity dispersion measurements \citep{Biviano06,WojtakLokas10}.
\cite{White10}, using high resolution N-body simulations, studied the stability of the cluster velocity dispersion
as a function of the number of subhalos used to estimate it.
They found that the results are generally stable once $\gtrsim 50$ subhalos are included; however, there is 
an intrinsic scatter of $\sim 10 \%$ between the line-of-sight velocity dispersion and the DM halo velocity dispersion,
mainly due to the halo triaxiality.
They also found that the line-of-sight velocity dispersion is biased low  when 
a small number of subhalos is used (increasing the scatter), but estimates with $N_{\textnormal{mem}} \gtrsim 20$ tend asymptotically to 
the true velocity dispersion of clusters.
Indeed, there is a general consensus in the literature that identifying on the order of 20 member galaxies is sufficient to derive 
the velocity dispersion of clusters  \citep{Beers90,Aguerri10,Hou12},
although there could be a $\sim 15\%$ underestimation if only brighter galaxies are used,
indicating that dynamical friction of brighter galaxies has a significant impact on the 
measured velocity dispersion \citep{Old13, Saro13}.

As mentioned in \S\ref{sec:results}, we have recovered a total of 1004 spectroscopically confirmed cluster members, which gives us an average
of $ \left<N_{\textnormal{mem}}\right> \sim$ 35 member galaxies per cluster. Therefore, we could assume that our estimates represent the ``true''
line-of-sight velocity dispersion of our clusters based on the studies above. However, we need to explore if these measurements are affected
by dynamical friction since our spectroscopic strategy was to prioritize brighter galaxies ahead of fainter ones.
In order to probe this possible systematic bias, we analyze the
behavior of the velocity dispersion as a function of the number of brighter cluster members of 6 clusters from our sample with 
$N_{\textnormal{mem}} \gtrsim 50$.
We rank the member galaxies according to their absolute $r'-$band magnitude and calculate the line-of-sight velocity dispersion using
the brightest galaxies from $N_{\textnormal{mem}}  = 8$ to $N_{\textnormal{mem}} = N_{total}$.
We plot the resultant velocity dispersions $\sigma({\small N_{\textnormal{mem}}})$ in Figure \ref{fig:veldisp_vs_Nmem_clusters} (top panel), normalized by
the final dispersion $\sigma_c$, obtained using all available cluster members. 
As showed by previous studies, the scatter between $\sigma({\small N_{\textnormal{mem}}})$ and $\sigma_c$ increases considerably for a lower
number of cluster members ($N_{\textnormal{mem}} \lesssim 15$), showing a clear underestimation of $\sigma$  when $N_{\textnormal{mem}} \lesssim 10$. 
For measurements using  $N_{\textnormal{mem}} \gtrsim 20$, the scatter is almost symmetric, i.e.
the bias introduced by using only a fraction of the total
number of member galaxies may underestimate or overestimate  
the ``true'' line-of-sight velocity dispersions, and whether the bias is from above or below
depends upon the cluster under consideration \citep[e.g.][]{White10}.
Therefore, estimates using $N_{\textnormal{mem}} \gtrsim 20$ are only slightly affected by dynamical friction;
pointing out that the most likely effect that is affecting our measurements should be associated to the lack of galaxy tracers.

To corroborate these conclusions, we repeat the previous process but choosing the member galaxies
randomly with respect to their absolute $r-$band magnitude.
The results are shown in the bottom panel of Figure \ref{fig:veldisp_vs_Nmem_clusters}. 
In this figure one can easily see that for estimates using $N_{\textnormal{mem}} \gtrsim 20$, the scatter around the 
``true'' line-of-sight velocity dispersion is symmetric and similar
to the scatter obtained when cluster members are sorted by their brightness.
Therefore, we can assume that velocity dispersion estimates are mainly affected by the lack 
of cluster members rather than by dynamical friction.  
Indeed, in both cases of Figure \ref{fig:veldisp_vs_Nmem_clusters},
the average scatter is of the order of $\sim10\%$ when $N_{\textnormal{mem}}$ is equal
to the average number of cluster members, i.e. $N_{\textnormal{mem}}=\left<N_{\textnormal{mem}}\right> = 35$. 
It is to be noted that this $10\%$ of scatter should be considered as the lower scatter that we 
should include in our velocity dispersion estimates in order to take into account the lack 
of cluster members.  
But we leave this discussion for future observational and simulation studies, where we could compare
results from a larger number of cluster members with specific simulations for the SL galaxy cluster population.

\subsection{Line-of-sight effect}

It is well known that virialized halos that host galaxy clusters are triaxial \citep{Thomas_Couchman92,Warren92,JingSuto02},
with the major axis approximately twice as long as the minor axes
which are approximately equal in size. This prolate shape could potentially lead to a bias in the
velocity dispersion estimates because the velocity tensor is quite anisotropic and generally well aligned
with the inertia tensor, with a typical 
misalignment angle of $\sim30\deg$ \citep{Tormen97,KasunEvrard05,White10}.
In other words, viewing along the major axis may contribute to a higher velocity dispersion, 
while the two minor axes to lower values; hence, the final measurements depend significantly on the chosen 
line-of-sight. 
In fact, numerical simulations have shown that, although the 3D velocity dispersion of DM particles
within $r_{200}$ is well correlated with $M_{200}$ and the galaxies show
little velocity bias compared to the DM particles, the line-of-sight
velocity dispersions show a considerably larger scatter and the mass estimates
from these measurements are biased with respect to the true values \citep{Evrard08, White10, Saro13}. 

Studies based on simulations of SL halos have shown that the most effective strong lenses are not more 
triaxial than the general halo population \citep{Hennawi07, Meneghetti10}; 
however, they are more likely to have their major axes aligned along the line-of-sight.
Therefore, we should assume that the velocity dispersion estimates for a sample of SL-selected clusters 
will be biased high with respect to the velocity dispersions
measured for clusters that are randomly oriented on the sky.
However, the intrinsic  misalignment between the halo positional ellipsoid and the velocity tensor
introduces an element of randomization in the orientation of the velocity tensor with respect
to the line-of-sight that reduces the impact of the orientation bias.
Specific predictions for the magnitude of this bias require the convolution of the probability distributions 
for the position orientation angle of the SL-selected clusters with their velocity 
principal axes.  Since  we do not have these probability distributions,
we leave this additional correction for future analyses and as in the previous section, 
we do not include this extra uncertainty in our estimates. 
But it is important to note that our line-of-sight velocity dispersion measurements should have, at least, 
a scatter of the same order as found by \cite{White10} for
the normal cluster population, i.e. a scatter of the order of $\sim 10\%$
between the line-of-sight and 3D velocity dispersions.

\section{Summary and conclusions} \label{sec:conclusion}

We have conducted a large spectroscopic follow-up program of 29 SL-selected galaxy clusters
discovered in the RCS-2 survey. Our spectroscopic analysis has revealed the nature of
52 gravitational arcs present in the core of our galaxy clusters, 
which correspond to 35 background sources at high-redshifts that are clearly 
distorted by the gravitational potential of these clusters. 
These lensed galaxies span a wide redshift range of $0.8 \le z \le 2.9$, 
with a median redshift of ${z}_s = 1.8 \pm 0.1$, that matches the expectations.
This dataset extends the number of galaxy clusters with spectroscopic confirmation 
of their SL features that are available to perform lensing reconstructions of their mass distribution,
especially at $z \gtrsim 0.2$.

This campaign has also yielded a total of 1004 spectroscopically confirmed cluster members 
that gives an average of $ \left<N_{\textnormal{mem}}\right> \sim$ 35 member galaxies per cluster. 
These data
allow us to obtain robust redshifts for each cluster and measure velocity dispersions
with relatively high confidence level, which are translated into dynamical masses by using
the \cite{Evrard08} $\sigma_{DM} - M_{200}$ scaling relation. 
The redshift and mass ranges of our SL sample are distributed from $0.22 \le z \le 1.01$ and 
$5  \times10^{13} \le M_{200}/h^{-1}_{70}M_{\odot} \le 1.9\times10^{15}$, respectively.
These wide redshift and mass ranges allow diverse kinds of studies: from the analysis 
of the relationship between the concentration of the cluster halos and their masses
to studies of galaxy evolution in cluster environments.

We have analyzed some effects that could affect our velocity dispersion measurements,
such as the presence of substructure in our clusters, the region where cluster members were selected,
the final number of confirmed members, and the line-of-sight effects. Our primary conclusions are:
\begin{itemize}
\item We found that 10 out of 24 of our clusters, where analysis is possible, 
show signs of dynamical substructure, which is consistent 
with previous optical and X-ray studies of local clusters.
The velocity distributions of 23 of these clusters are consistent with a uni-modal distribution 
and we therefore assumed that Eq. \ref{eq:Mass_Evrard08} is applicable to these SL clusters in our sample.

\item We sampled cluster member galaxies within the inner $\sim 0.9$ $r_{200}$ of the clusters, which did 
not bias our results due to the flatness of the velocity dispersion profiles from $\sim 0.8$ $r_{200}$.
Furthermore, since we mainly sampled cluster members within the virial radius,
the probability to include interlopers is approximately $30\%$ lower than for sampling 
up to twice $r_{200}$ \citep{Old13, Saro13}.

\item We have found that using $N_{\textnormal{mem}} \gtrsim 20$, our velocity dispersion estimates 
are mainly affected by the lack of galaxy tracers rather than by dynamical friction. 
We found that in both cases, sorting the cluster members by their brightness or randomly,
the scatter is symmetric and of the order of $\sim10\%$ 
when $N_{\textnormal{mem}} = \left<N_{\textnormal{mem}}\right> = 35$. 
However, it is only a first guess of the magnitude of this potential bias. 
A better understanding of this effect needs deeper studies of the galaxy kinematics in SL clusters.

\end{itemize}

Summing up, we have found that our velocity dispersion measurements should be affected by
the lack of cluster members as well as by the scatter between the line-of-sight and 3D velocity dispersion. 
However,  specific predictions for the magnitude of these biases require an improvement in both simulation and observational studies, 
with larger simulations oriented to the SL galaxy cluster population and more spectroscopic information of the cluster members.
Even though our measurements may be biased or may have large uncertainties, which are translated into large errors in the mass estimates,
they serve as a base for a better understanding of the SL cluster properties.
Furthermore, these dynamical masses can be used as \textit{priors} for mass reconstruction studies,
that combined with SL signatures yield one of the most robust measurements of the mass distribution of SL clusters.

A complete characterization of the properties and biases of this population 
is crucial for taking full advantage of future SL samples 
coming from a new era of large area deep imaging surveys (e.g. PanSTARRS, LSST, DES).
The data and analysis presented in this work represent the first steps in this direction and
also pave the way for  multi-wavelength studies, i.e. when SL information is combined with WL, dynamical masses, X-ray, and SZe data.
These kinds of studies will allow us to quantify the biases between the different observable masses
in order to fully exploit the additional information provided by SL signatures, which can then be  
intelligently applied to scaling relations and mass estimates for the general cluster population.

\acknowledgments

M. Carrasco gives particular thanks to Matthias Bartelmann, Matteo Maturi, Agnese Fabris, Jorge Gorz\'alez, and Cristobal Sif\'on for useful discussions and collaborations in this work. 
M. Carrasco'  research is supported by  the Transregional  Collaborative  Research  Centre  TRR  33. 
M. Carrasco acknowledges the support of ESO through the ESO Studentship and DAA through the PUC-HD Graduate Exchange Fellowship.
We thank to ESO and its support staff for the implementation of the Programmes 081.A-0561,  083.A-0491, and 084.A-0317. 
Support for L.F. Barrientos and T. Anguita is provided by the Ministry of Economy, Development, and Tourism's Millennium Science Initiative through grant IC120009, awarded to The 
Millennium Institute of Astrophysics, MAS. L. F. Barrientos' research is supported by proyecto FONDECYT 1120676. T. Anguita acknowledges support by proyecto FONDECYT 11130630.
DGG acknowledges financial support from the National Research Foundation (NRF) of South Africa.



\clearpage
\appendix

\section{Velocity histograms} \label{appendix:vel_hist}

\begin{figure*}[h!]
\begin{center}
\begin{tabular}{c c c}
\includegraphics[width=45mm,angle=90, trim= 0mm 18mm 0mm 26mm,clip]{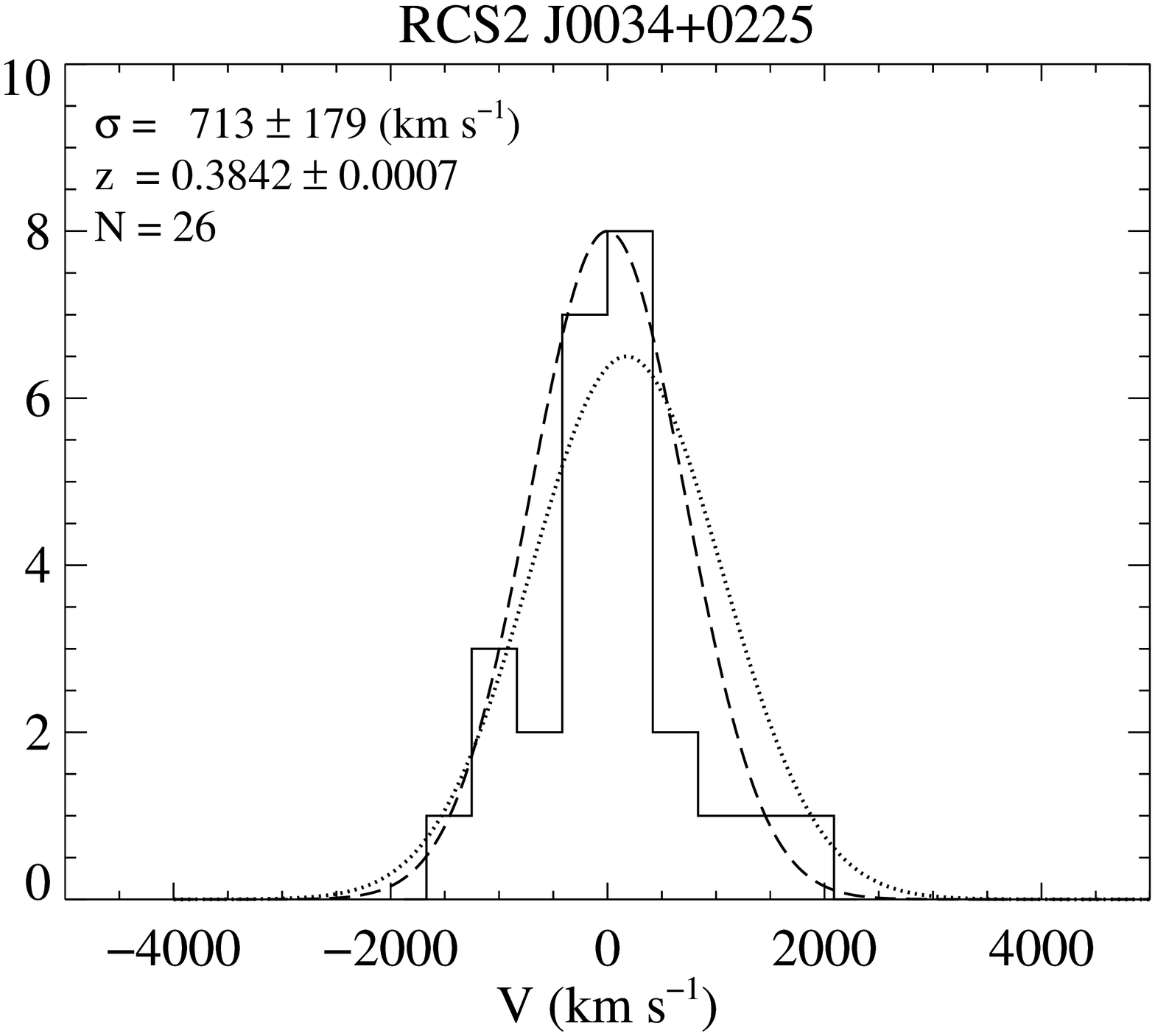} &
\includegraphics[width=45mm,angle=90, trim= 0mm 18mm 0mm 26mm,clip]{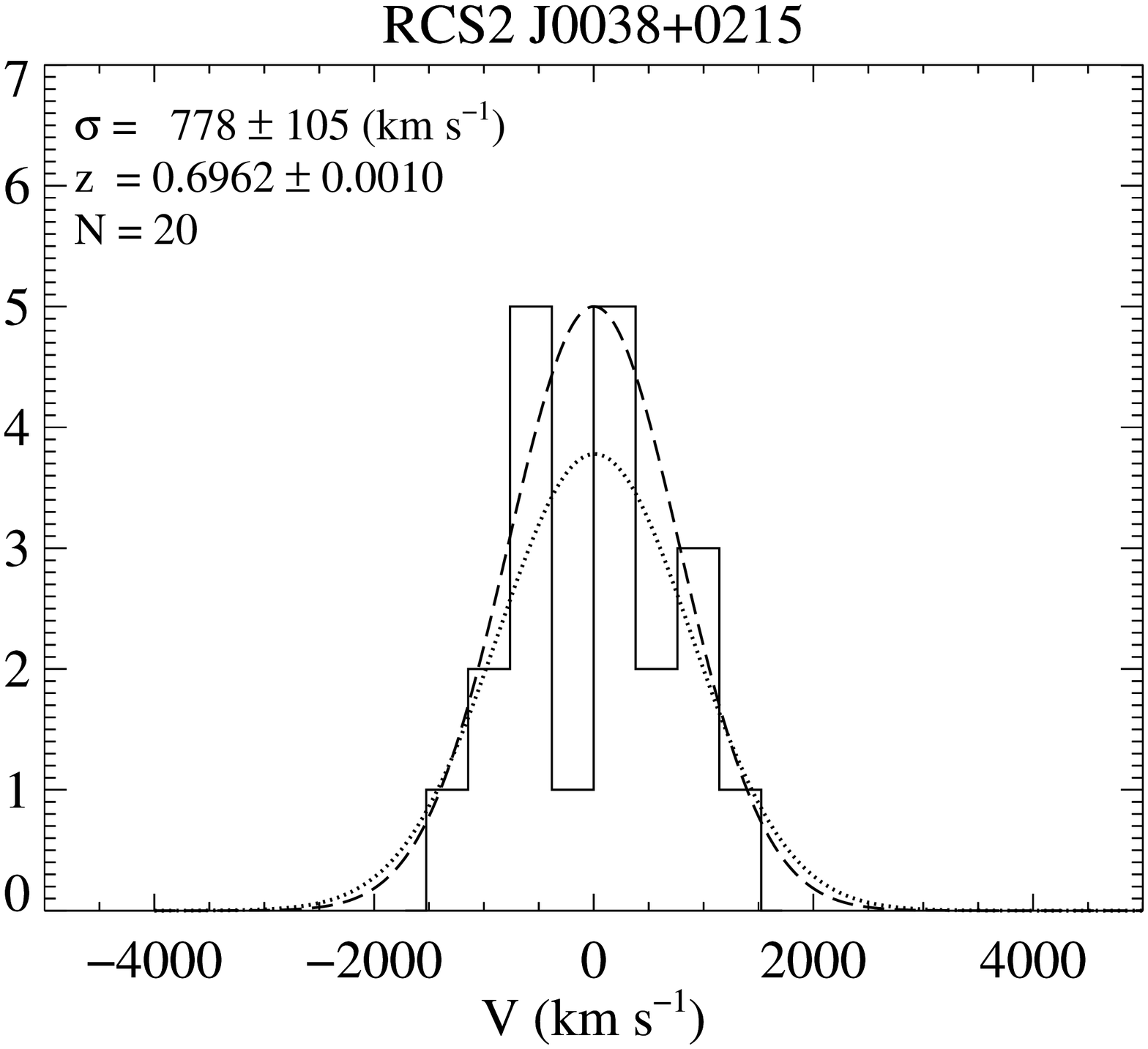} &
\includegraphics[width=45mm,angle=90, trim= 0mm 18mm 0mm 26mm,clip]{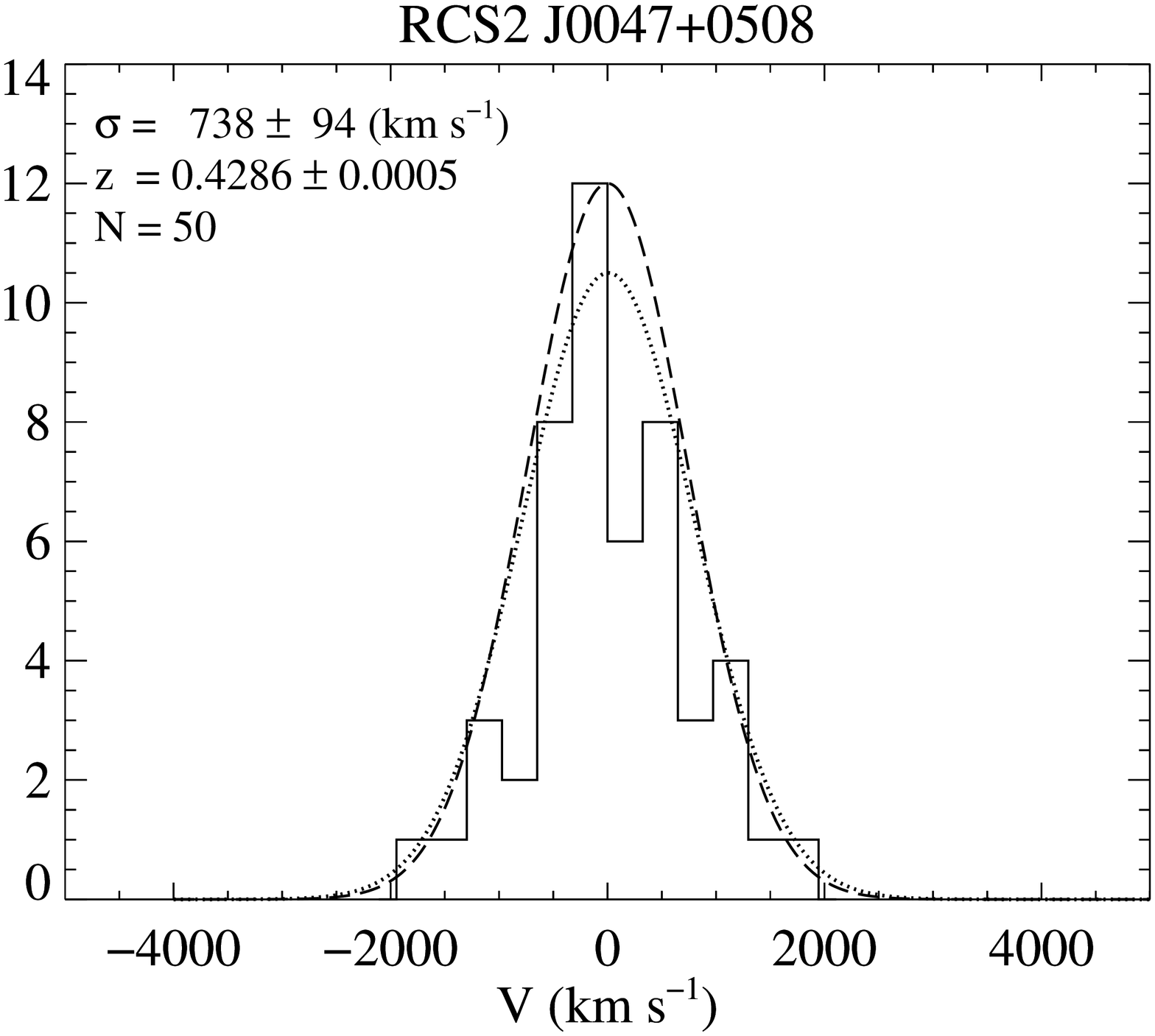} \\
\includegraphics[width=45mm,angle=90, trim= 0mm 18mm 0mm 26mm,clip]{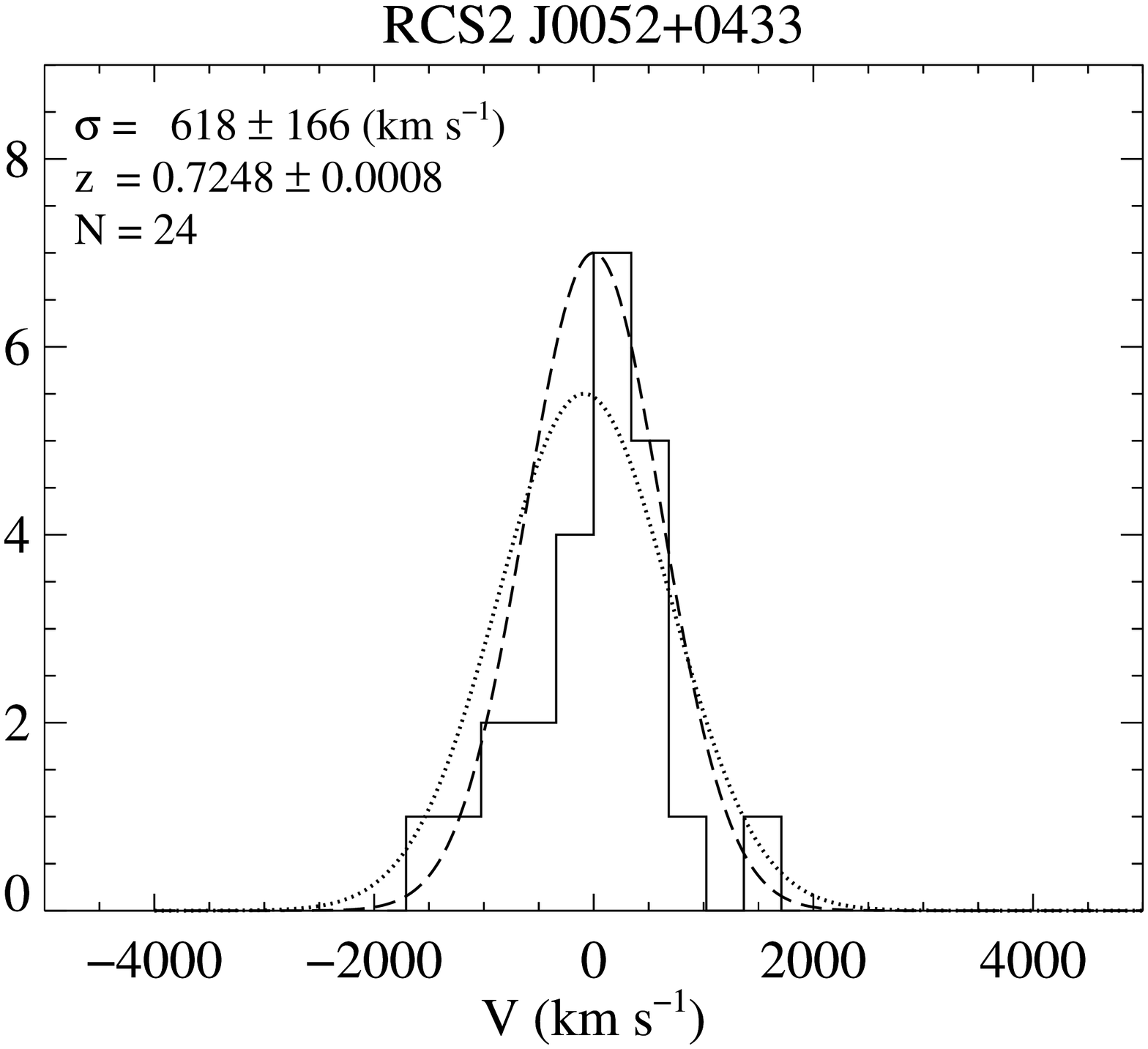} &
\includegraphics[width=45mm,angle=90, trim= 0mm 18mm 0mm 26mm,clip]{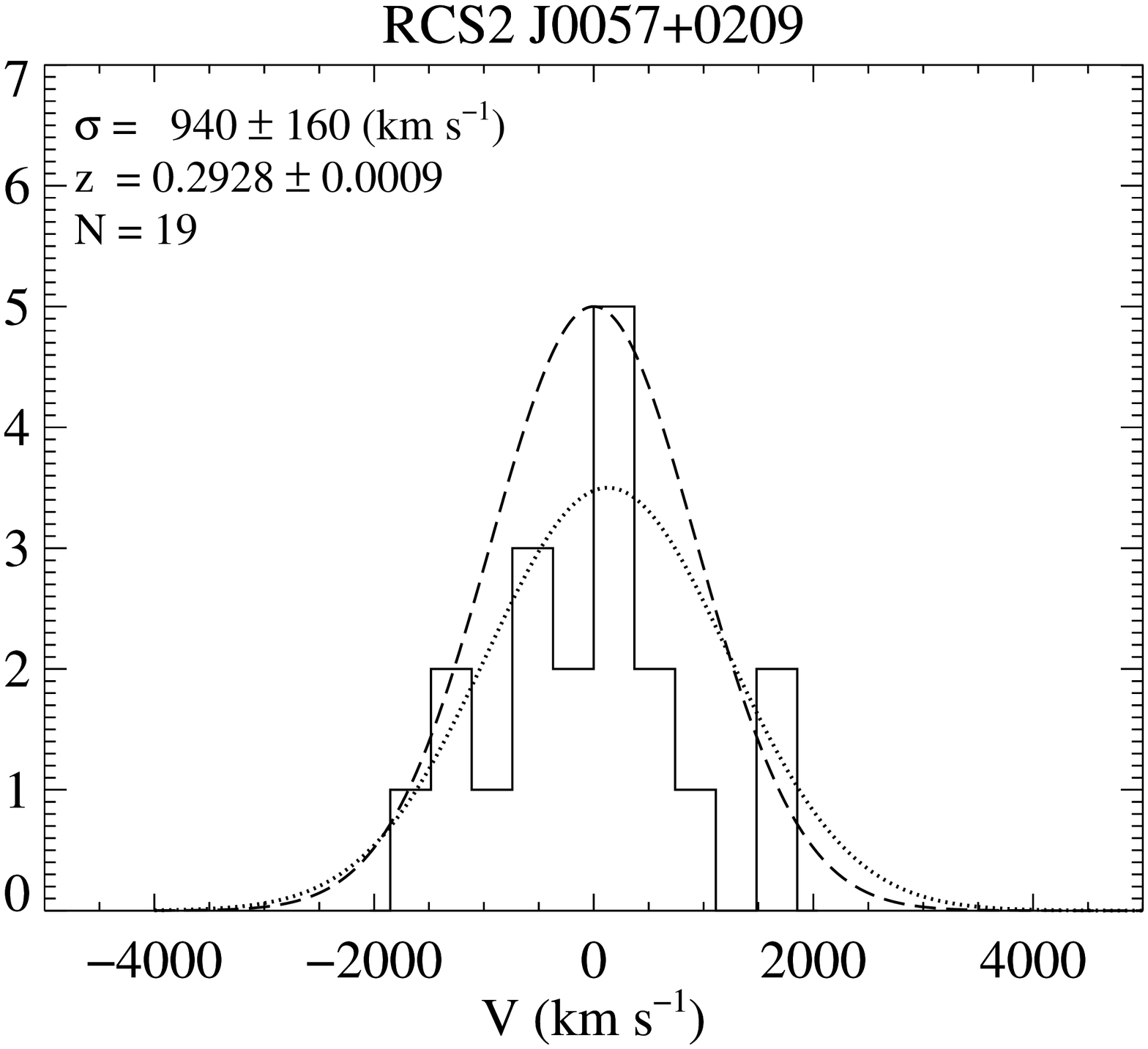} &
\includegraphics[width=45mm,angle=90, trim= 0mm 18mm 0mm 26mm,clip]{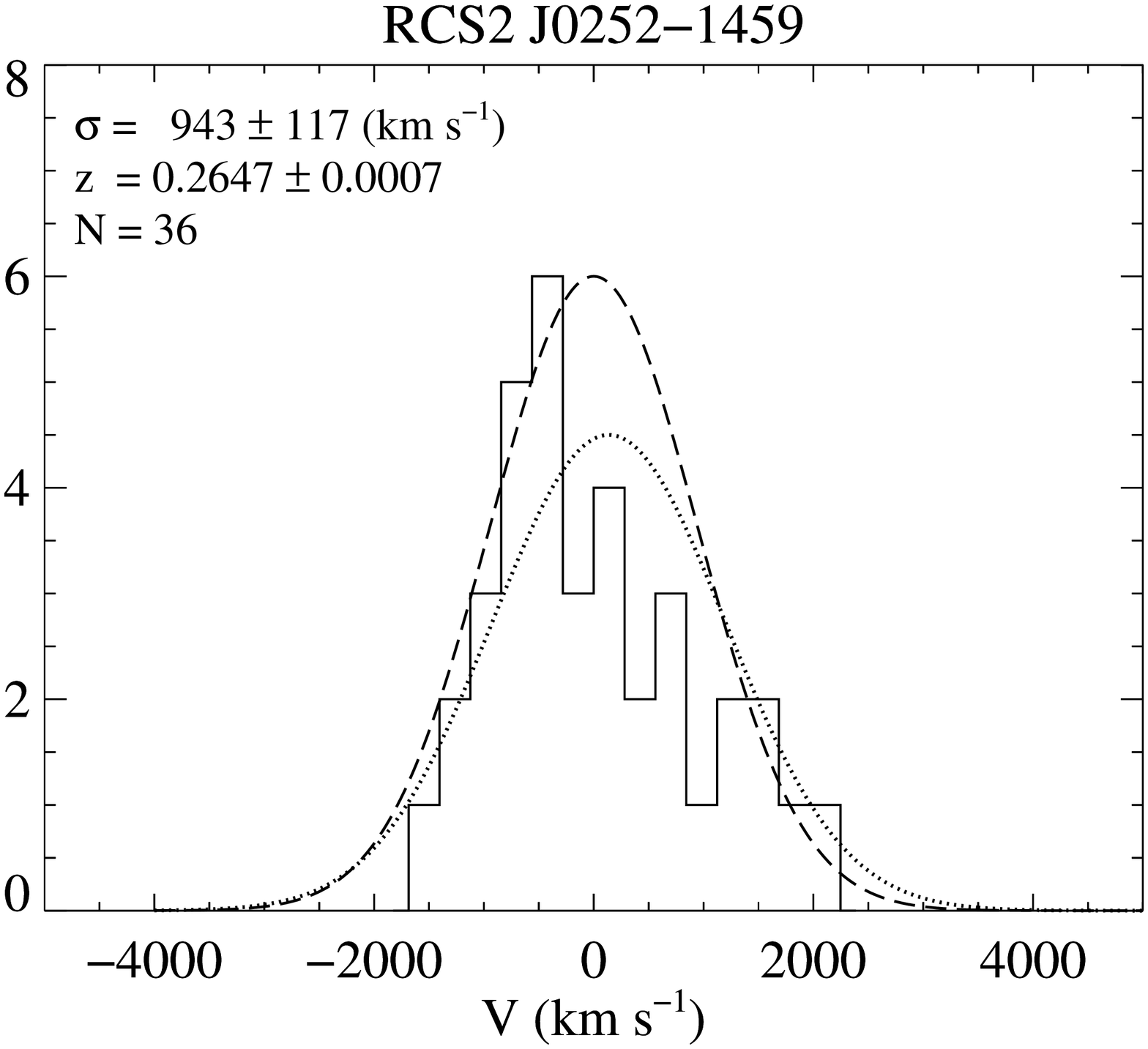} \\
\includegraphics[width=45mm,angle=90, trim= 0mm 18mm 0mm 26mm,clip]{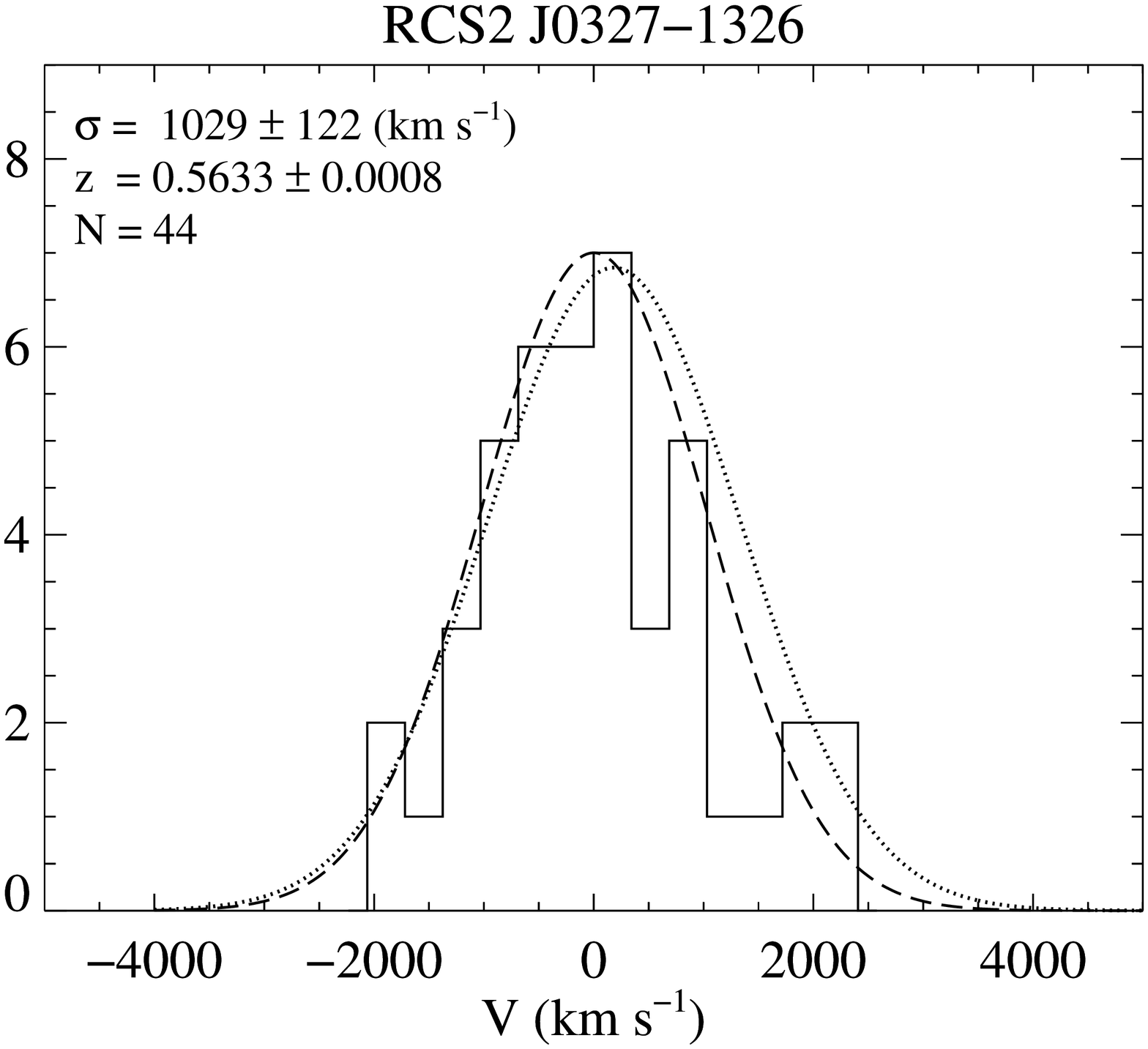} &
\includegraphics[width=45mm,angle=90, trim= 0mm 18mm 0mm 26mm,clip]{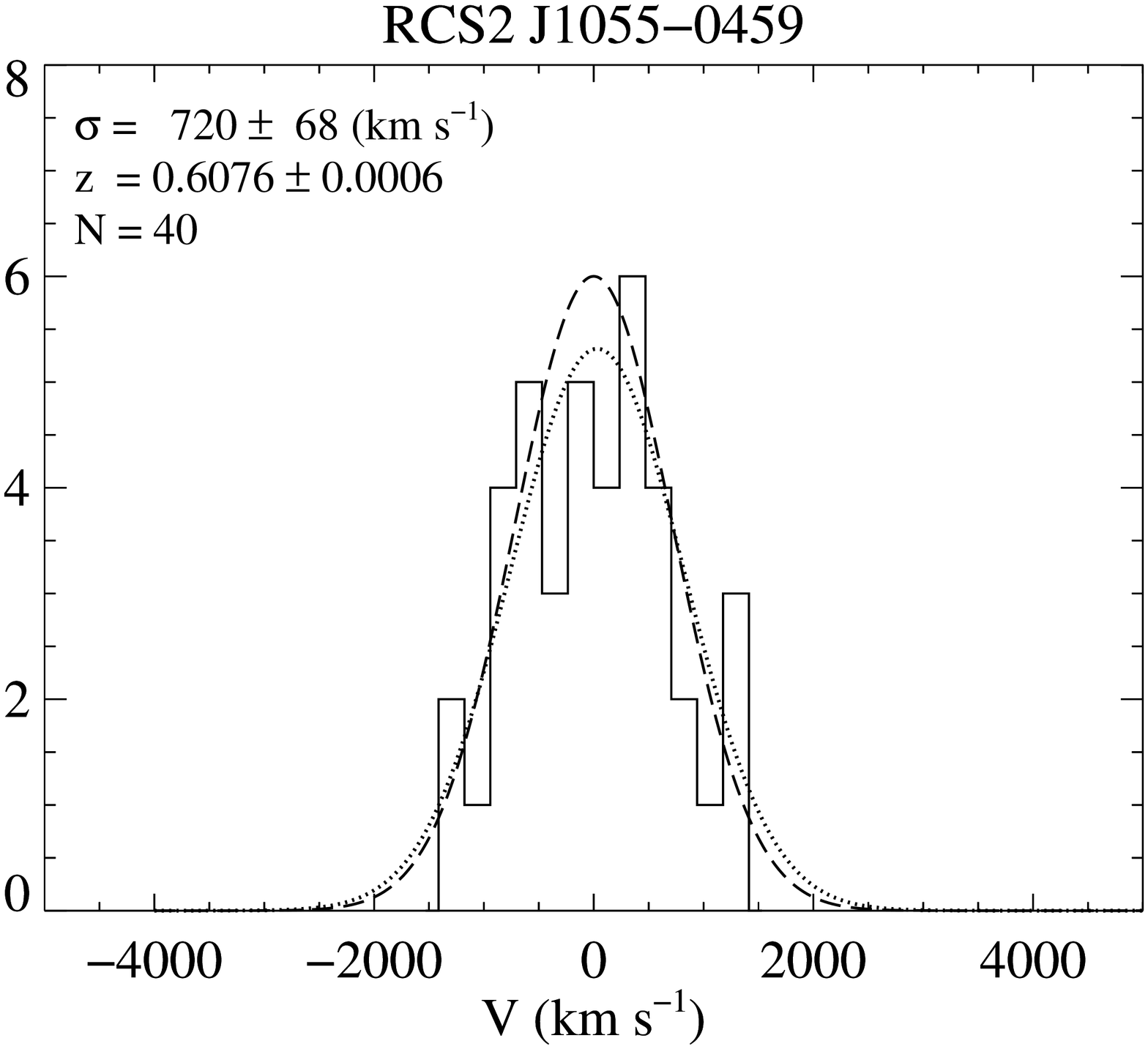} &
\includegraphics[width=45mm,angle=90, trim= 0mm 18mm 0mm 26mm,clip]{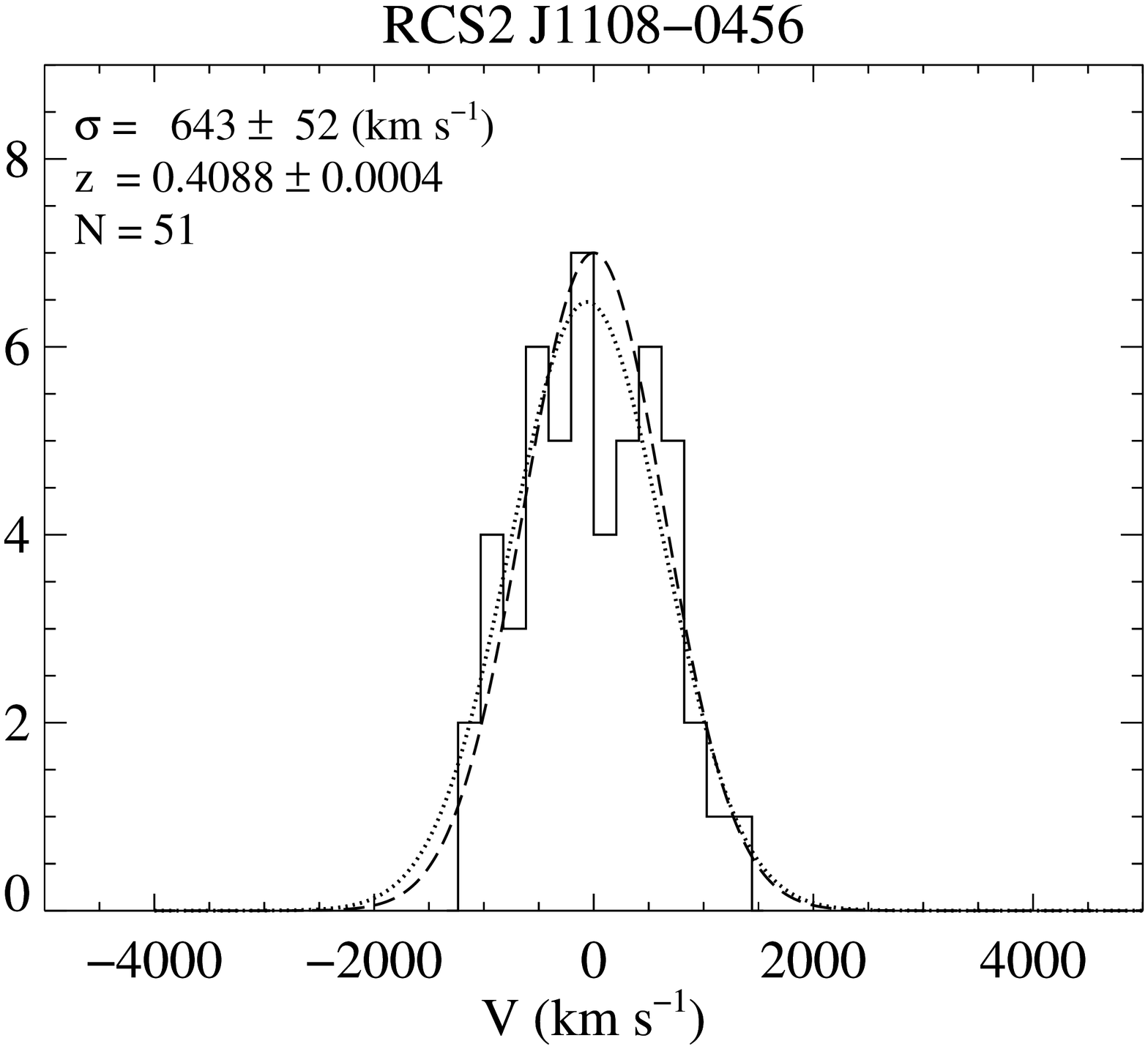} \\
\includegraphics[width=45mm,angle=90, trim= 0mm 18mm 0mm 26mm,clip]{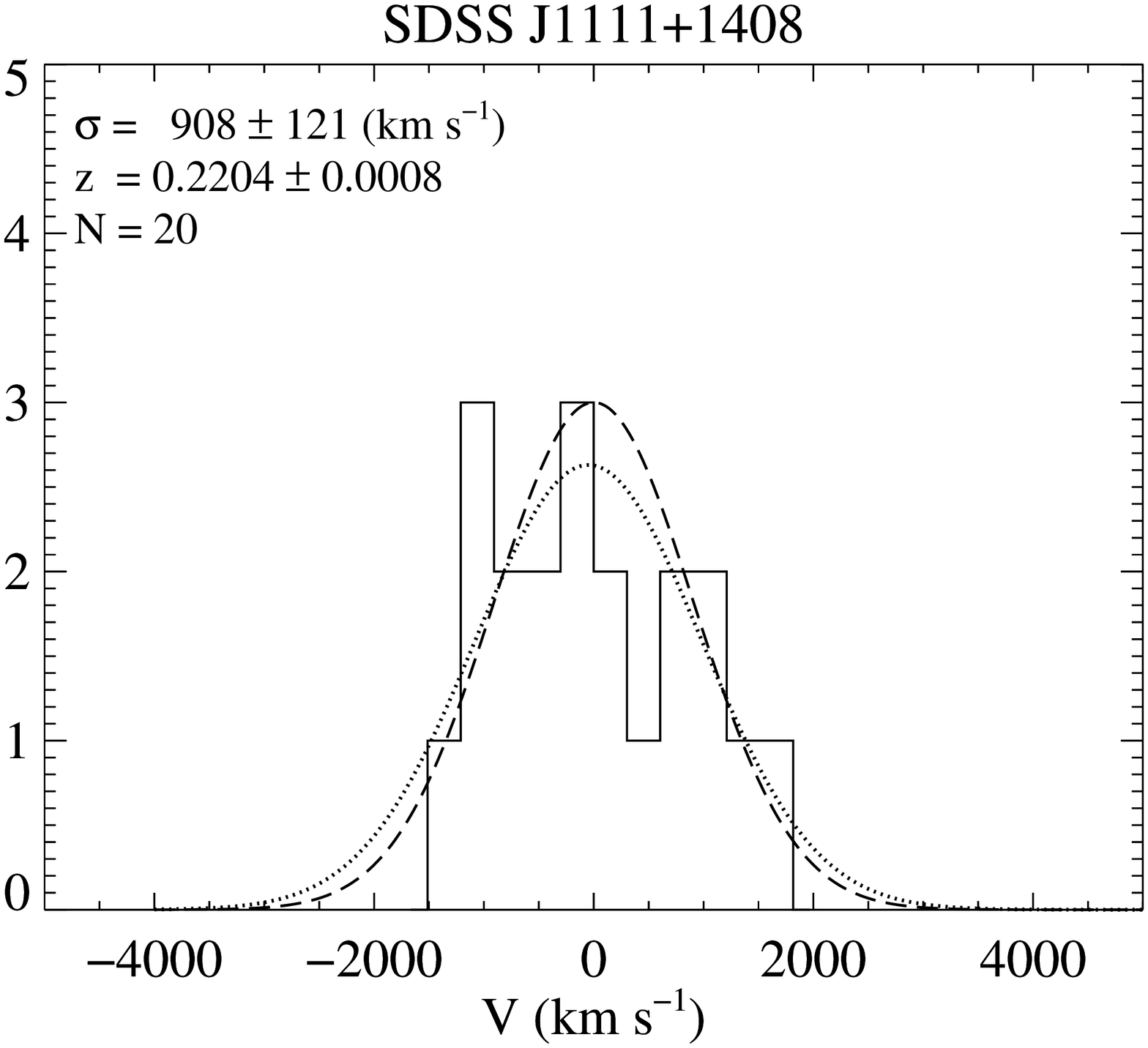} &
\includegraphics[width=45mm,angle=90, trim= 0mm 18mm 0mm 26mm,clip]{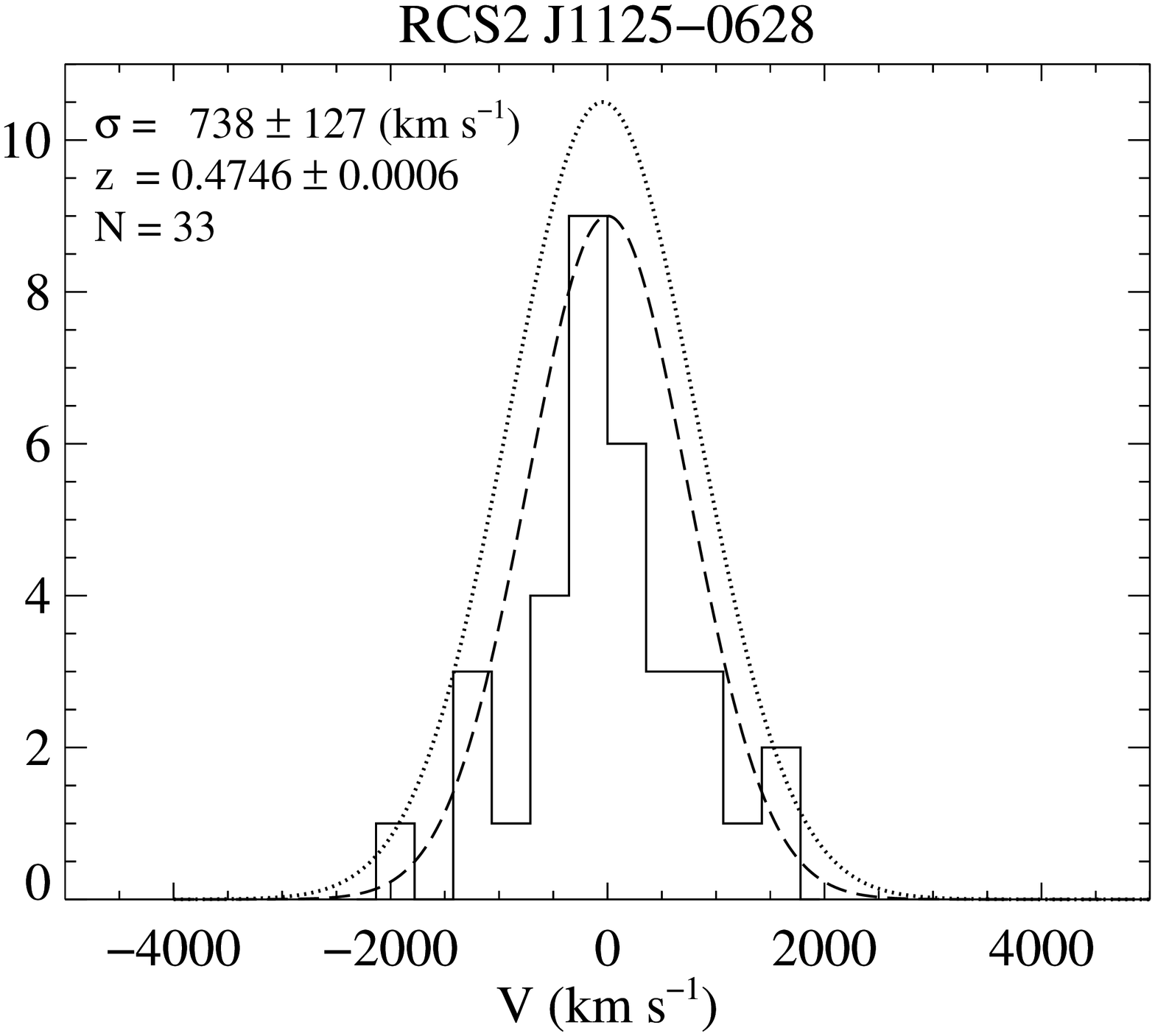} &
\includegraphics[width=45mm,angle=90, trim= 0mm 18mm 0mm 26mm,clip]{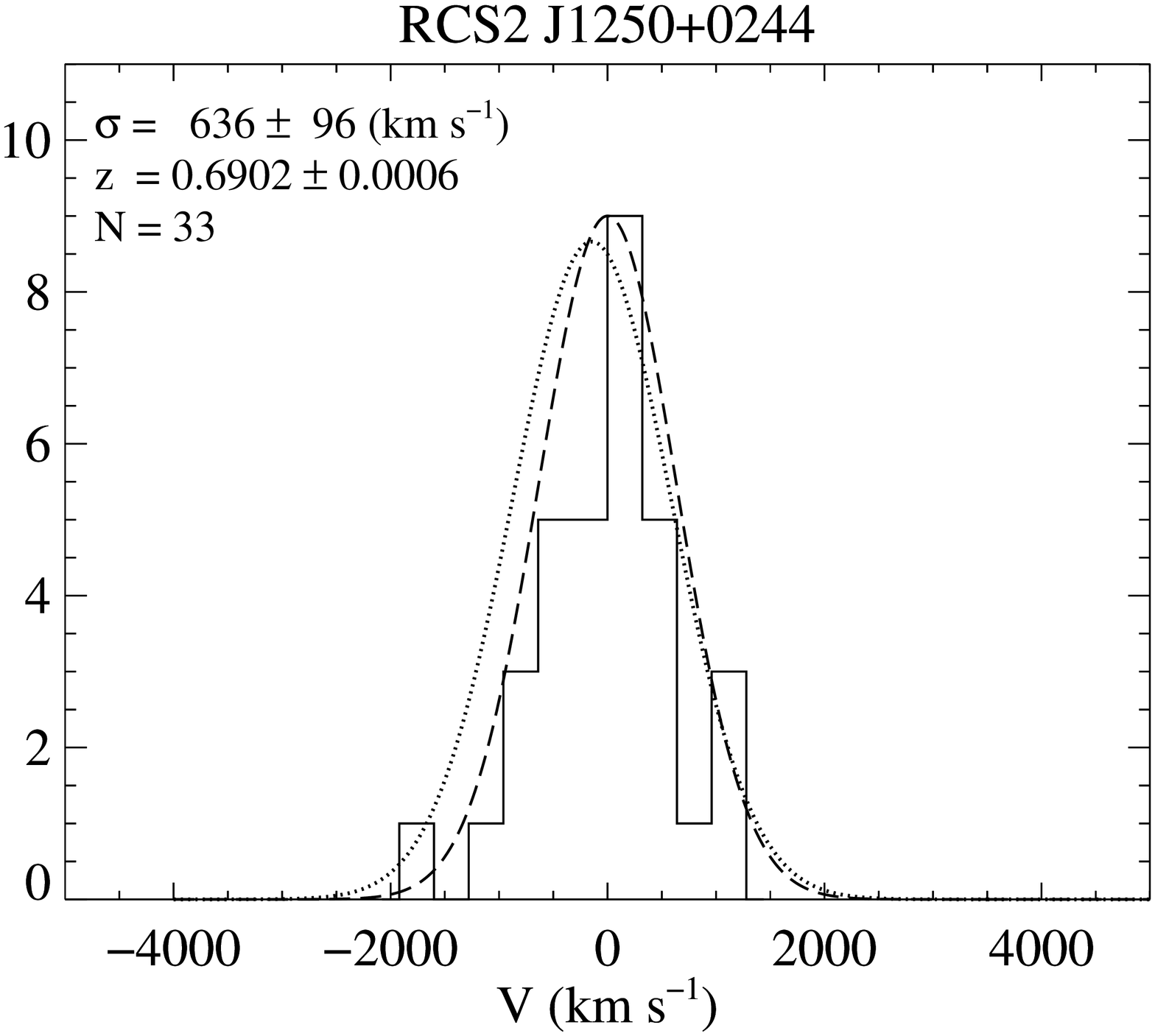} \\

\end{tabular}
\caption{\label{fig:vel_hist_1}
Velocity histograms for 12 out of 23 of our clusters with $N_{\textnormal{mem}} \gtrsim 20$ showing a uni-modal distribution. 
Each panel is labeled with the name of the corresponding cluster,
its rest-frame velocity dispersion, redshift, and the number of cluster members spectroscopically confirmed.  
Dashed lines correspond to Gaussian (reference) distributions with the mean and variance  equal to the mean velocity and velocity dispersion squared of each cluster derived from the bi-weight estimator analysis. 
Dotted lines correspond to the best-fit Gaussian curves to the velocity histograms, where the best-fitting Gaussian parameters are consistent with the mean velocity and velocity dispersion squared within $1\sigma$.
Also, note that the y-axes differ between plots. 
} 
\end{center}
\end{figure*}

\begin{figure*}[h!]
\begin{center}
\begin{tabular}{c c c}

\includegraphics[width=45mm,angle=90, trim= 0mm 18mm 0mm 26mm,clip]{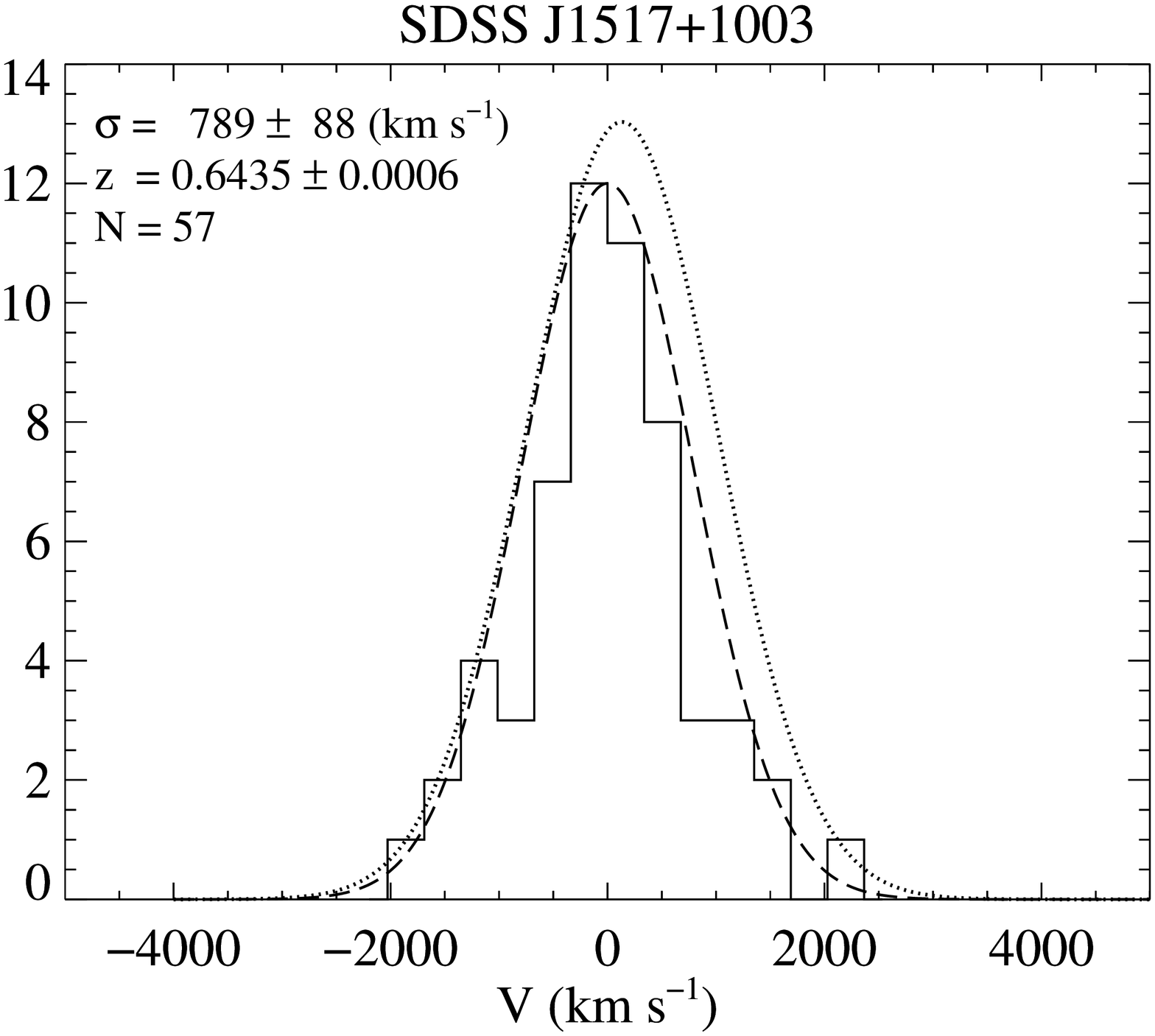} &
\includegraphics[width=45mm,angle=90, trim= 0mm 18mm 0mm 26mm,clip]{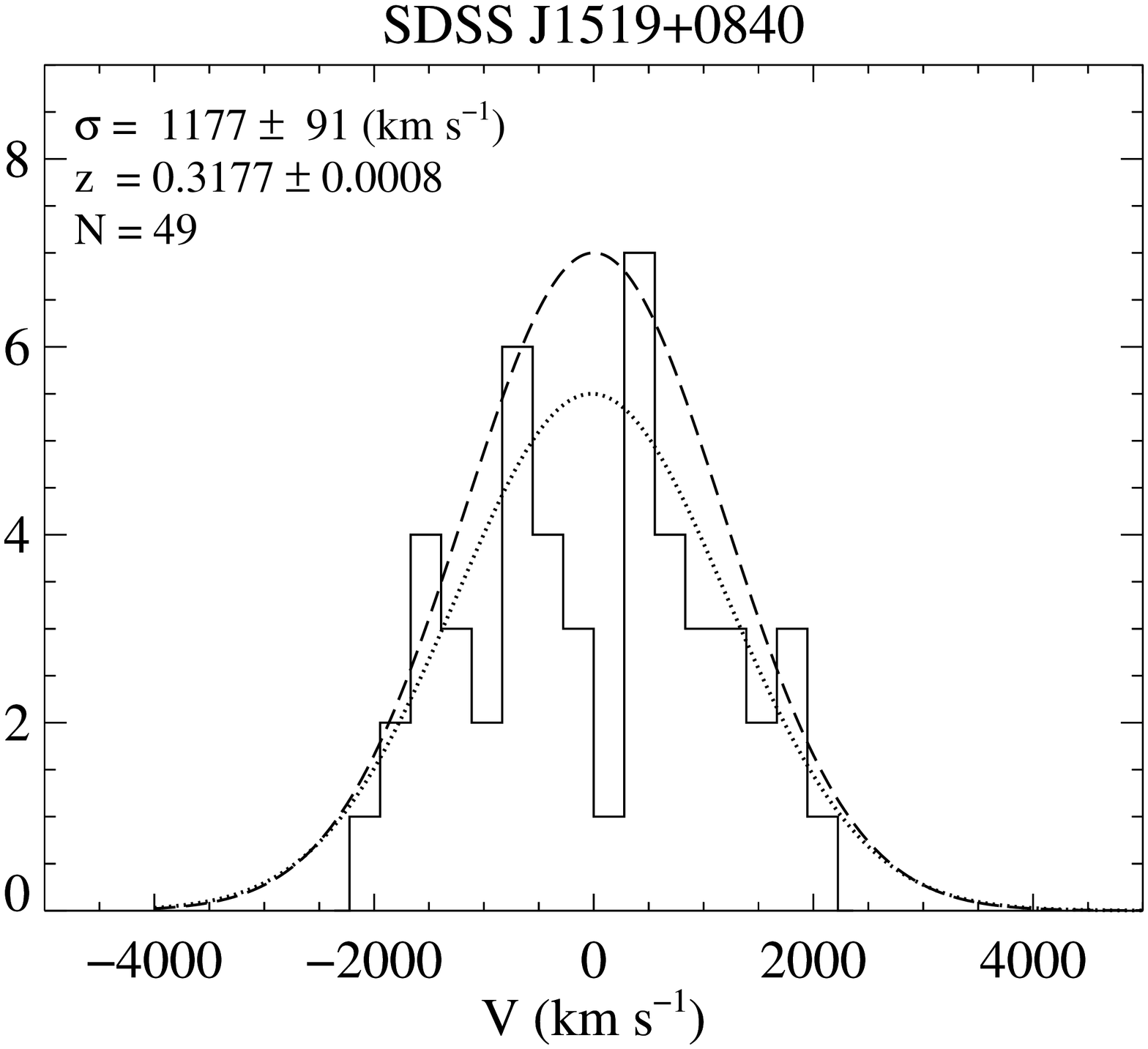} &
\includegraphics[width=45mm,angle=90, trim= 0mm 18mm 0mm 26mm,clip]{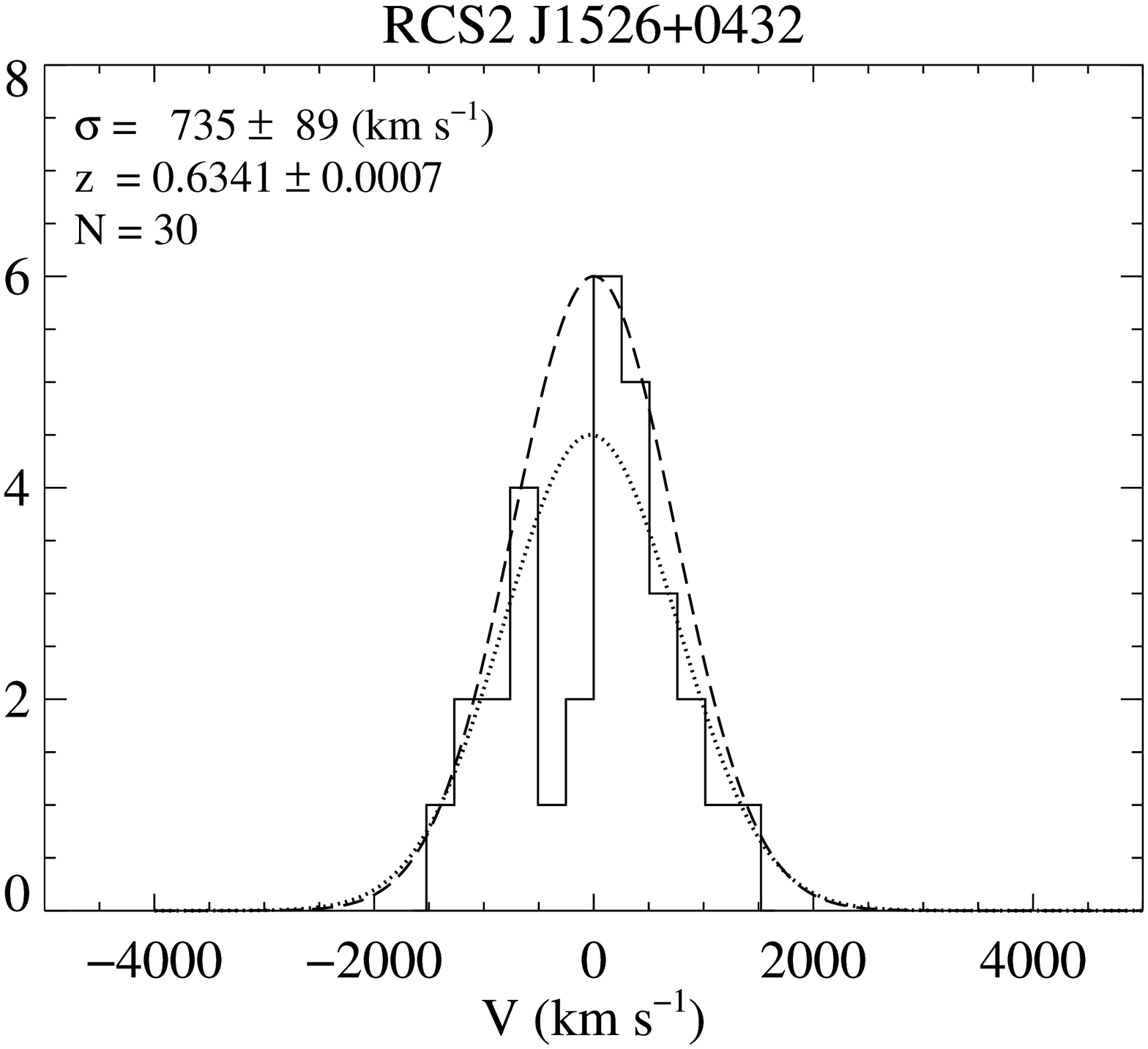} \\
\includegraphics[width=45mm,angle=90, trim= 0mm 18mm 0mm 26mm,clip]{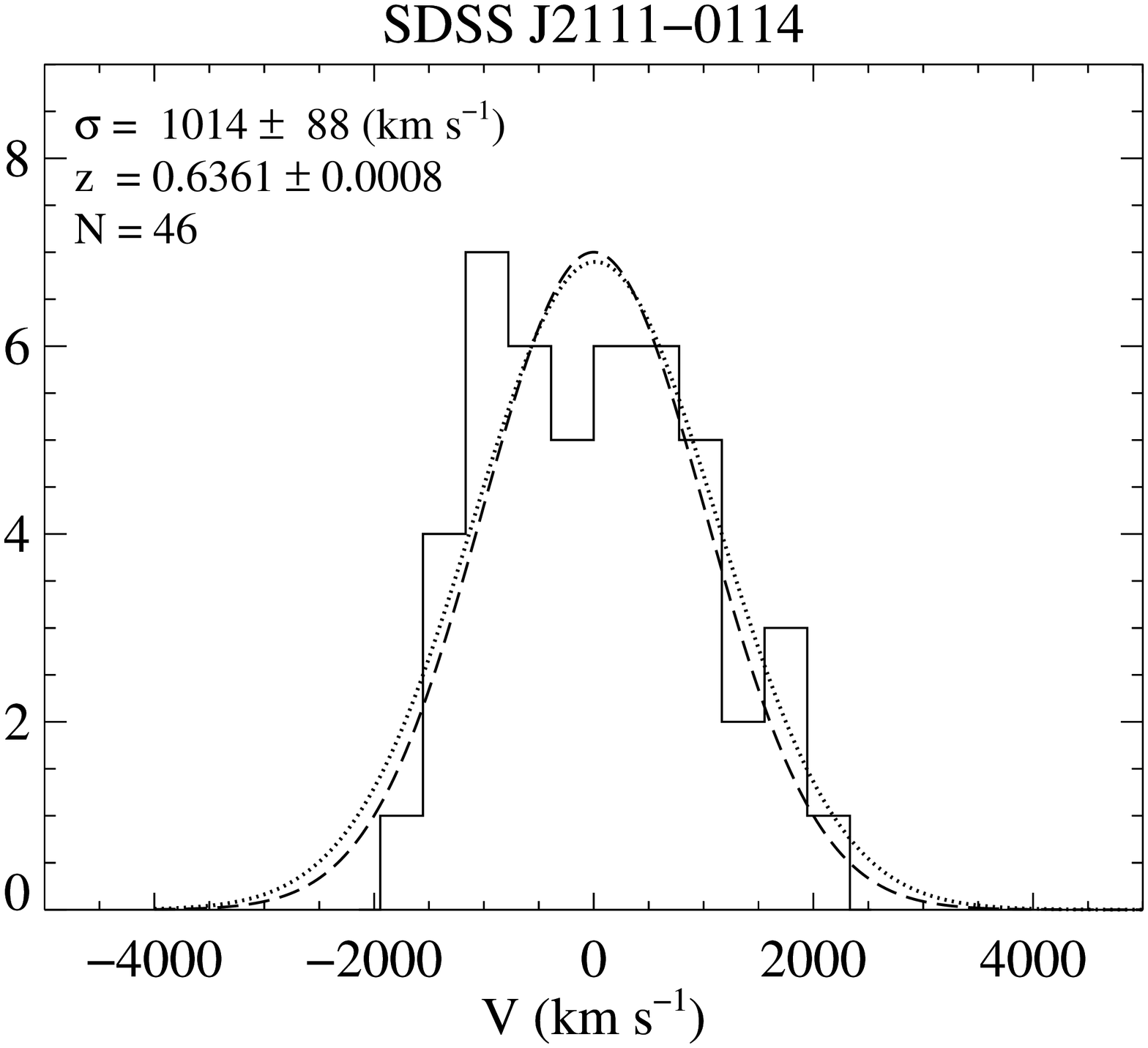} &
\includegraphics[width=45mm,angle=90, trim= 0mm 18mm 0mm 26mm,clip]{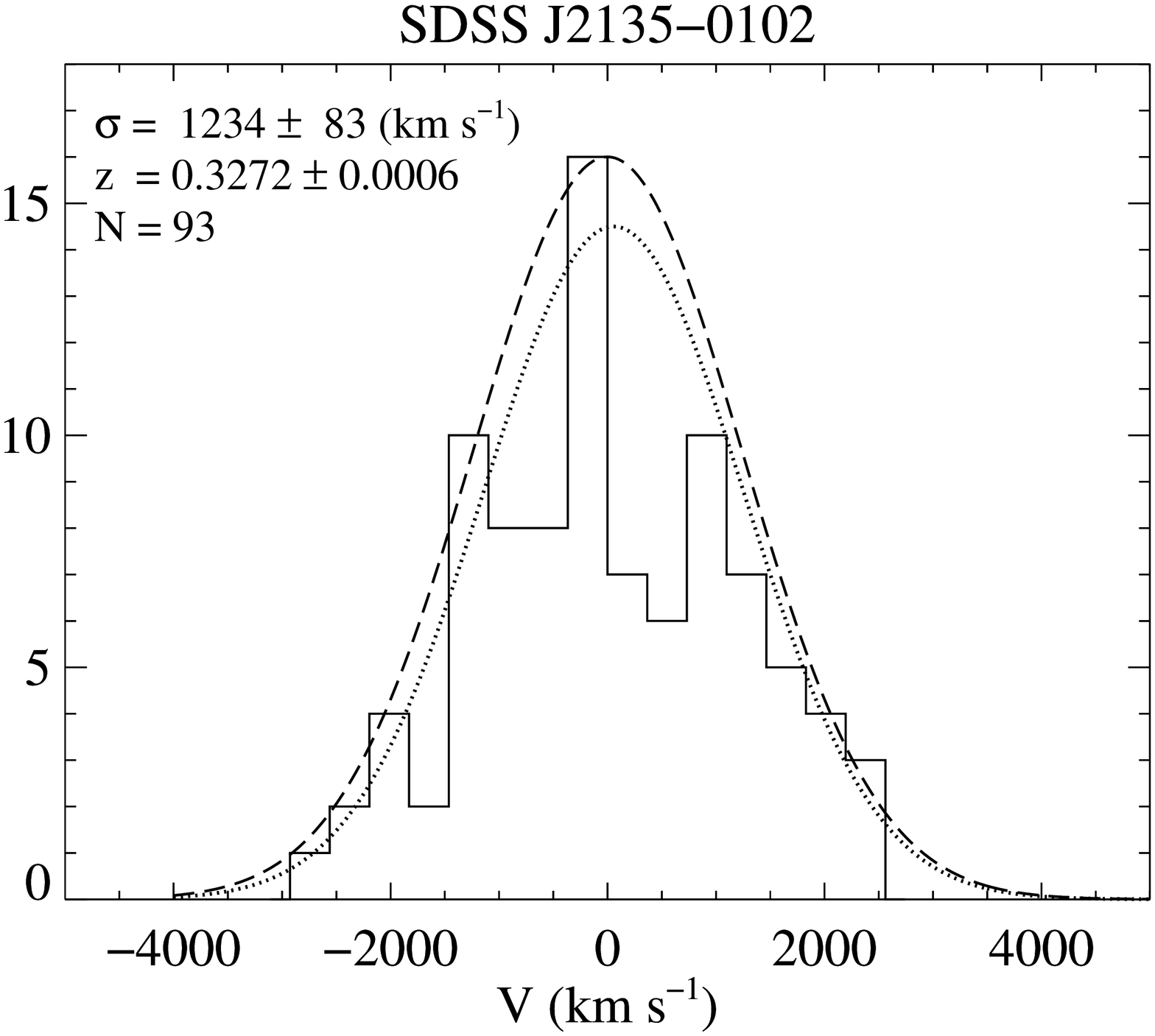} &
\includegraphics[width=45mm,angle=90, trim= 0mm 18mm 0mm 26mm,clip]{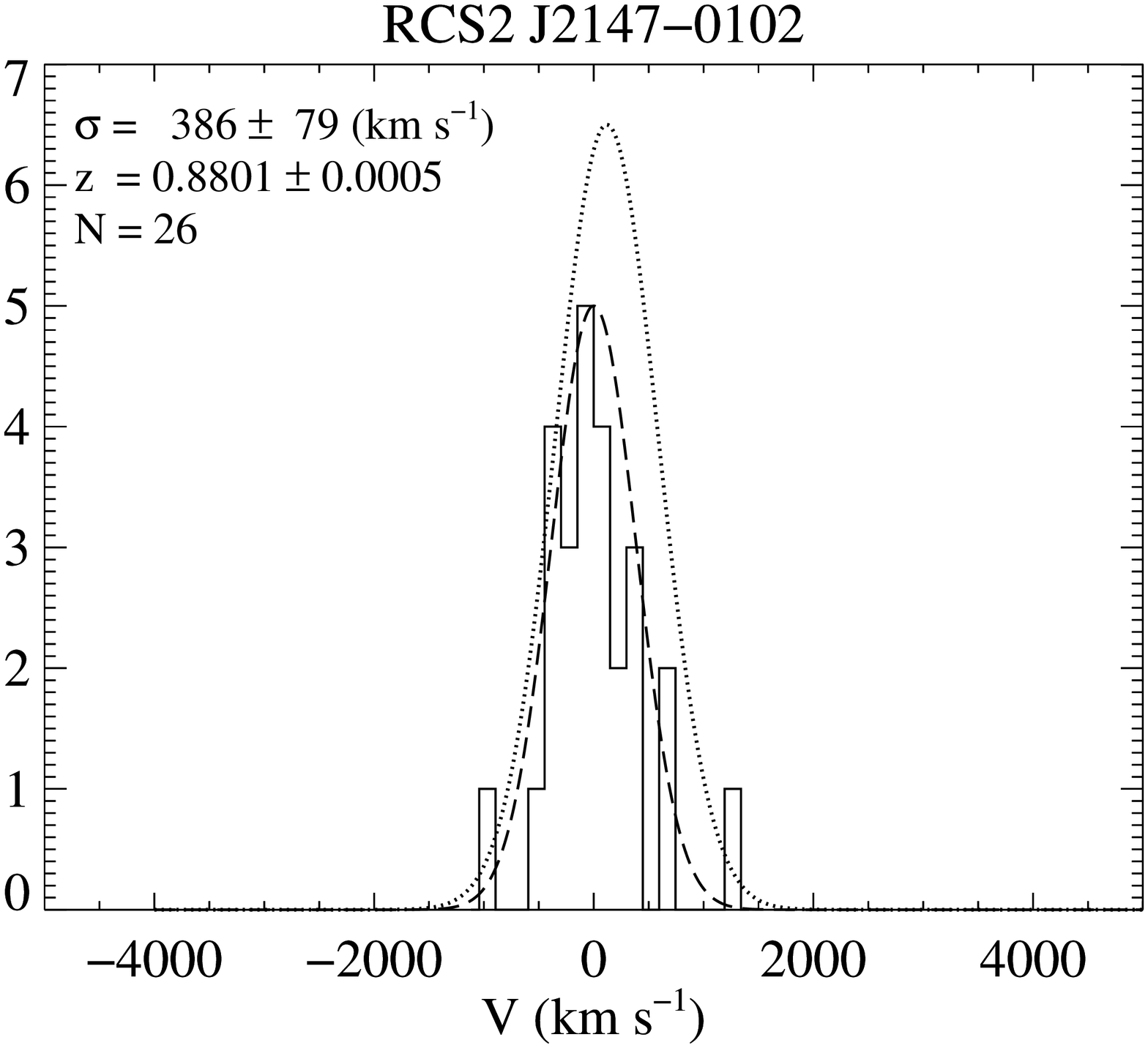} \\
\includegraphics[width=45mm,angle=90, trim= 0mm 18mm 0mm 26mm,clip]{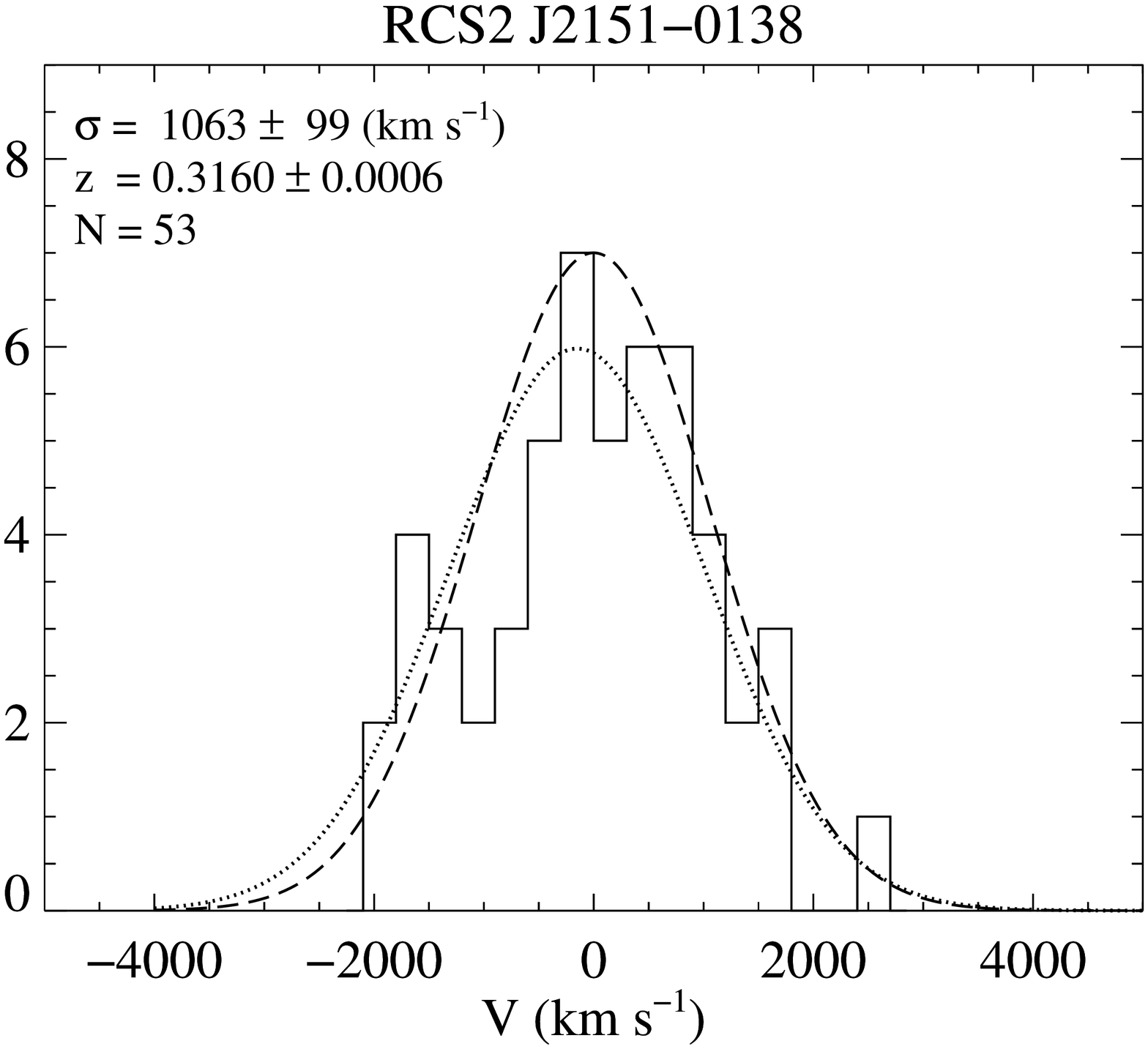} &
\includegraphics[width=45mm,angle=90, trim= 0mm 18mm 0mm 26mm,clip]{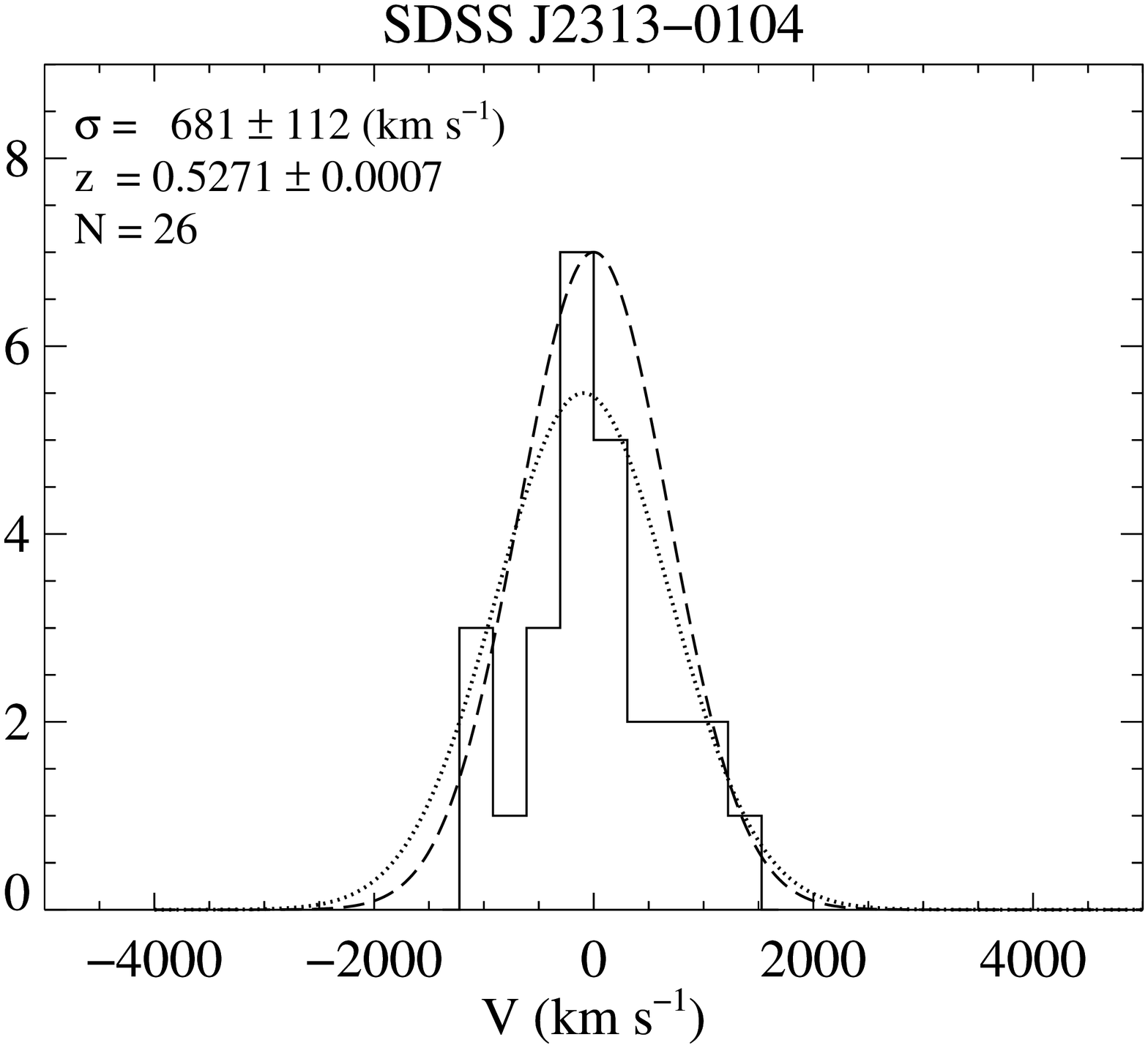} &
\includegraphics[width=45mm,angle=90, trim= 0mm 18mm 0mm 26mm,clip]{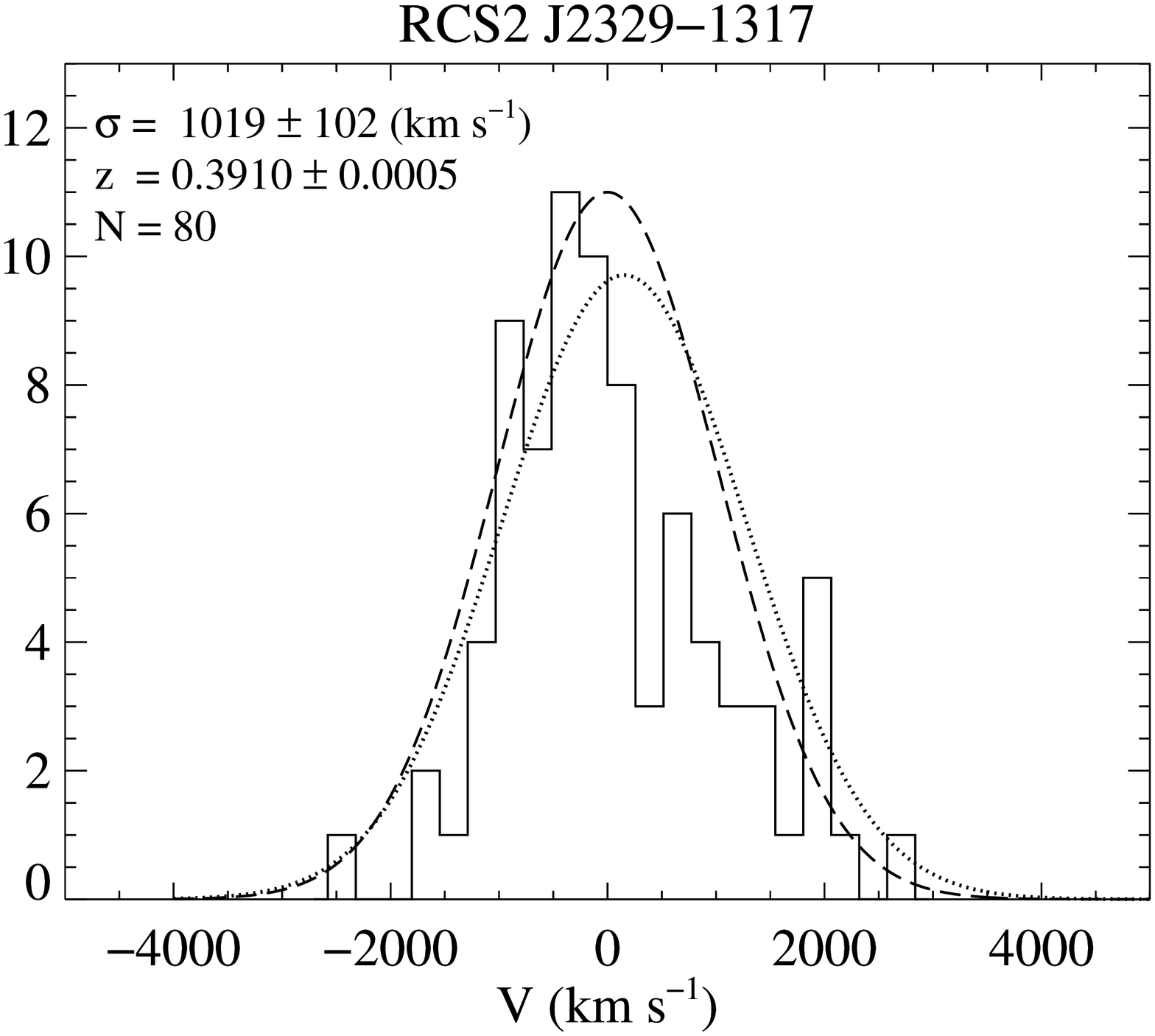} \\
\includegraphics[width=45mm,angle=90, trim= 0mm 18mm 0mm 26mm,clip]{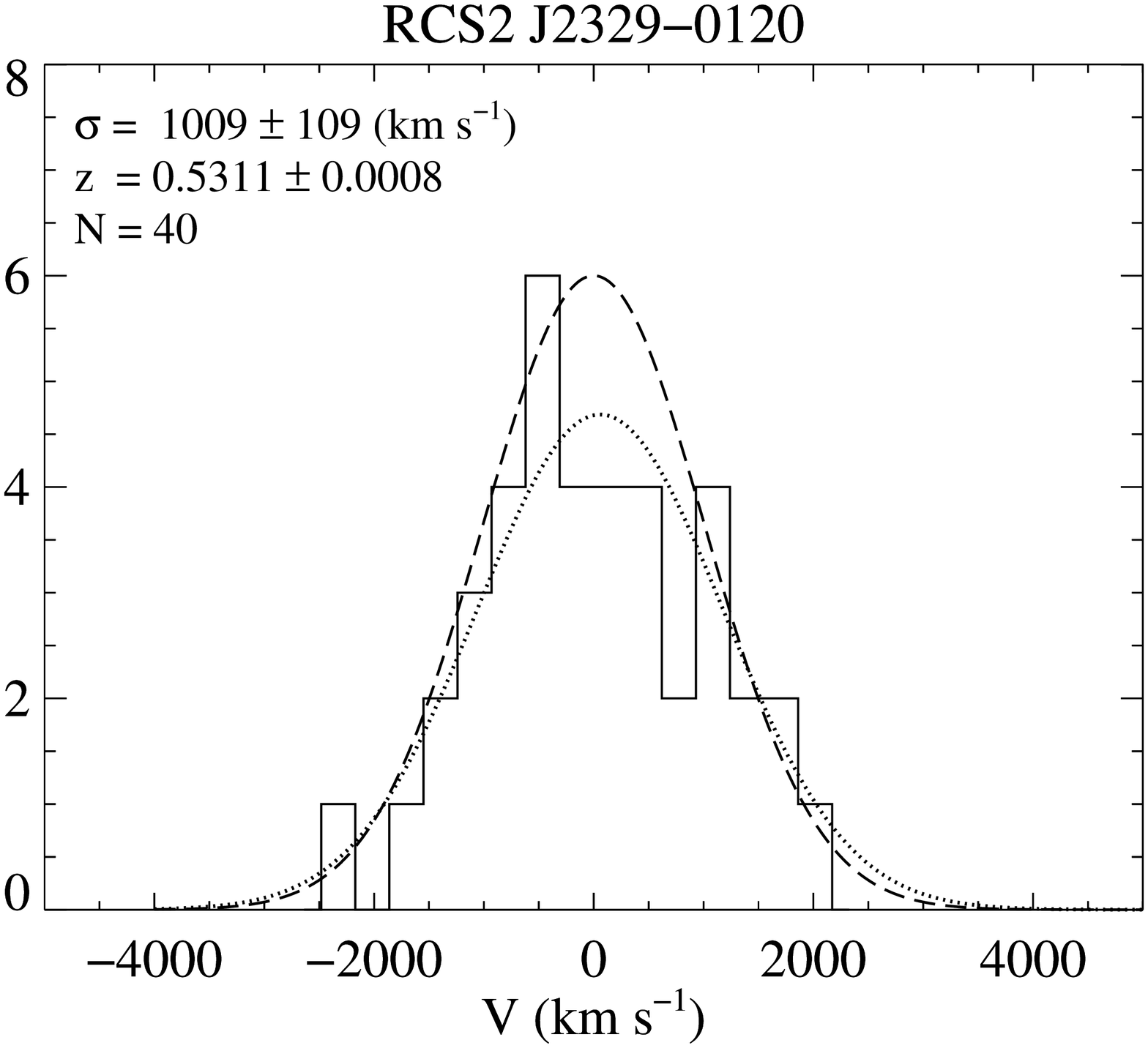} &
\includegraphics[width=45mm,angle=90, trim= 0mm 18mm 0mm 26mm,clip]{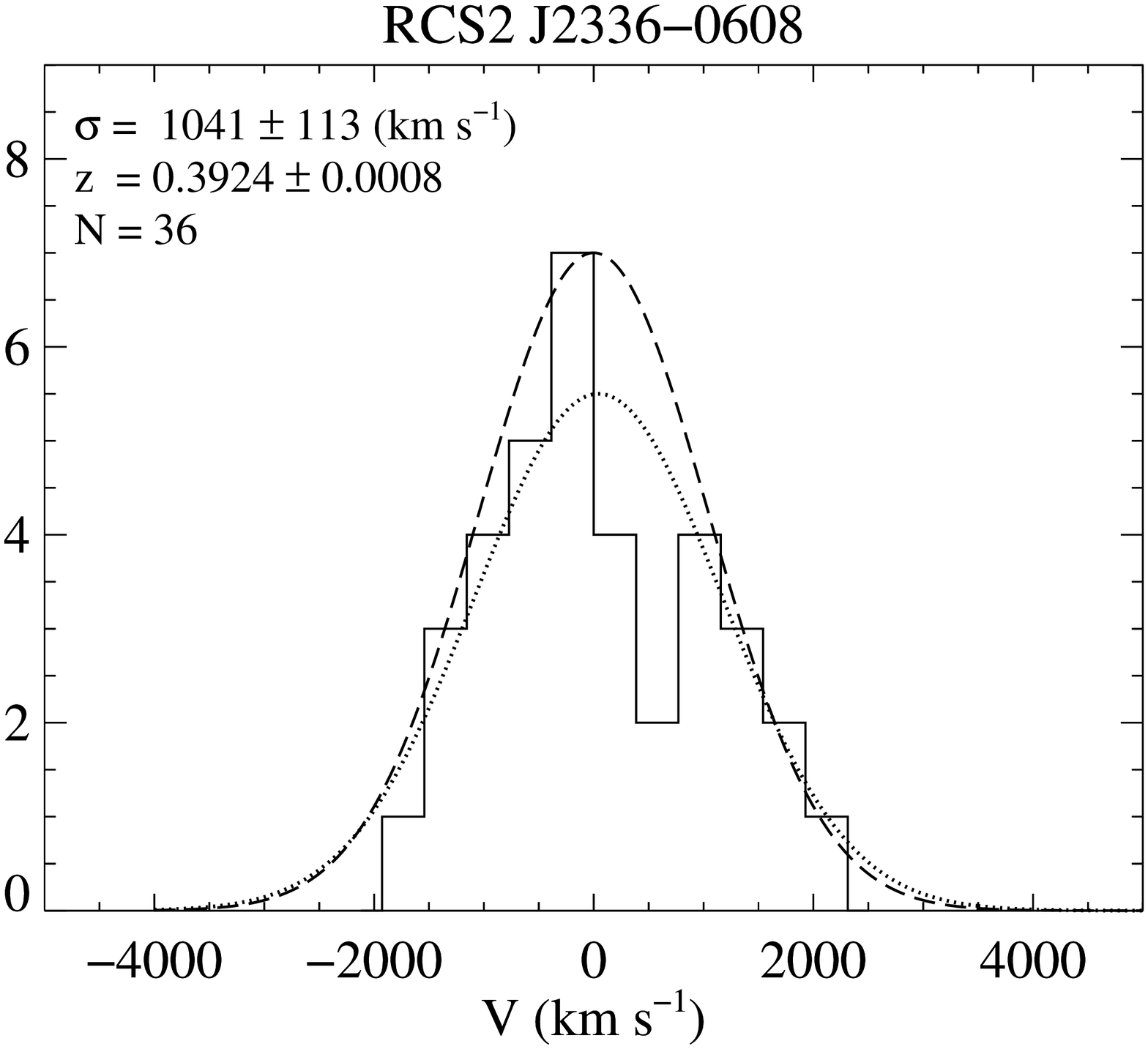} &
\end{tabular}
\caption{ \label{fig:vel_hist_2}
Velocity histograms for 11 out of 23 of our clusters with $N_{\textnormal{mem}} \gtrsim 20$ showing a uni-modal distribution. 
The histograms are displayed in the same fashion as in Figure \ref{fig:vel_hist_1}.
Also, note that the y-axes differ between plots.
} 
\end{center}
\end{figure*}

\begin{figure*}[h!]
\begin{center}
\begin{tabular}{c c c}
\includegraphics[width=45mm,angle=90, trim= 0mm 18mm 0mm 23mm,clip]{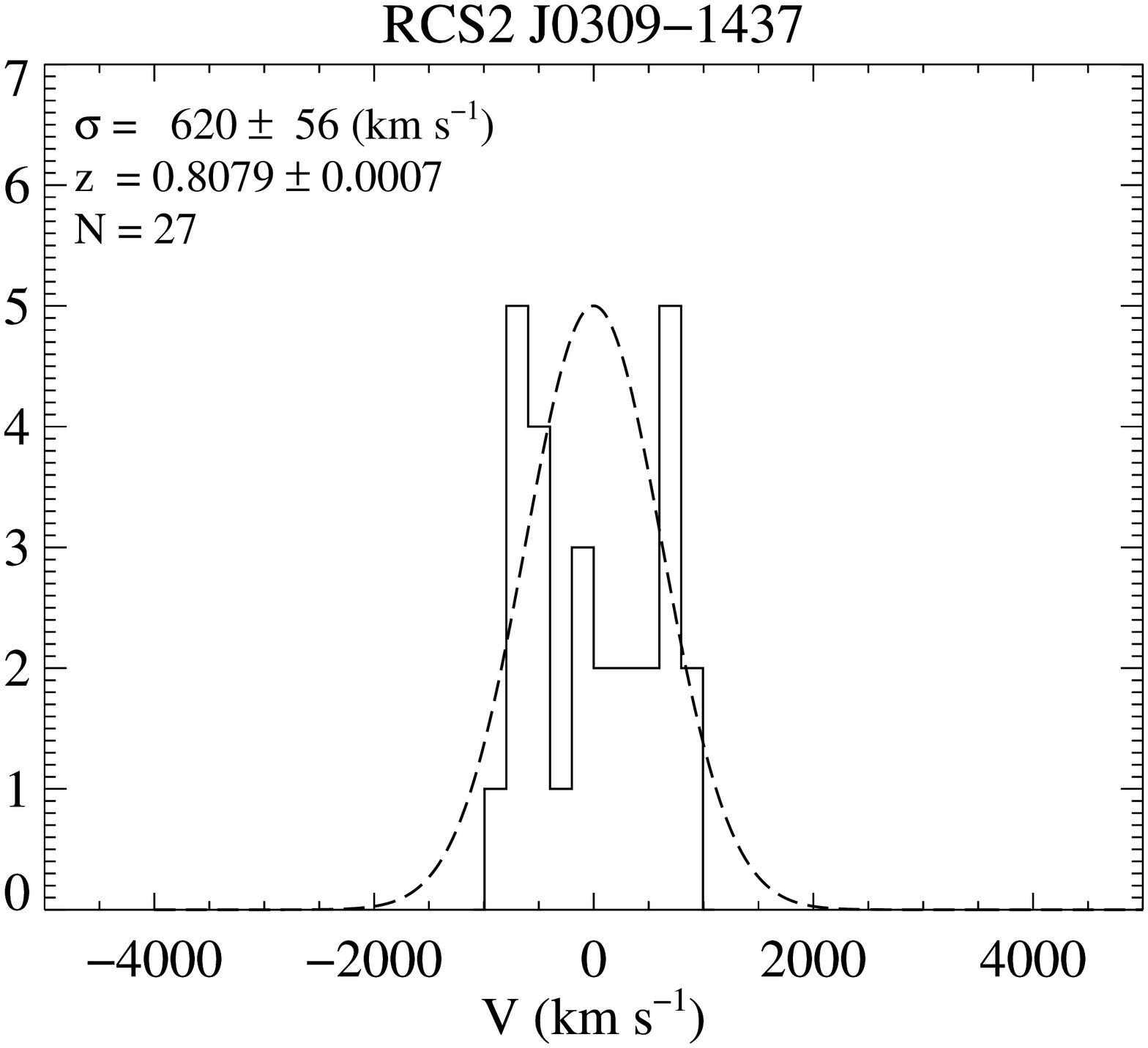} &
\includegraphics[width=45mm,angle=90, trim= 0mm 18mm 0mm 23mm,clip]{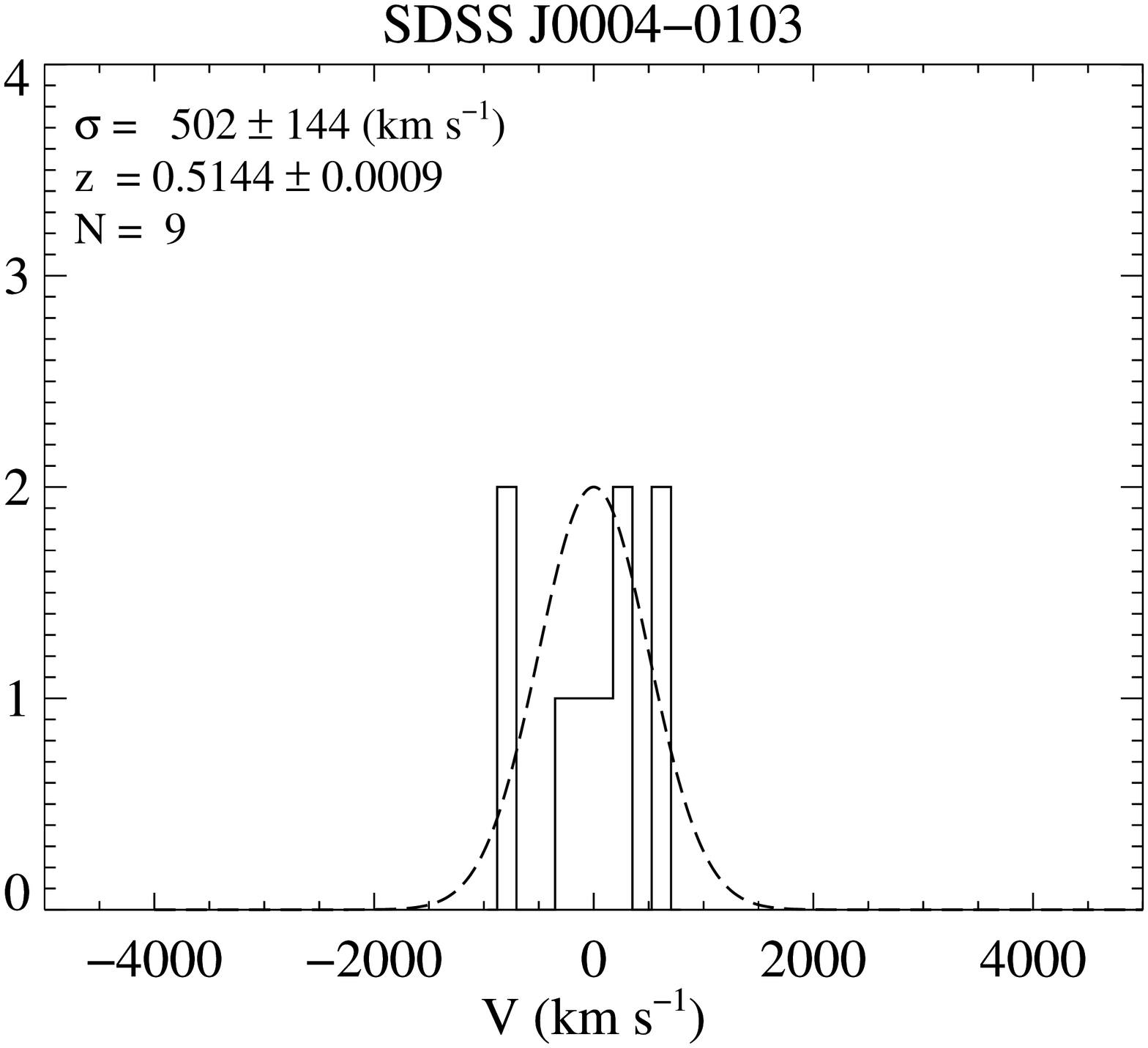} &
\includegraphics[width=45mm,angle=90, trim= 0mm 18mm 0mm 23mm,clip]{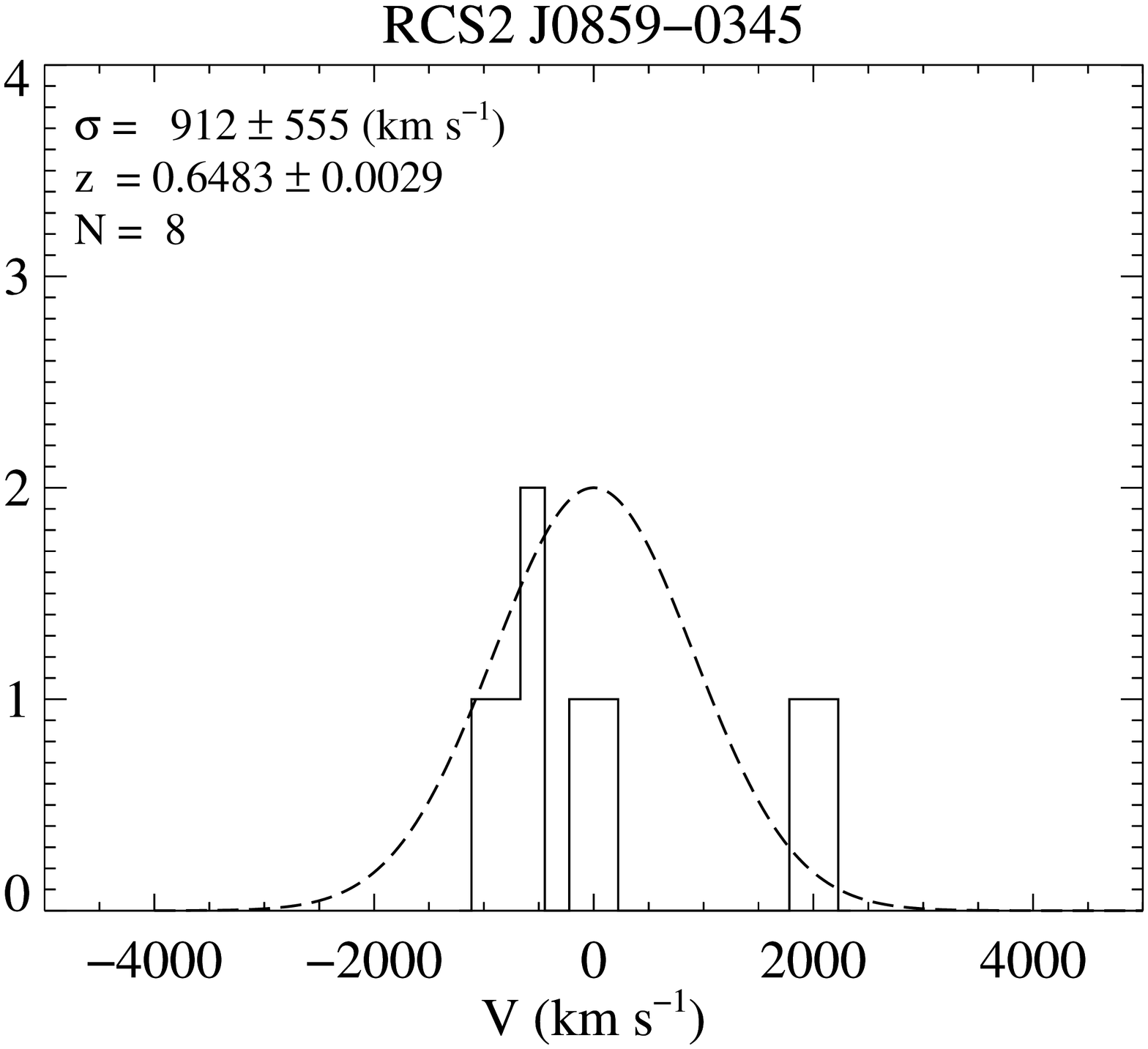} \\
\includegraphics[width=45mm,angle=90, trim= 0mm 18mm 0mm 23mm,clip]{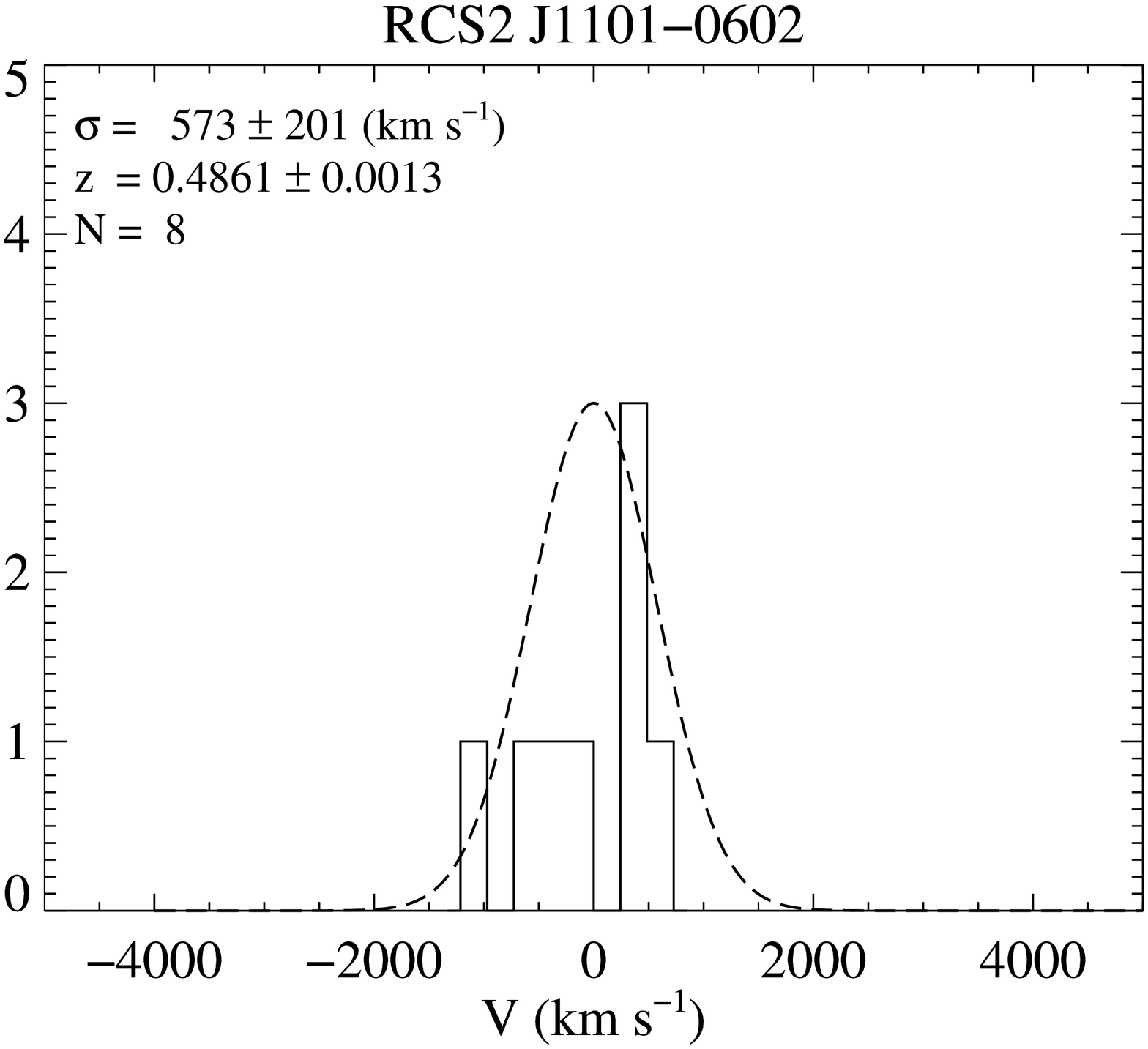} &
\includegraphics[width=45mm,angle=90, trim= 0mm 18mm 0mm 23mm,clip]{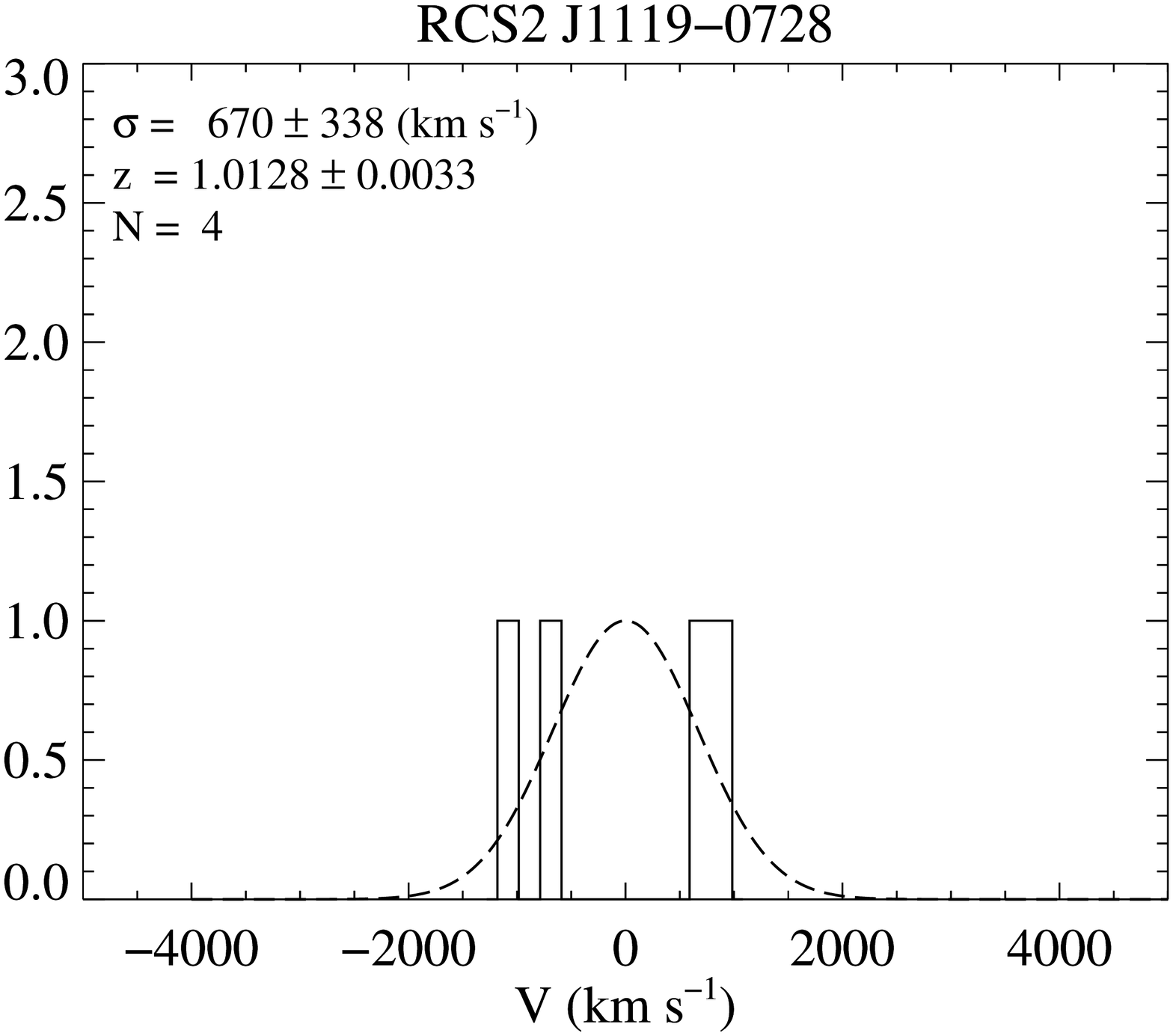} &
\includegraphics[width=45mm,angle=90, trim= 0mm 18mm 0mm 23mm,clip]{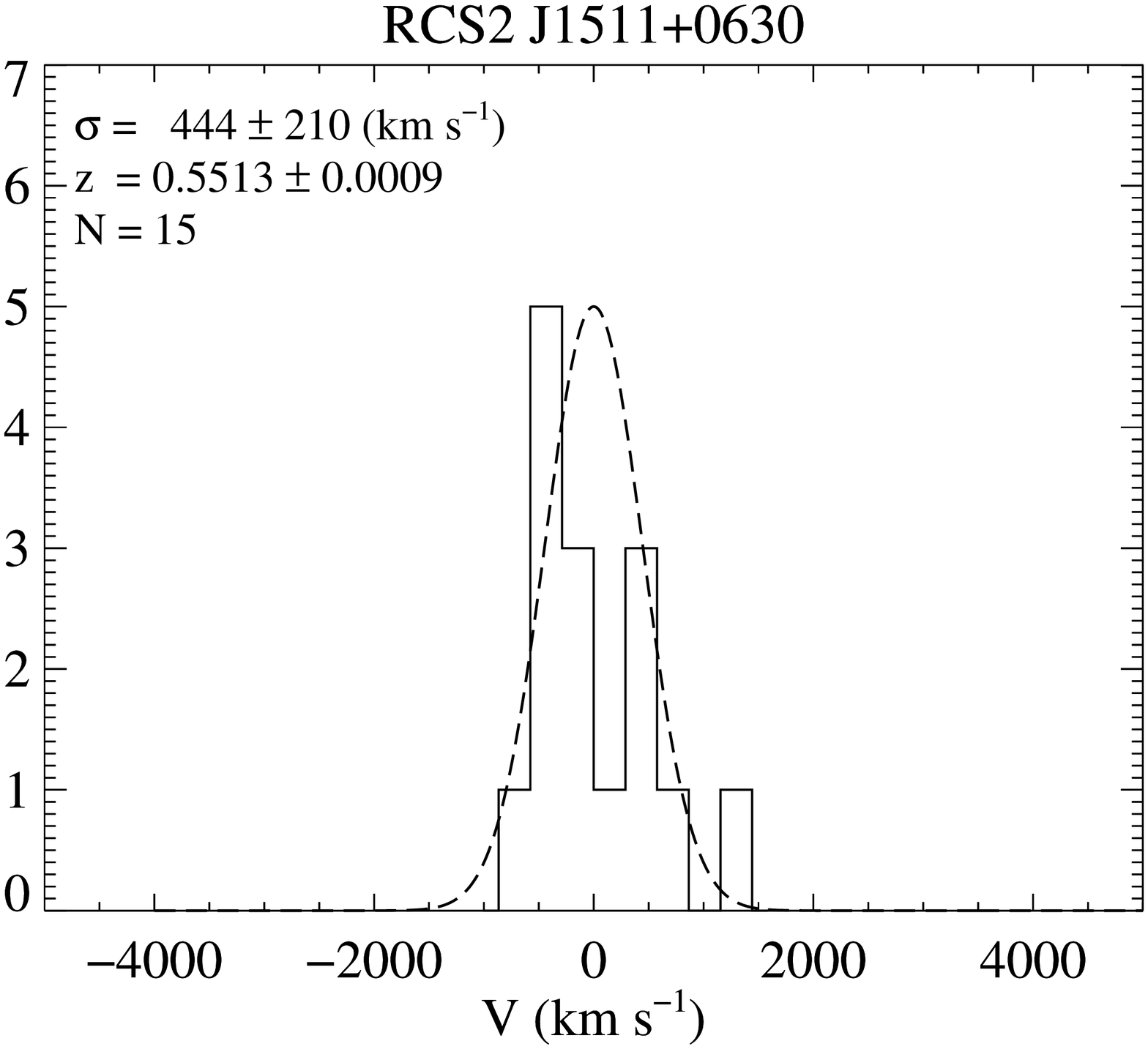} \\
\end{tabular}
\caption{ \label{fig:vel_hist_3}
The upper left panel shows the velocity histogram of RCS2 J0309$-$1437, which is the only system in our sample (with $N_{\textnormal{mem}} \gtrsim 20$)
where the KS test suggests that its velocity distribution is not consistent with a uni-modal distribution. 
The other panels show the velocity histograms for five clusters with $N_{\textnormal{mem}} \lesssim 20$, 
where the analysis of their velocity data is not possible due to the few number of spectroscopically confirmed cluster members. 
The histograms are displayed in the same fashion as in Figure \ref{fig:vel_hist_1}.
Dashed lines correspond to Gaussian distributions with the mean and variance  equal to the first guess of the mean velocity and velocity dispersion squared of each cluster.
Also, note that the y-axes differ between plots.
} 
\end{center}
\end{figure*}

\clearpage

\section{Strong lensing galaxy clusters} \label{appendix:SL_clusters}

\begin{figure*}[h!]
\begin{center}
\begin{tabular}{c c}
\includegraphics[width=87mm,angle=0, trim= 0mm 0mm 0mm 0mm,clip]{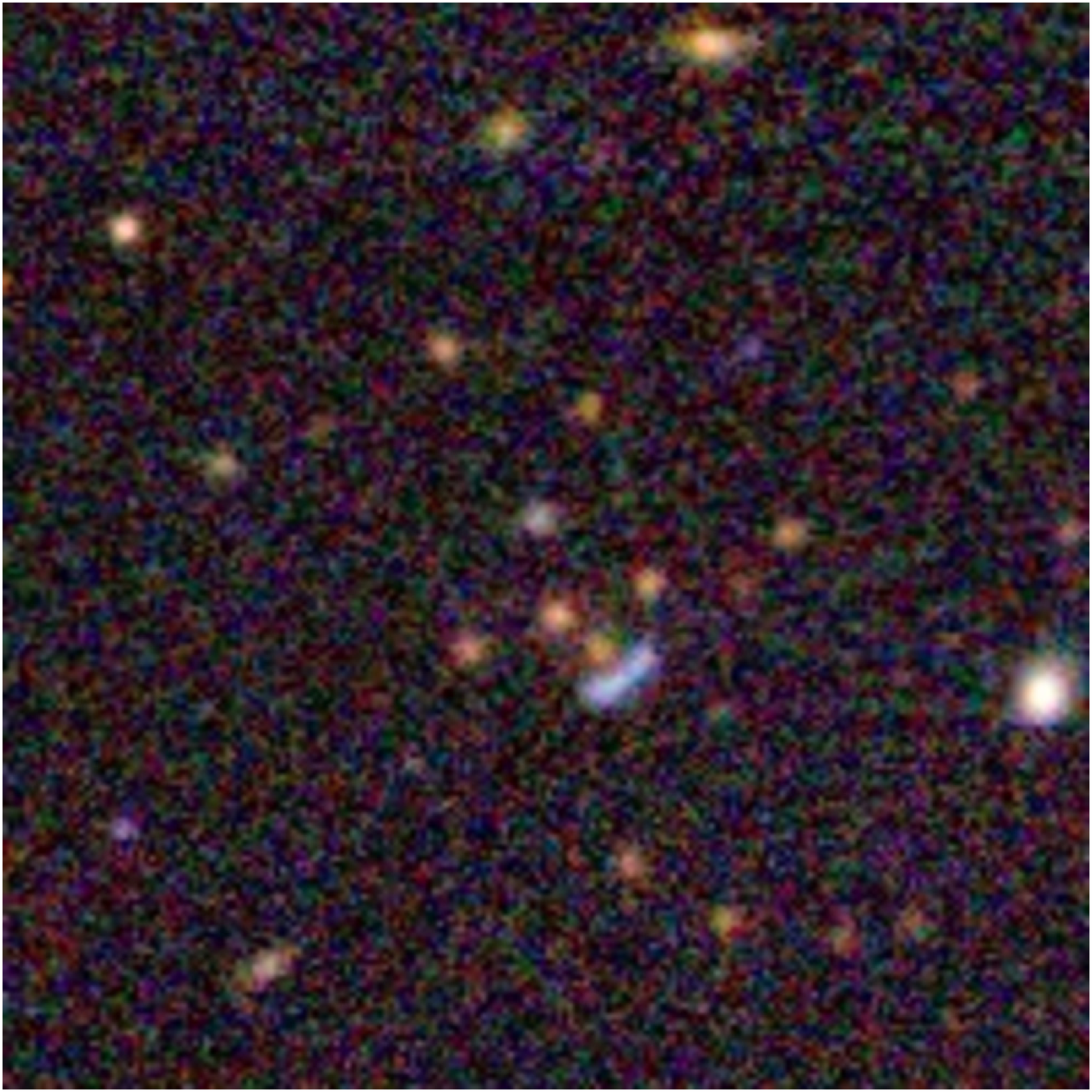} & 
\includegraphics[width=87mm,angle=0, trim= 0mm 0mm 0mm 0mm,clip]{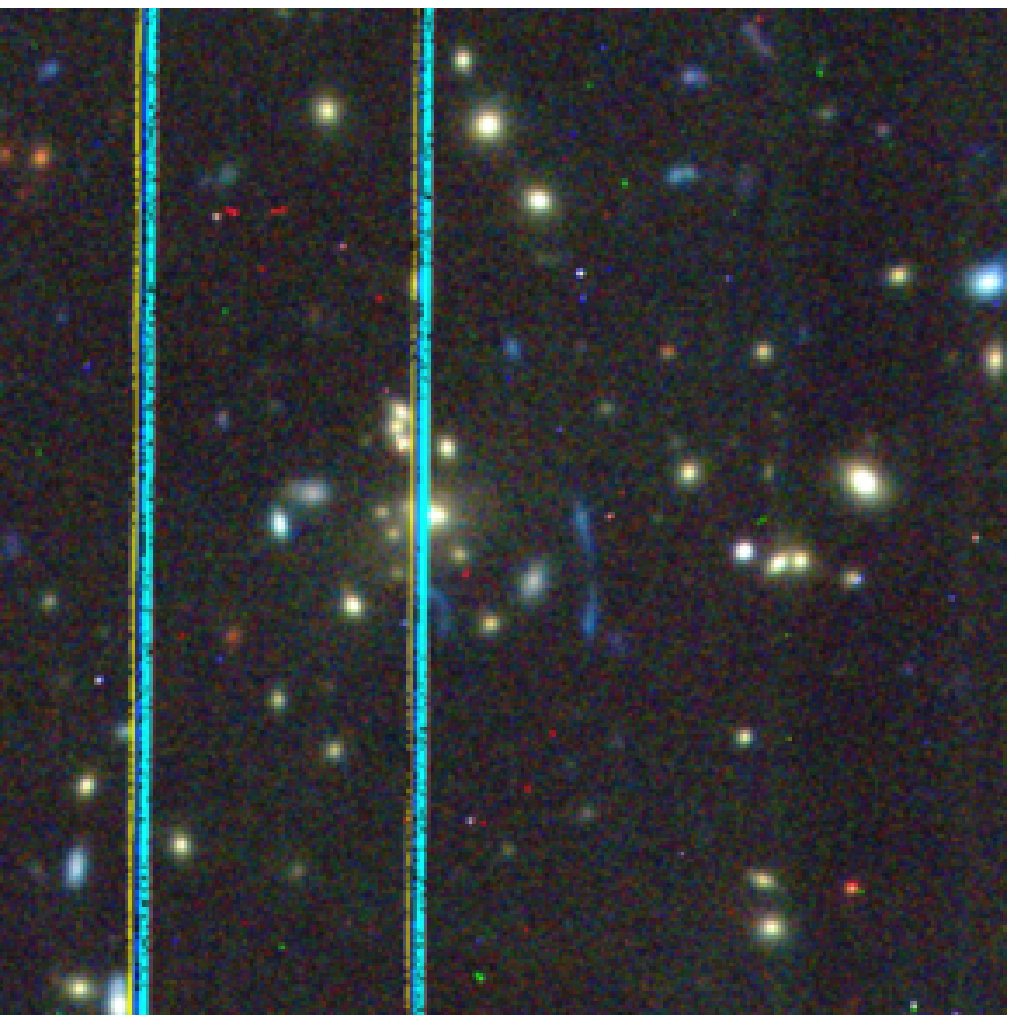} \\[3ex]
\includegraphics[width=87mm,angle=0, trim= 0mm 0mm 0mm 0mm,clip]{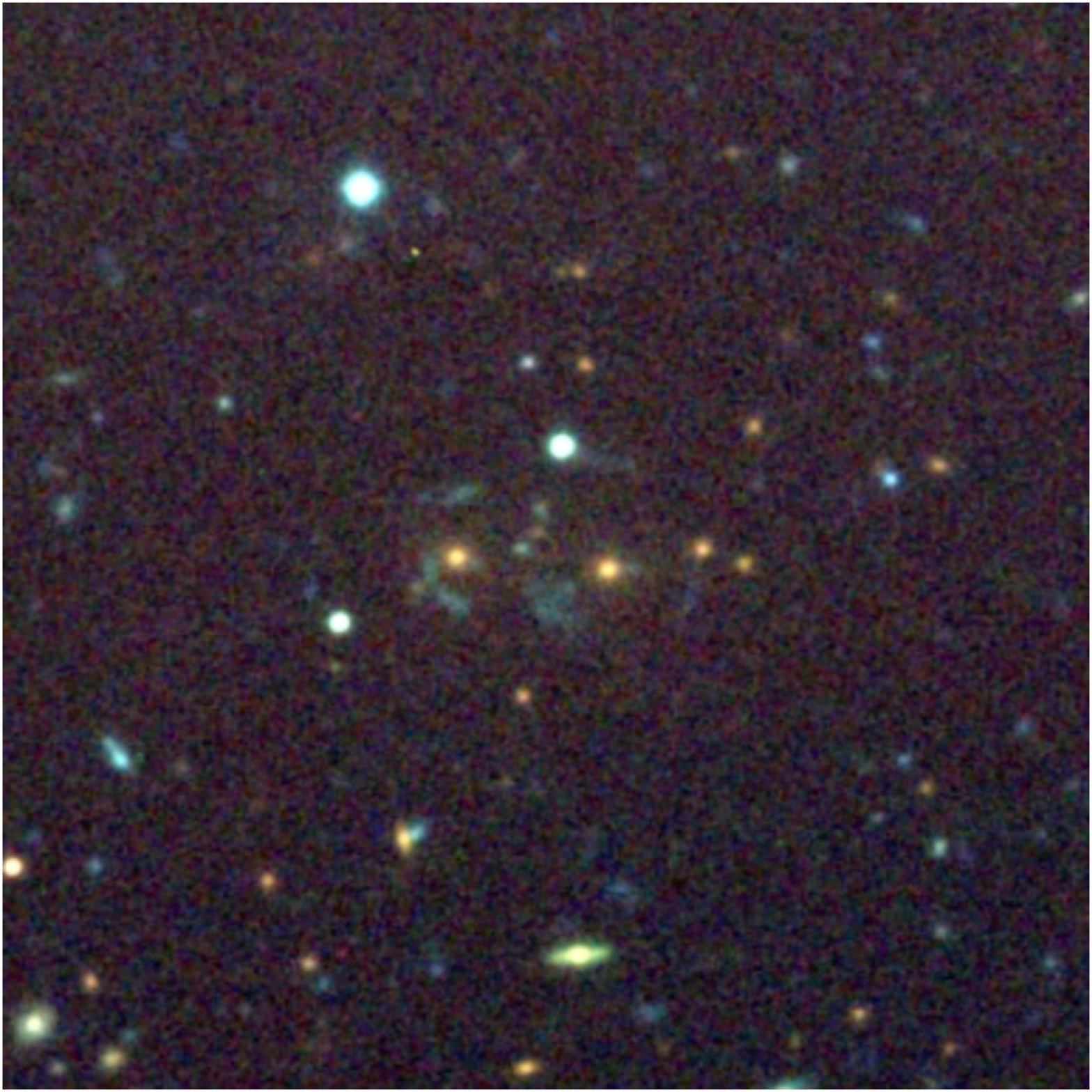} &
\includegraphics[width=87mm,angle=0, trim= 0mm 0mm 0mm 0mm,clip]{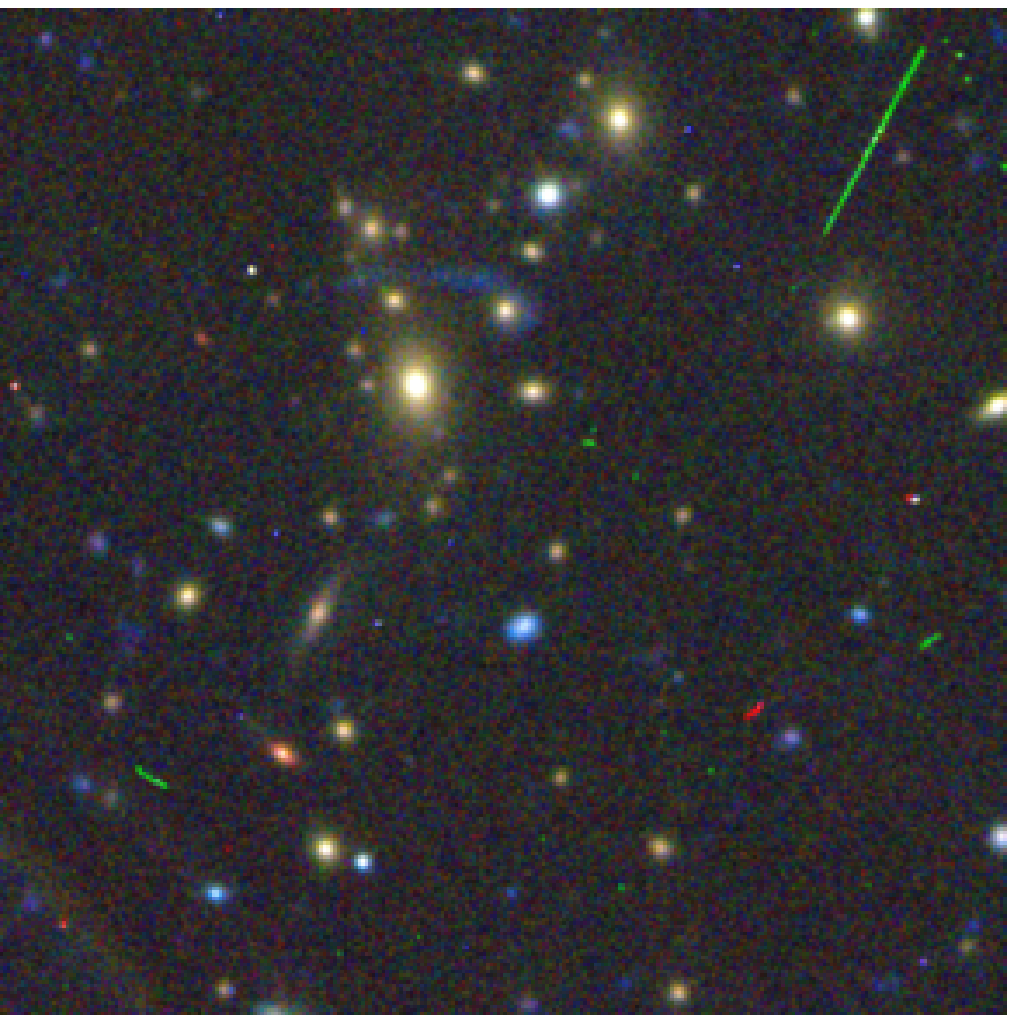} \\
\end{tabular}
\put(-435, 240){\bf \large \color{white} SDSS J0004$-$0103}
   \put(-370, 88){\bf \color{yellow} S1.1}
\put(-175, 240){\bf \large \color{white} RCS2 J0034+0225}
   \put(-102, 128){\bf \color{yellow} S1.1}
   \put(-102, 105){\bf \color{yellow} S1.2}
\put(-435, -20){\bf \large \color{white} RCS2 J0038+0215}
    \put(-413,-155){\bf \color{yellow} S1.1}
    \put(-413,-108){\bf \color{yellow} S1.2}
\put(-175, -20){\bf \large \color{white} RCS2 J0047+0508}
    \put(-140,-62){\bf \color{yellow} S1.1}
    \put(-190,-200){\bf \color{yellow} S2.1}

\caption{ \label{fig:SLclusters_fov_01}
From the top-left to bottom-right panels we show the SL-selected galaxy clusters
SDSS J0004$-$0103, RCS2 J0034+0225, RCS2 J0038+0215, and RCS2 J0047+0508. 
Lensed galaxies are labeled with markers that match the labels in Table \ref{table:lensed_galaxy}, where their spectroscopic redshifts are shown. 
Objects labeled with the same identifier but different decimal numbers (S1.1, S1.2, etc) have the same redshifts within the errors, 
and we therefore assume that they correspond to multiple images of the same background source. 
Color composite images are made from \textit{grz} imaging obtained from the RCS-2 survey with CFHT/MegaCam.
All images cover a field of view of 75\arcsec$\times$75\arcsec.} 
\end{center}
\end{figure*}

\begin{figure*}[h!]
\begin{center}
\begin{tabular}{c c}
\includegraphics[width=87mm,angle=0, trim= 0mm 0mm 0mm 0mm,clip]{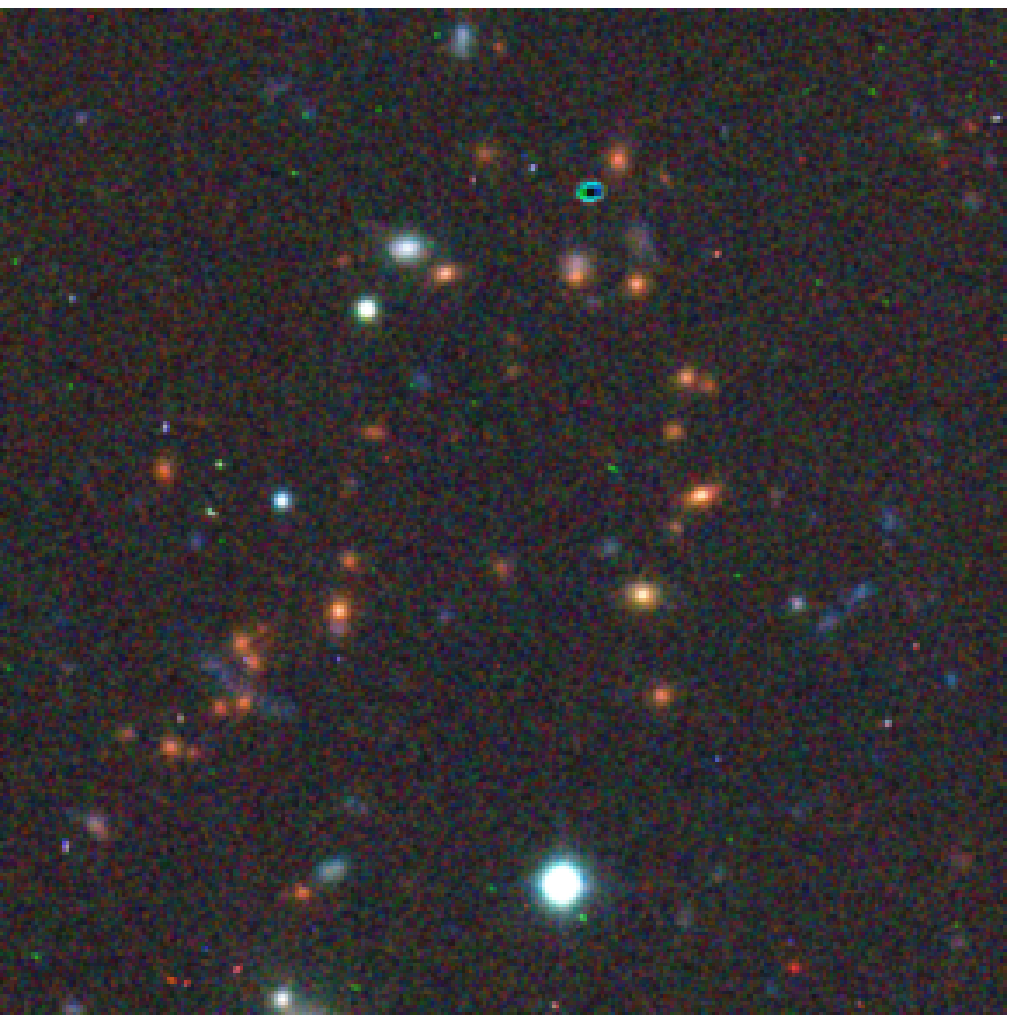} &
\includegraphics[width=87mm,angle=0, trim= 0mm 0mm 0mm 0mm,clip]{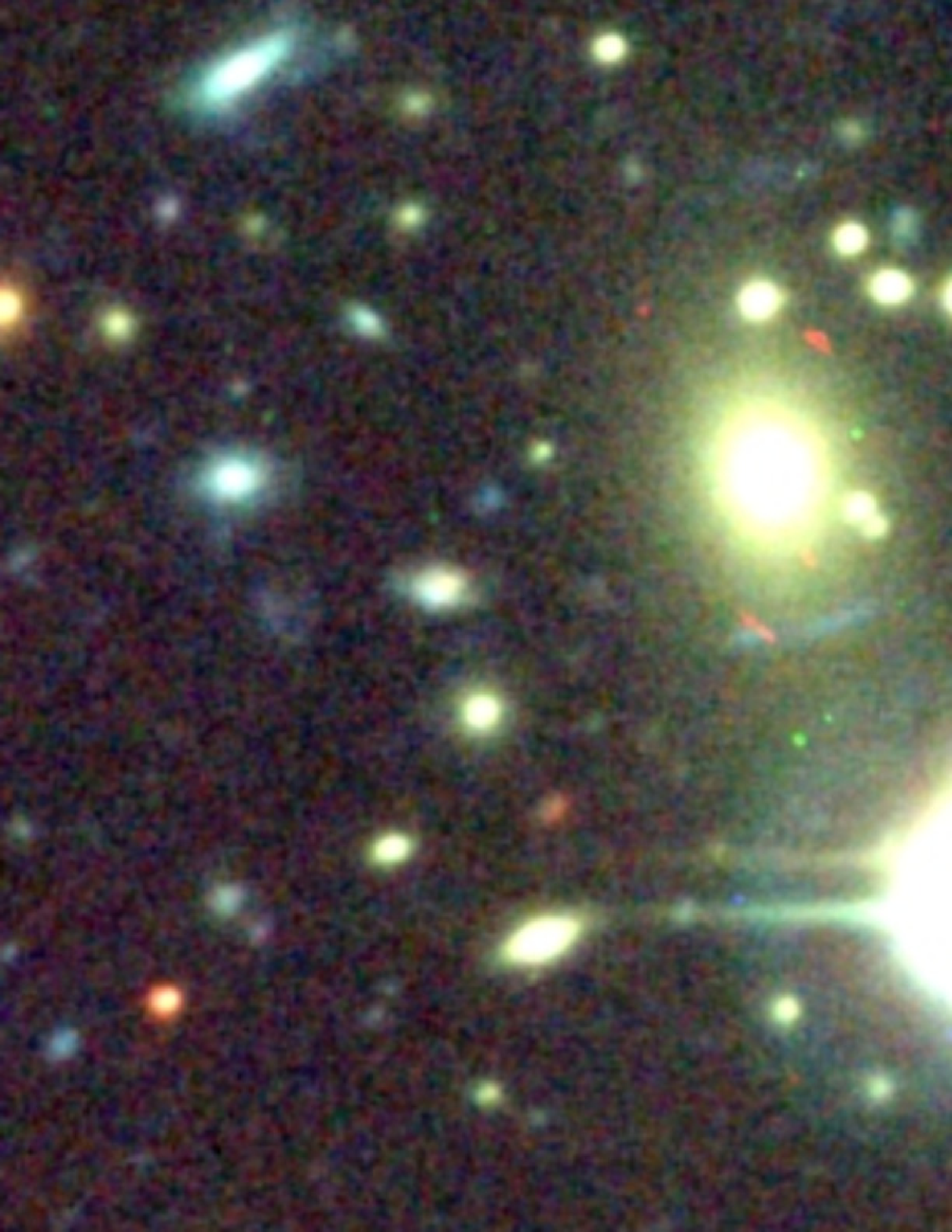} \\[3ex]
\includegraphics[width=87mm,angle=0, trim= 0mm 0mm 0mm 0mm,clip]{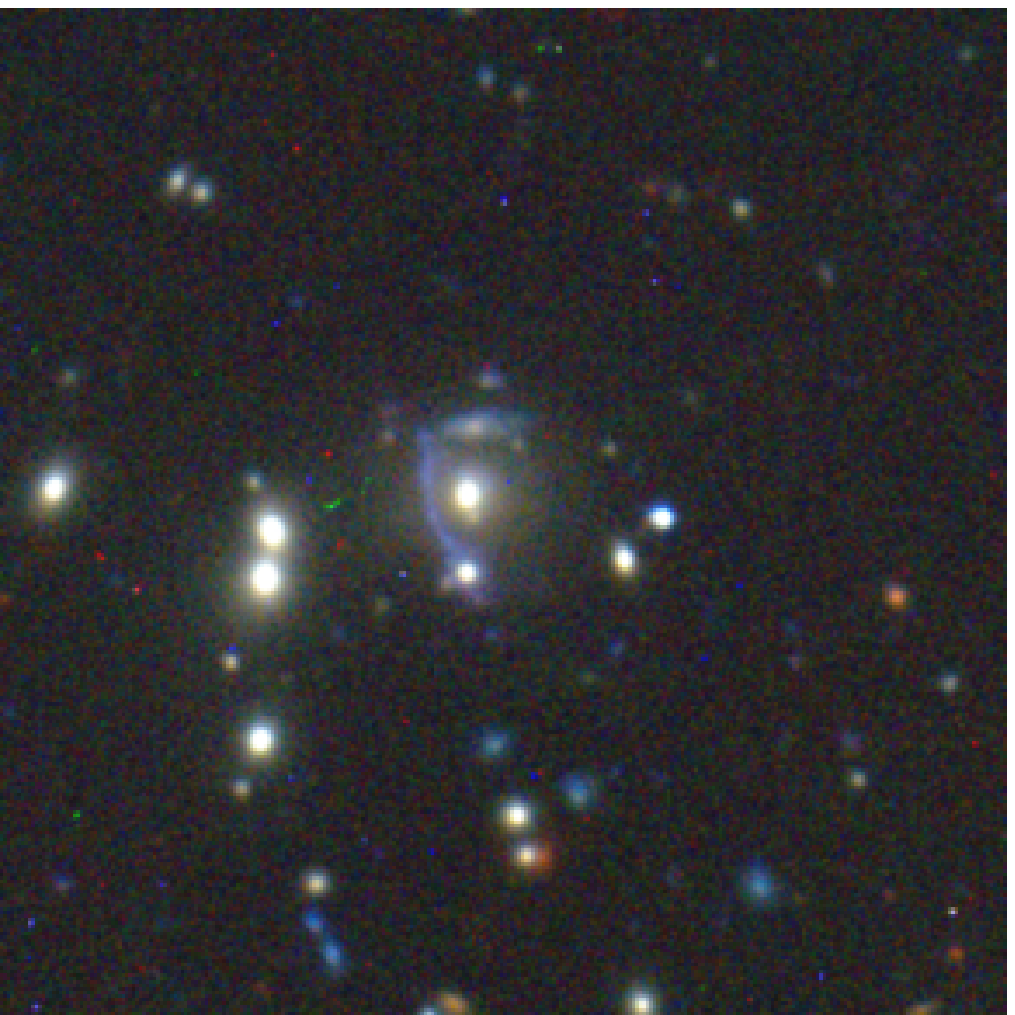} &
\includegraphics[width=87mm,angle=0, trim= 0mm 0mm 0mm 0mm,clip]{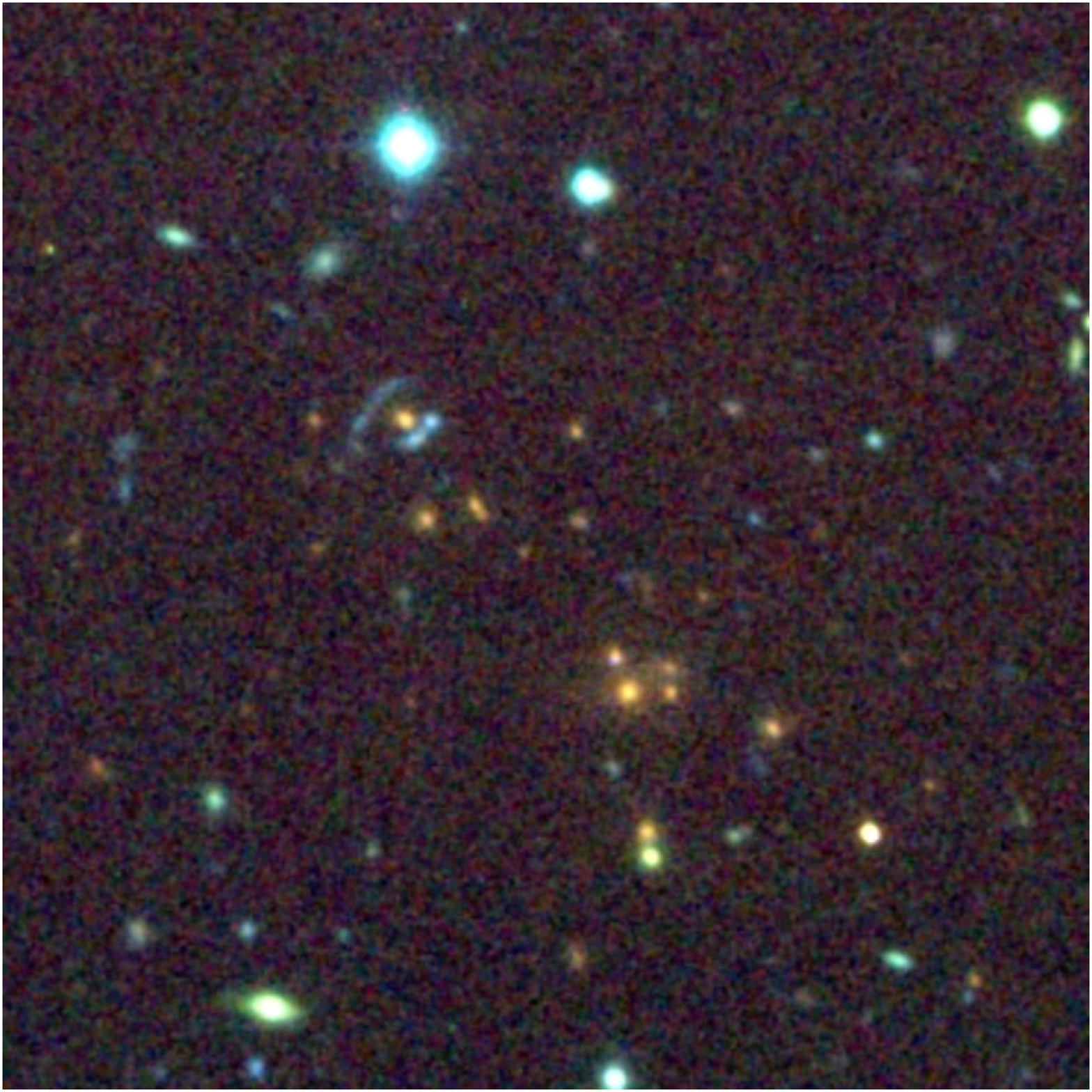} \\
\end{tabular}
\put(-435, 240){\bf \large \color{white} RCS2 J0052+0433}
    \put(-292, 106){\bf \color{yellow} S1.1}
    \put(-444, 71){\bf \color{yellow} S2.2}
    \put(-484, 98){\bf \color{yellow} S2.1}
\put(-175, 240){\bf \large \color{white} RCS2 J0057+0209}
    \put(-120, 94){\bf \color{yellow} S1.1}
    \put(-75, 109){\bf \color{yellow} S1.2}
\put(-435, -20){\bf \large \color{white} RCS2 J0252$-$1459}
    \put(-427,-125){\bf \color{yellow} S1.1}
\put(-175, -20){\bf \large \color{white} RCS2 J0309$-$1437}
    \put(-150,-100){\bf \color{yellow} S1.1}
    \put(-186,-110){\bf \color{yellow} S2.1}
\caption{ \label{fig:SLclusters_fov_02}
From the top-left to bottom-right panels we show the SL-selected galaxy clusters
RCS2 J0052+0433, RCS2 J0057+0209, RCS2 J0252$-$1459, and RCS2 J0309$-$1437. 
Lensed galaxies are labeled in the same manner as in Figure \ref{fig:SLclusters_fov_01}.
All images cover a field of view of 75\arcsec$\times$75\arcsec.} 
\end{center}
\end{figure*}

\begin{figure*}[h!]
\begin{center}
\begin{tabular}{c c}
\includegraphics[width=87mm,angle=0, trim= 0mm 0mm 0mm 0mm,clip]{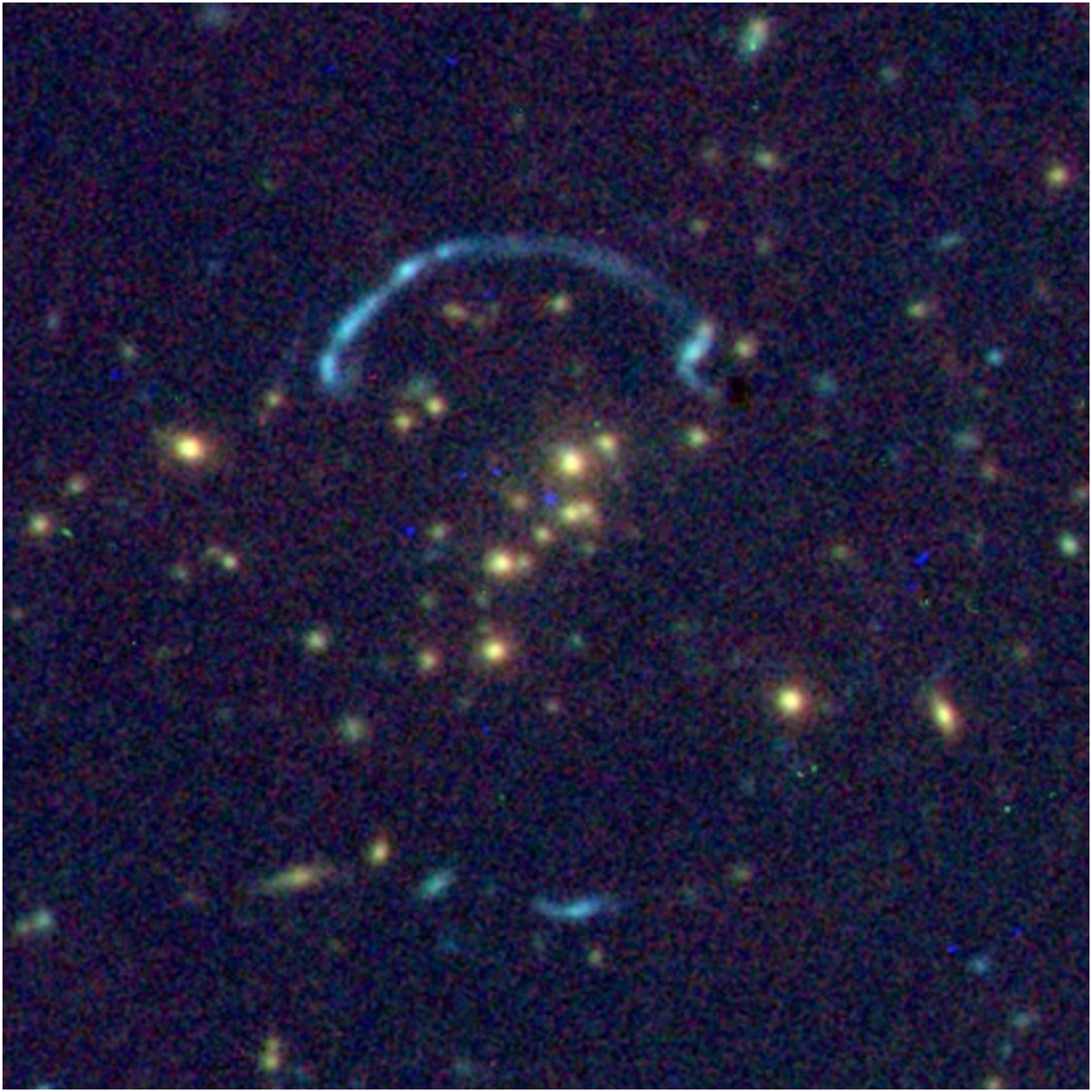} &
\includegraphics[width=87mm,angle=0, trim= 0mm 0mm 0mm 0mm,clip]{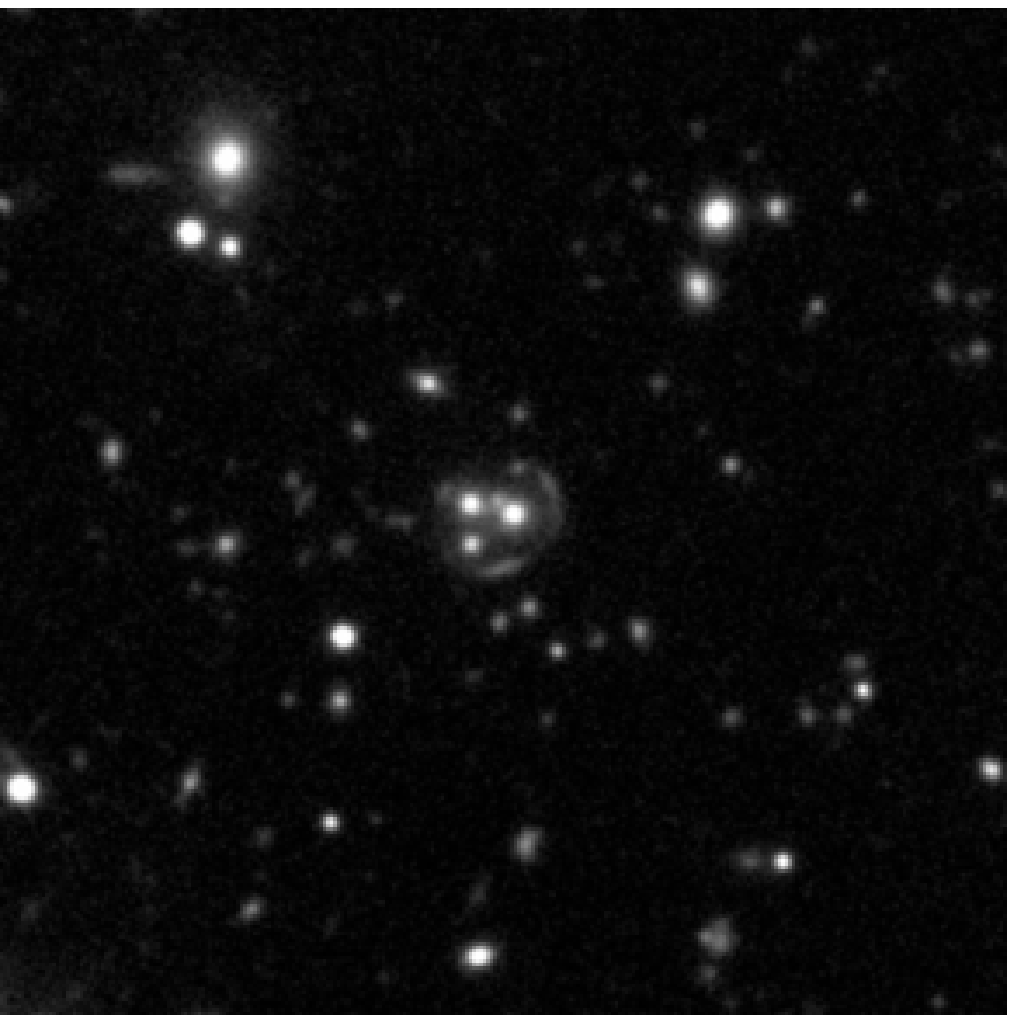} \\[3ex]
\includegraphics[width=87mm,angle=0, trim= 0mm 0mm 0mm 0mm,clip]{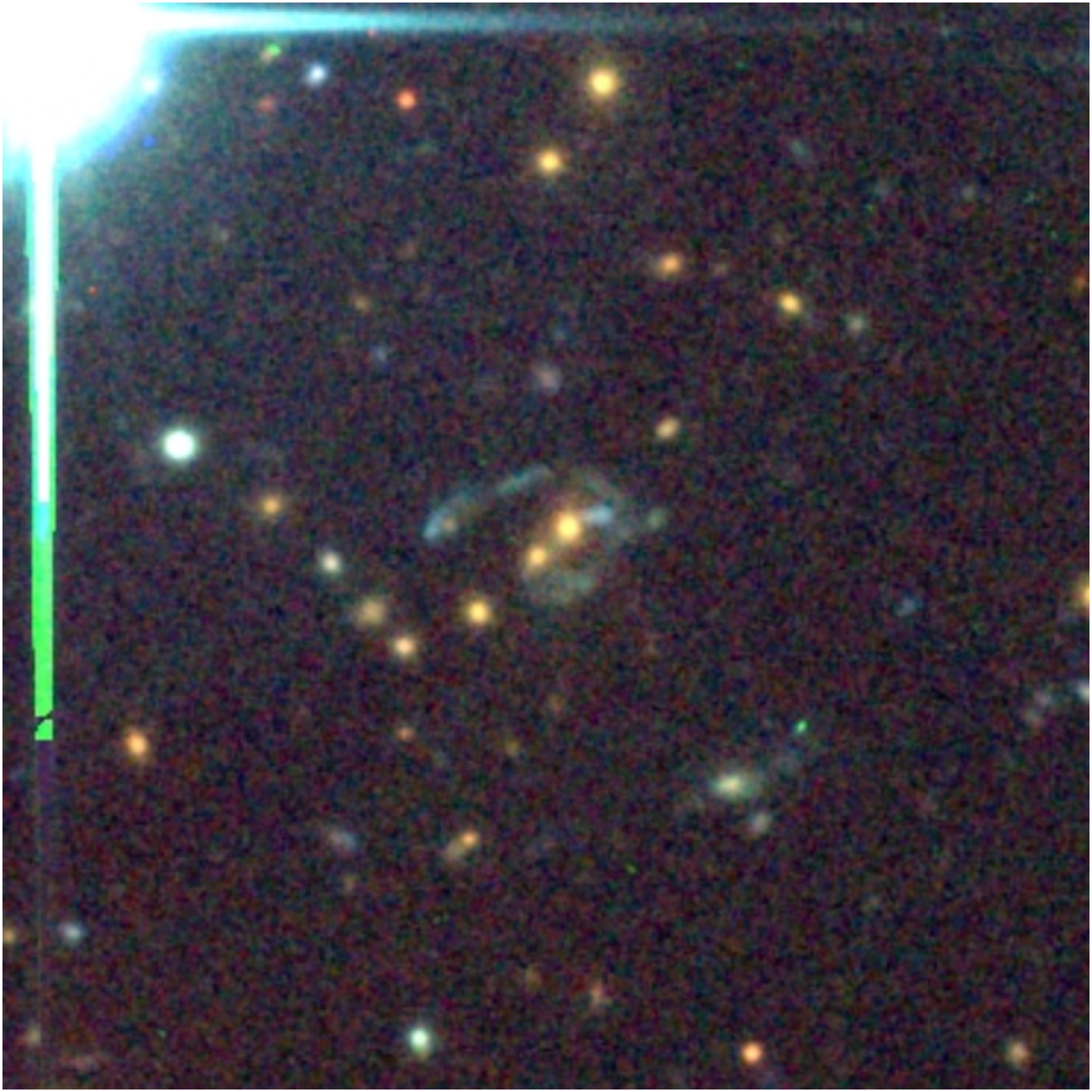} &
\includegraphics[width=87mm,angle=0, trim= 0mm 0mm 0mm 0mm,clip]{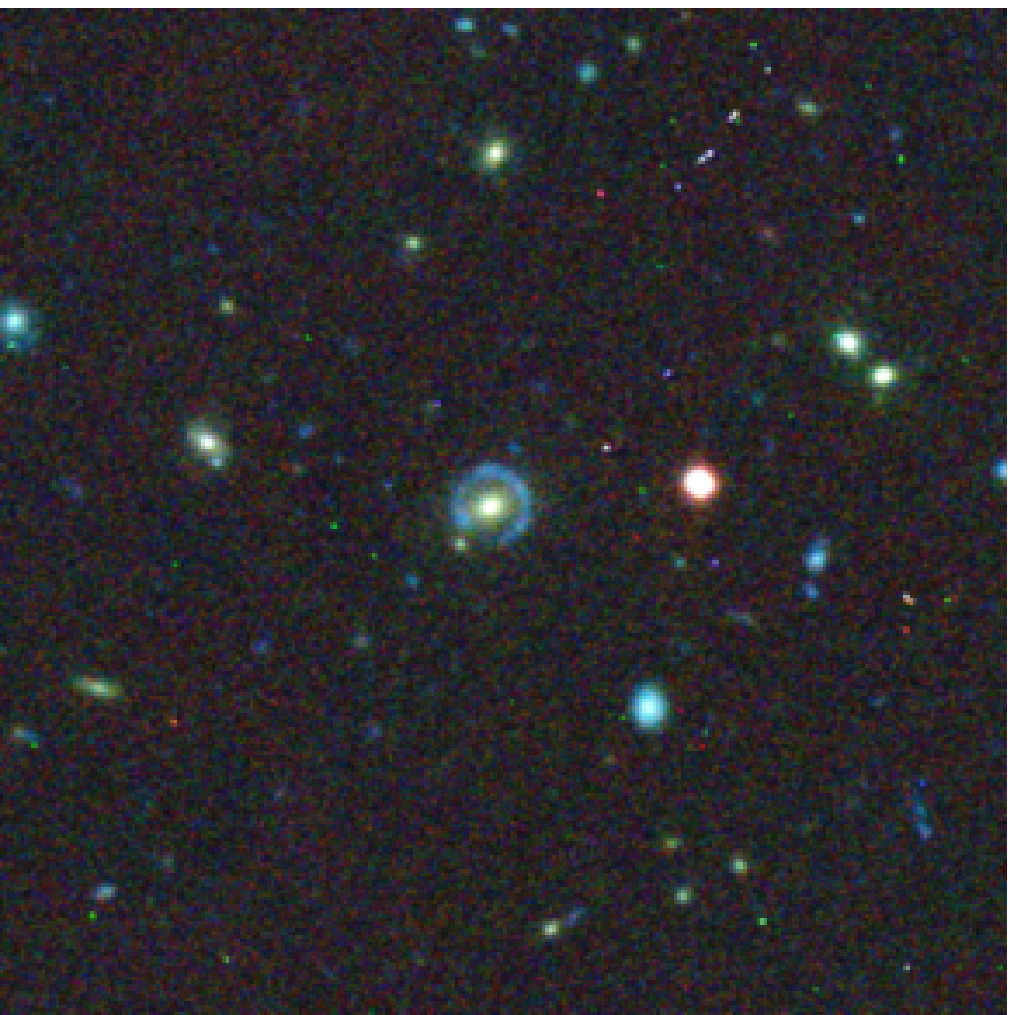} \\
\end{tabular}
\put(-435, 240){\bf \large \color{white} RCS2 J0327$-$1326}
    \put(-383, 40){\bf \color{yellow} S1.1}
    \put(-345, 175){\bf \color{yellow} S1.2}
    \put(-450, 188){\bf \color{yellow} S1.3}
\put(-175, 240){\bf \large \color{white} RCS2 J0859$-$0345}
    \put(-110, 135){\bf \color{yellow} S1.1}
\put(-435, -20){\bf \large \color{white} RCS2 J1055$-$0459}
    \put(-434,-120){\bf \color{yellow} S1.1}
    \put(-397,-104){\bf \color{yellow} S1.2}
\put(-175, -20){\bf \large \color{white} RCS2 J1101$-$0602}
    \put(-118,-122){\bf \color{yellow} S1.1}
\caption{ \label{fig:SLclusters_fov_03}
From the top-left to bottom-right panels we show the SL-selected galaxy clusters
RCS2 J0327$-$1326, RCS2 J0859$-$0345, RCS2 J1055$-$0459, and RCS2 J1101$-$0602. 
Lensed galaxies are labeled in the same manner as in Figure \ref{fig:SLclusters_fov_01}.
All images cover a field of view of 75\arcsec$\times$75\arcsec. 
} 
\end{center}
\end{figure*}

\begin{figure*}[h!]
\begin{center}
\begin{tabular}{c c}
\includegraphics[width=87mm,angle=0, trim= 0mm 0mm 0mm 0mm,clip]{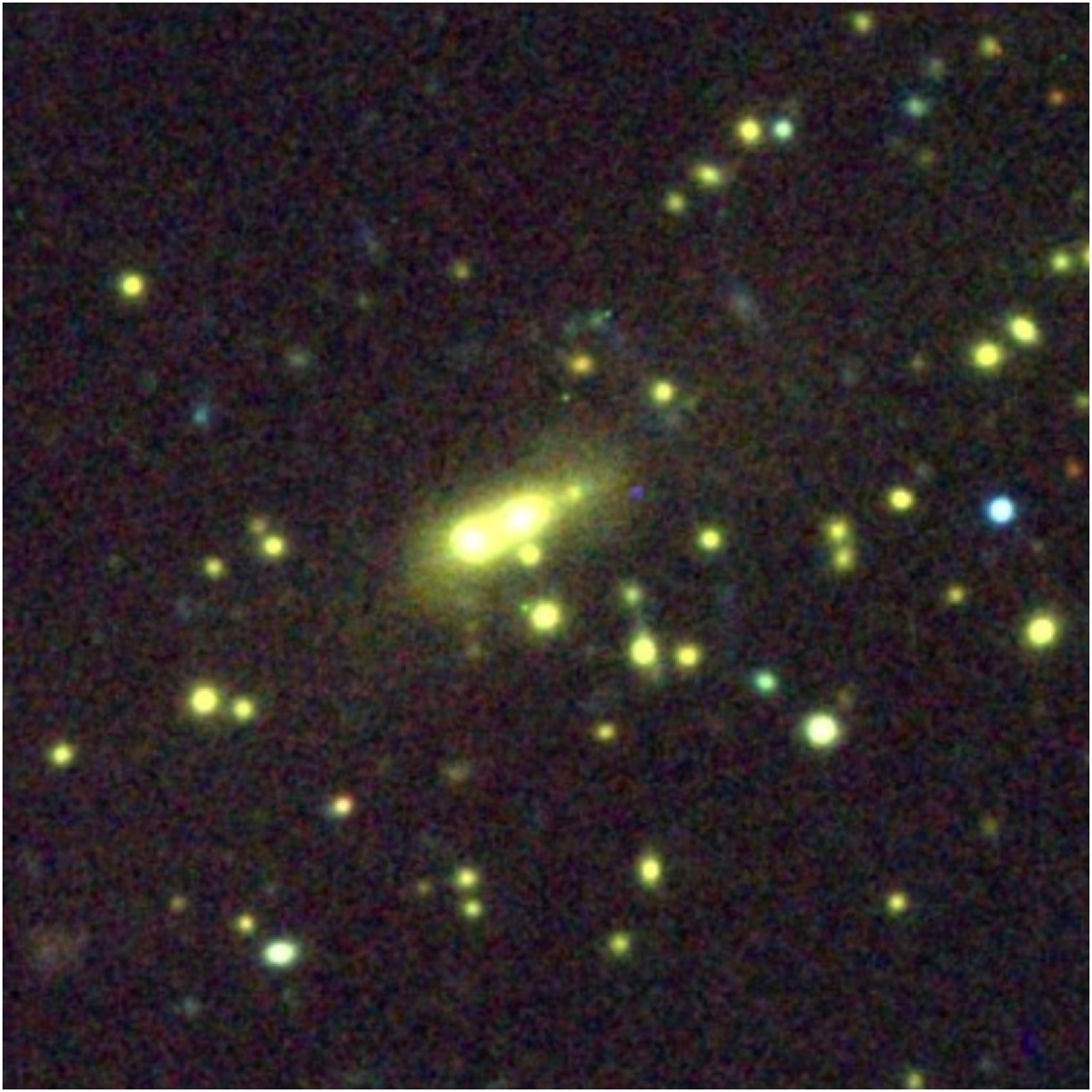} &
\includegraphics[width=87mm,angle=0, trim= 0mm 0mm 0mm 0mm,clip]{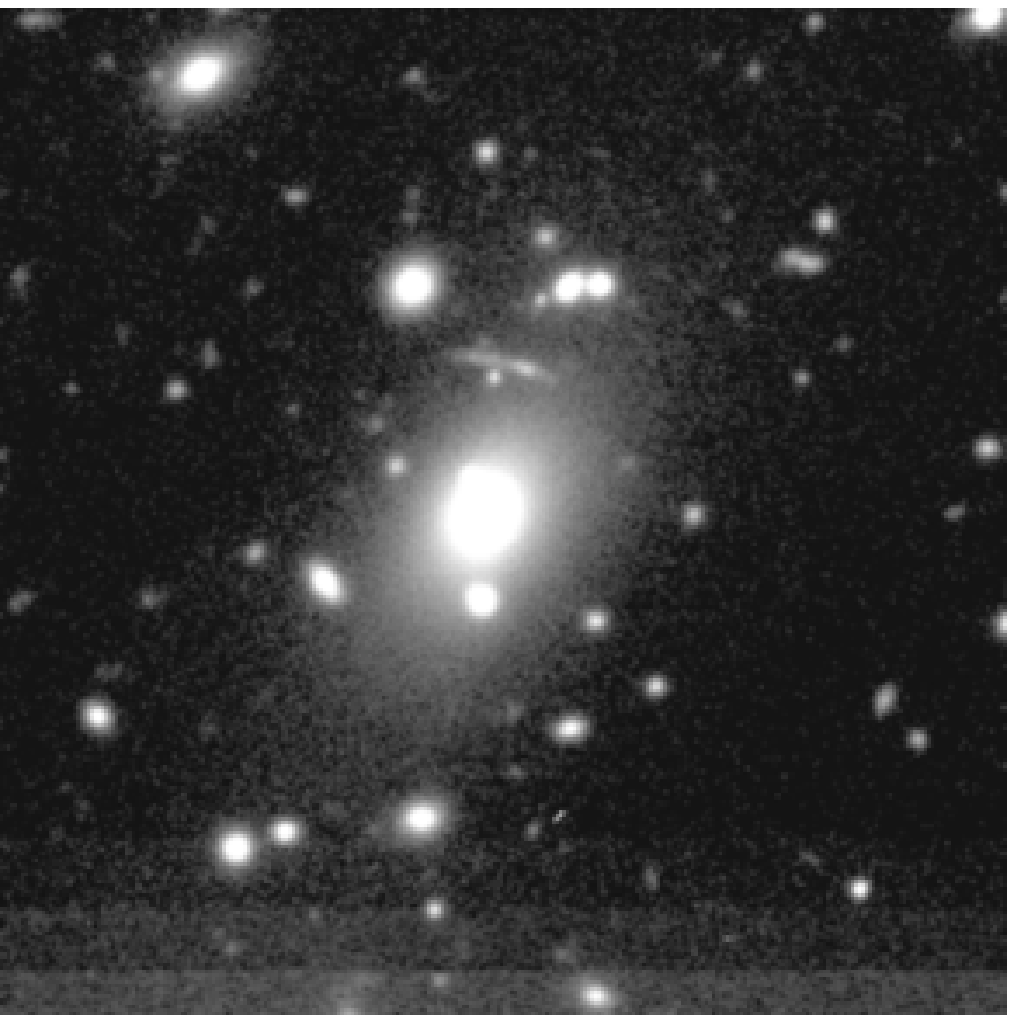} \\[3ex]
\includegraphics[width=87mm,angle=0, trim= 0mm 0mm 0mm 0mm,clip]{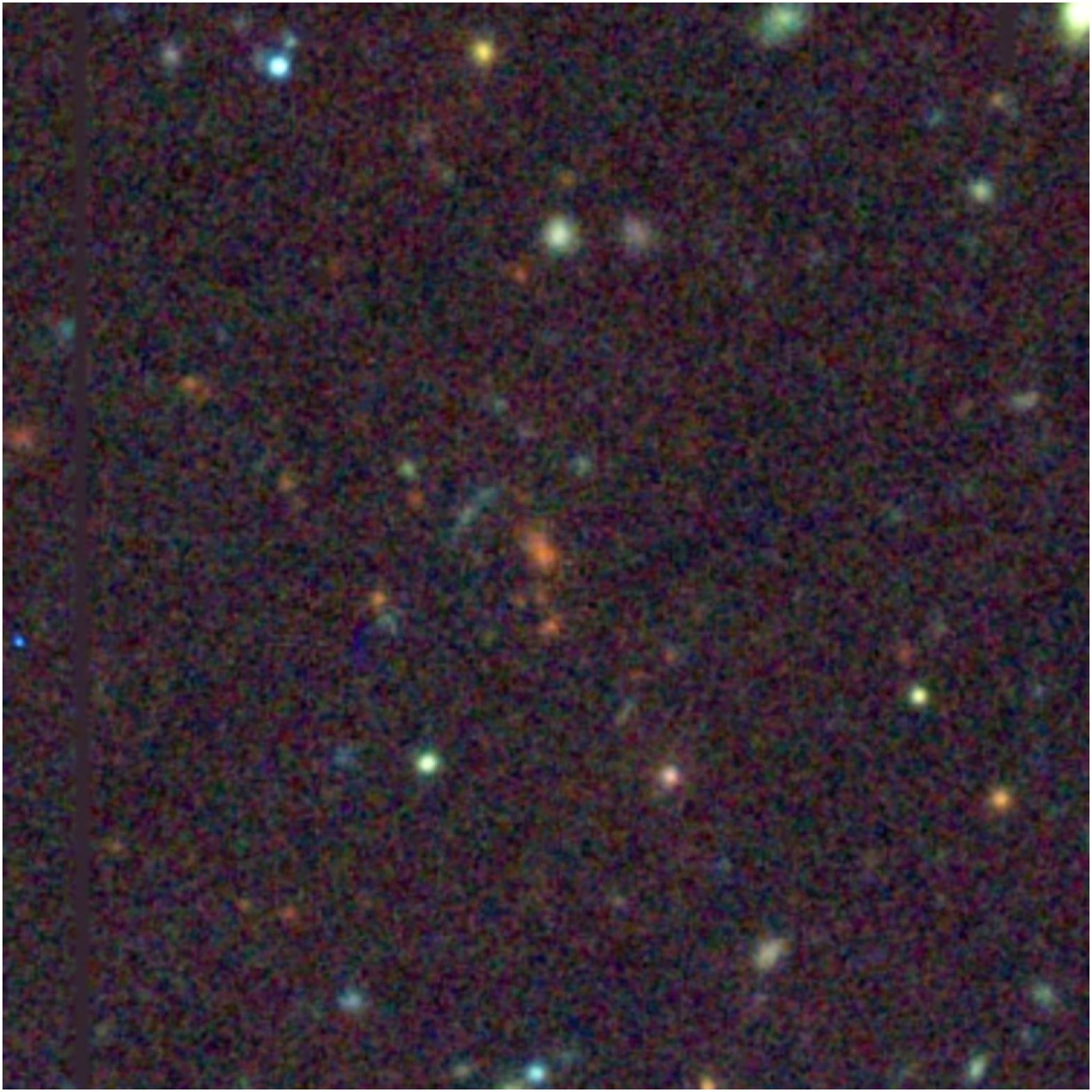} &
\includegraphics[width=87mm,angle=0, trim= 0mm 0mm 0mm 0mm,clip]{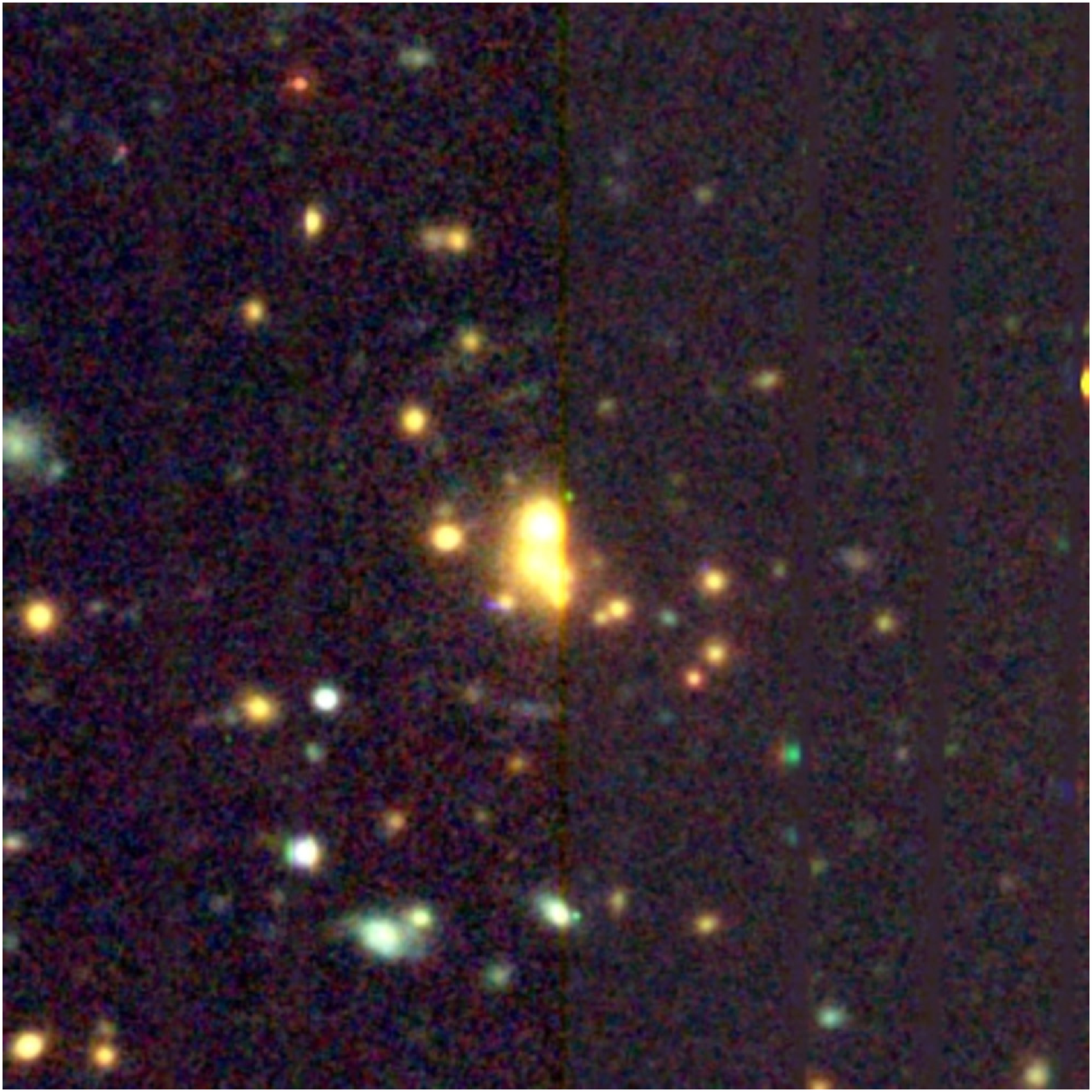} \\
\end{tabular}
\put(-435, 240){\bf \large \color{white} RCS2 J1108$-$0456}
    \put(-367, 183){\bf \color{yellow} S1.1}
    \put(-448, 200){\bf \color{yellow} S2.1}
\put(-175, 240){\bf \large \color{white} SDSS J1111+1408}
    \put(-127, 175){\bf \color{yellow} S1.1}
\put(-435, -20){\bf \large \color{white} RCS2 J1119$-$0728}
    \put(-411,-109){\bf \color{yellow} S1.1}
\put(-175, -20){\bf \large \color{white} RCS2 J1125$-$0628}
    \put(-162,-170 ){\bf \color{yellow} S1.1}
    \put(-136,-176){\bf \color{yellow} S1.2}
\caption{ \label{fig:SLclusters_fov_04}
From the top-left to bottom-right panels we show the SL-selected galaxy clusters
RCS2 J1108$-$0456, SDSS J1111+1408, RCS2 J1119$-$0728, and RCS2 J1125$-$0628. 
Lensed galaxies are labeled in the same manner as in Figure \ref{fig:SLclusters_fov_01}.
All images cover a field of view of 75\arcsec$\times$75\arcsec.} 
\end{center}
\end{figure*}

\begin{figure*}[h!]
\begin{center}
\begin{tabular}{c c}
\includegraphics[width=87mm,angle=0, trim= 0mm 0mm 0mm 0mm,clip]{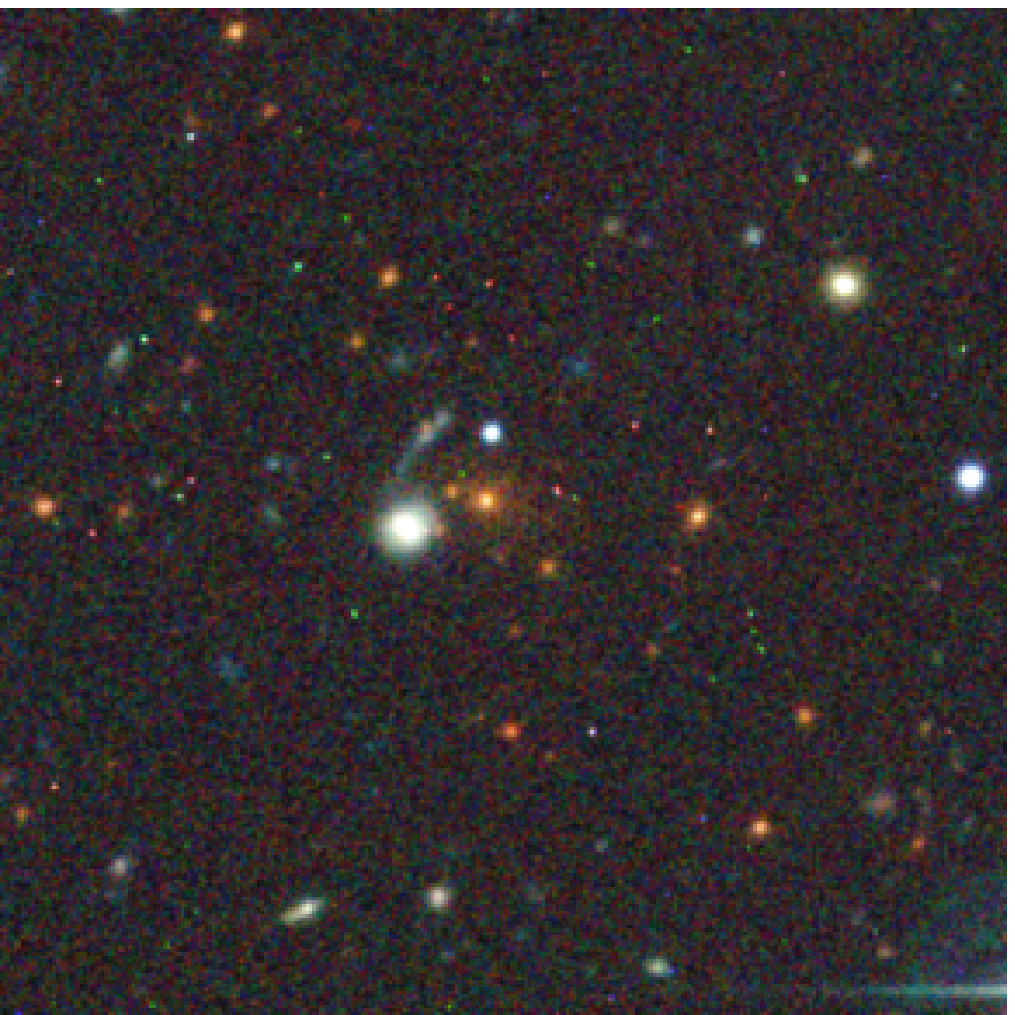} &
\includegraphics[width=87mm,angle=0, trim= 0mm 0mm 0mm 0mm,clip]{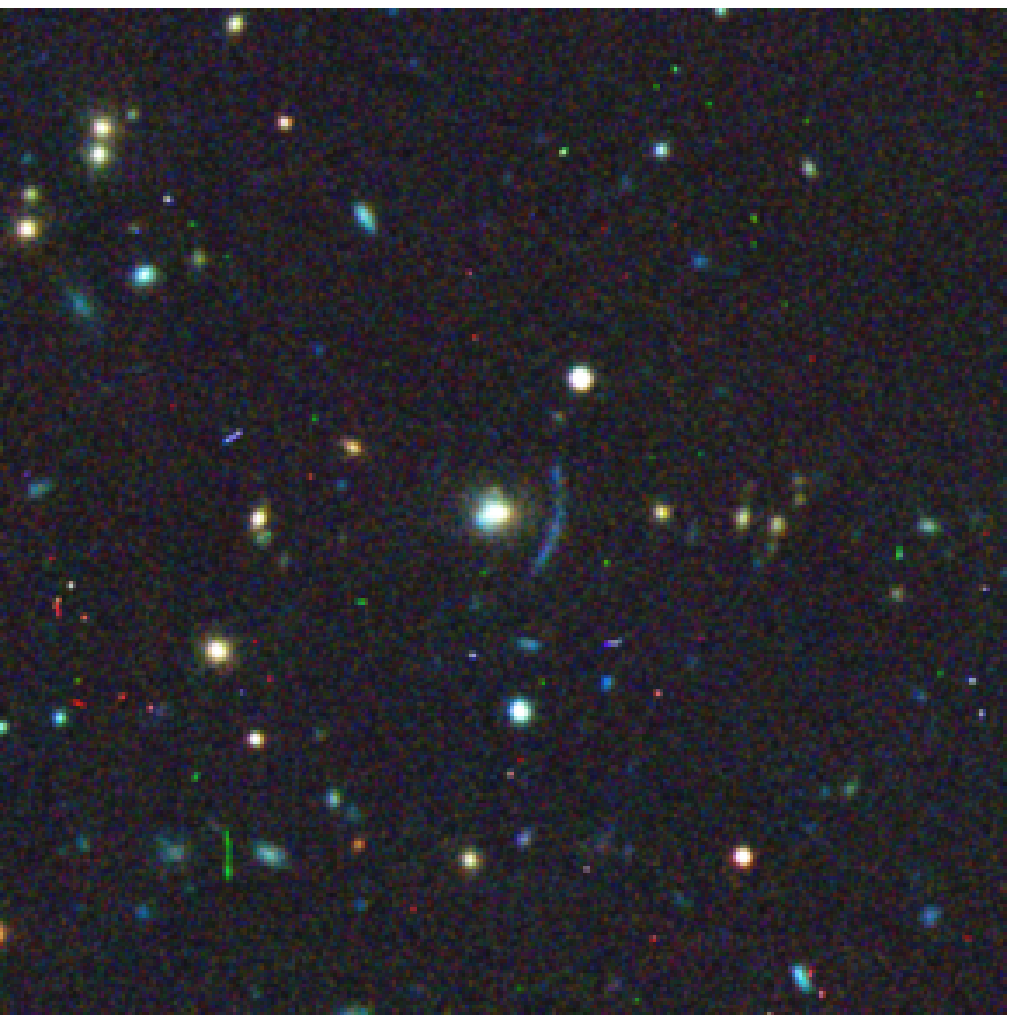} \\[3ex]
\includegraphics[width=87mm,angle=0, trim= 0mm 0mm 0mm 0mm,clip]{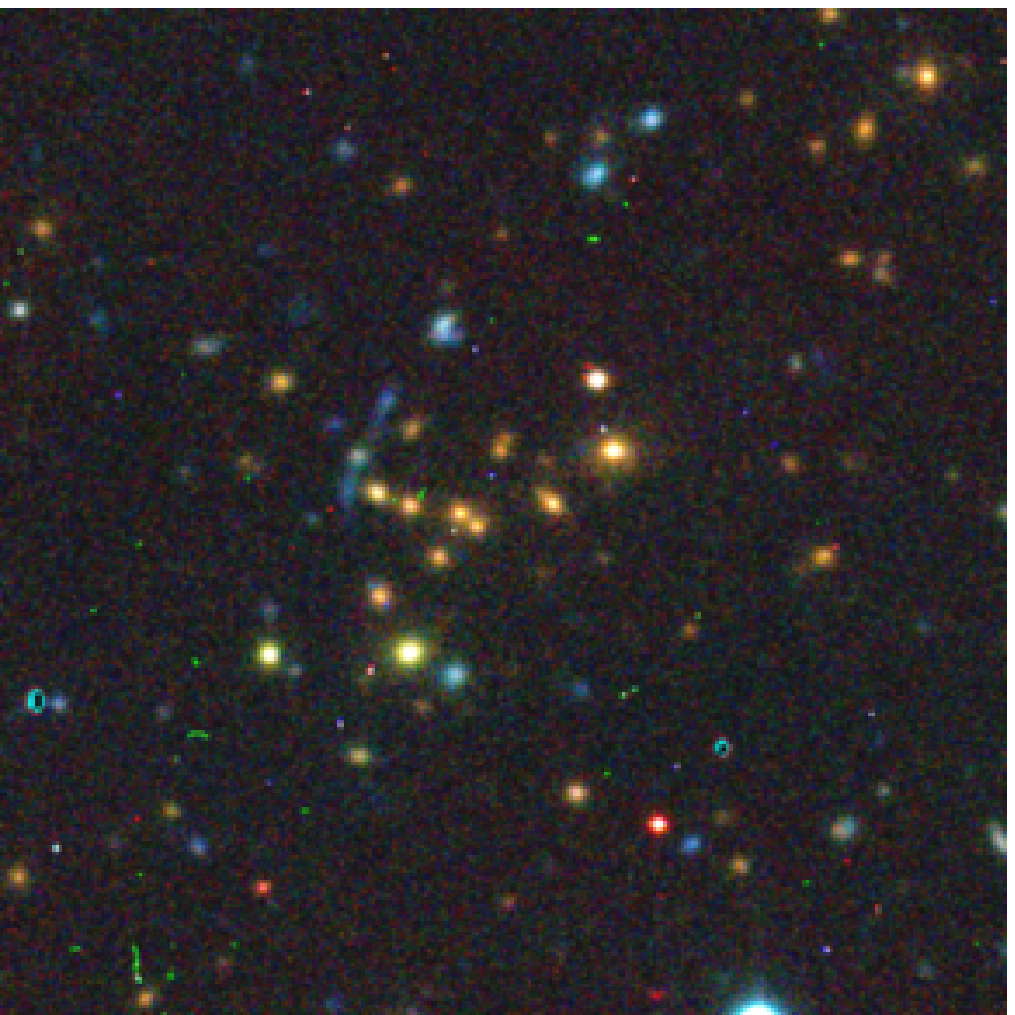} &
\includegraphics[width=87mm,angle=0, trim= 0mm 0mm 0mm 0mm,clip]{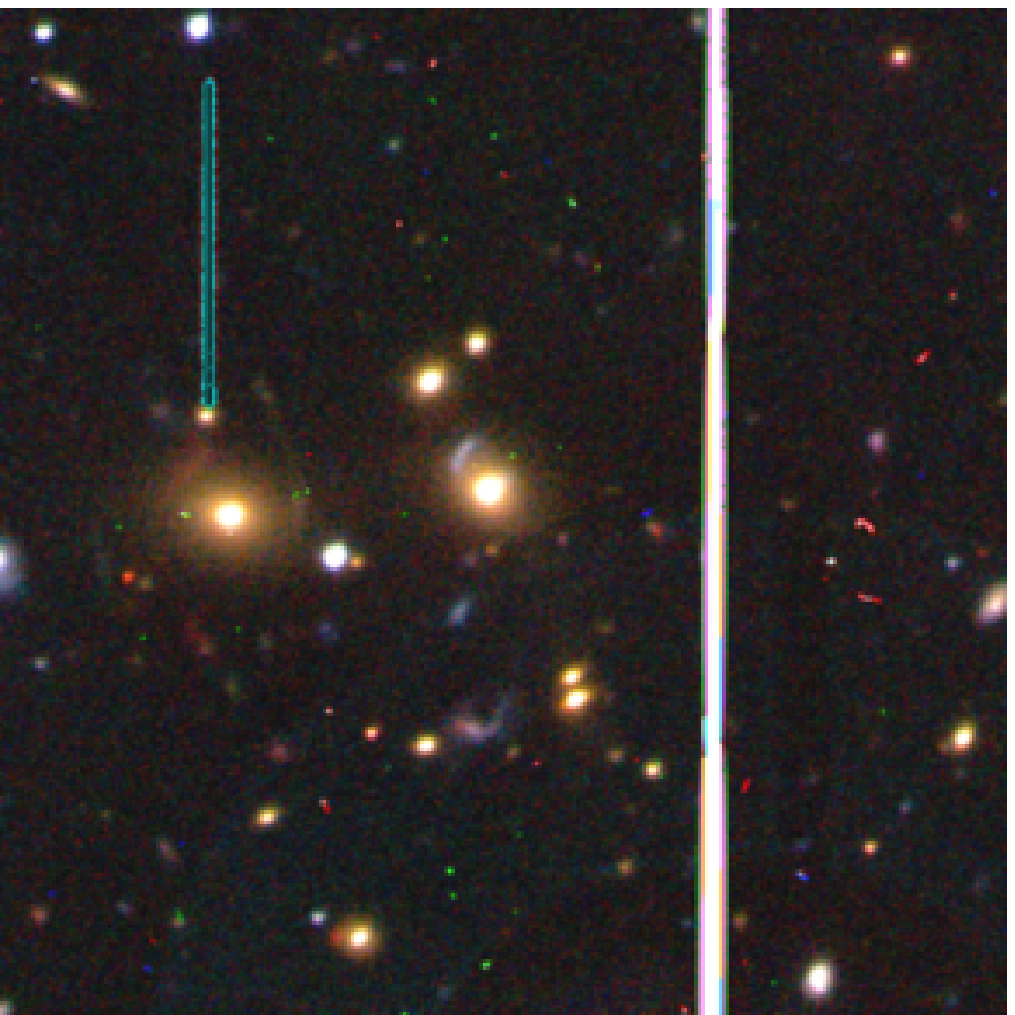} \\
\end{tabular}
\put(-435, 240){\bf \large \color{white} RCS2 J1250+0244}
    \put(-420, 160){\bf \color{yellow} S1.1}
\put(-175, 240){\bf \large \color{white} RCS2 J1511+0630}
    \put(-110, 130){\bf \color{yellow} S1.1}
\put(-435, -20){\bf \large \color{white} SDSS J1517+1003}
    \put(-418,-92){\bf \color{yellow} S1.1}
    \put(-435,-135){\bf \color{yellow} S1.2}
\put(-175, -20){\bf \large \color{white} SDSS J1519+0840}
    \put(-143,-104){\bf \color{yellow} S1.1}
    \put(-148,-165){\bf \color{yellow} S2.1}
\caption{ \label{fig:SLclusters_fov_05}
From the top-left to bottom-right panels we show the SL-selected galaxy clusters
RCS2 J1250+0244, RCS2 J1511+0630, SDSS J1517+1003, and SDSS J1519+0840. 
Lensed galaxies are labeled in the same manner as in Figure \ref{fig:SLclusters_fov_01}.
All images cover a field of view of 75\arcsec$\times$75\arcsec.} 
\end{center}
\end{figure*}

\begin{figure*}[h!]
\begin{center}
\begin{tabular}{c c}
\includegraphics[width=87mm,angle=0, trim= 0mm 0mm 0mm 0mm,clip]{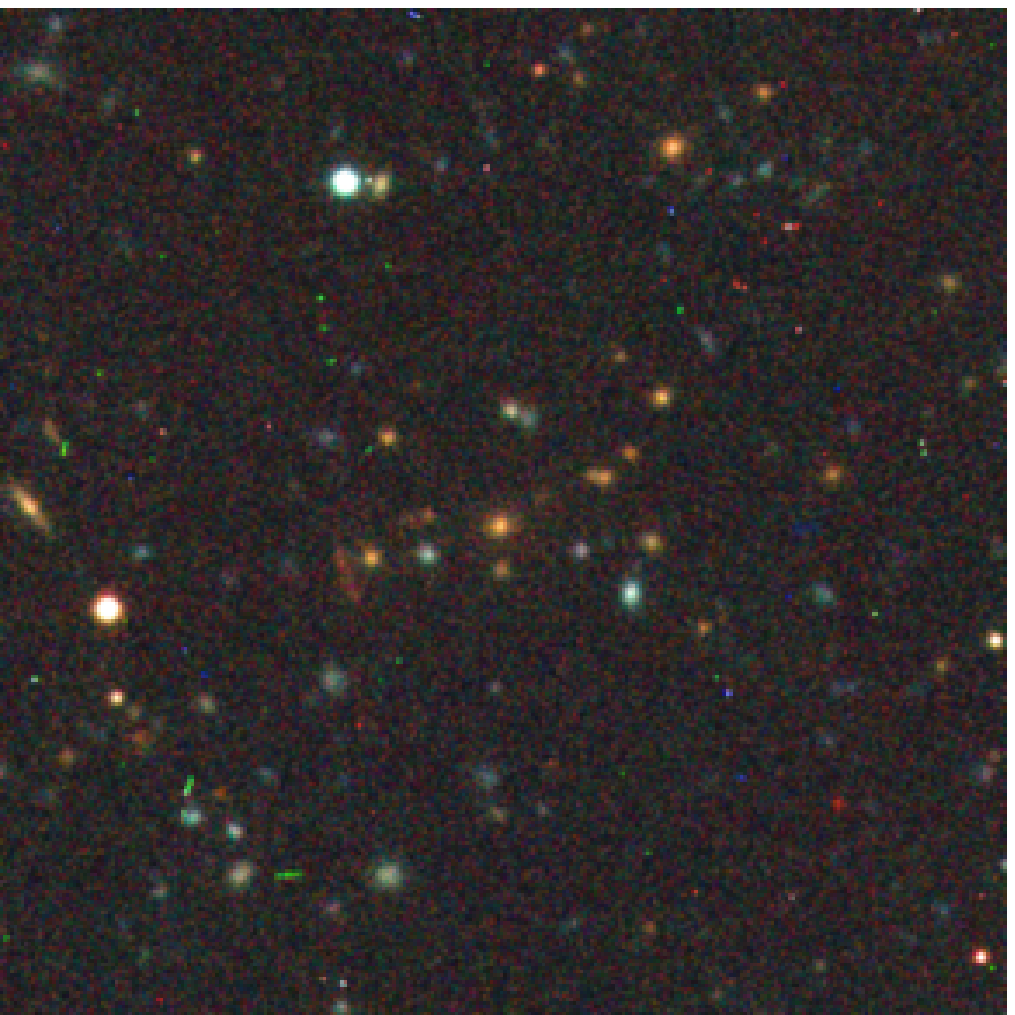} &
\includegraphics[width=87mm,angle=0, trim= 0mm 0mm 0mm 0mm,clip]{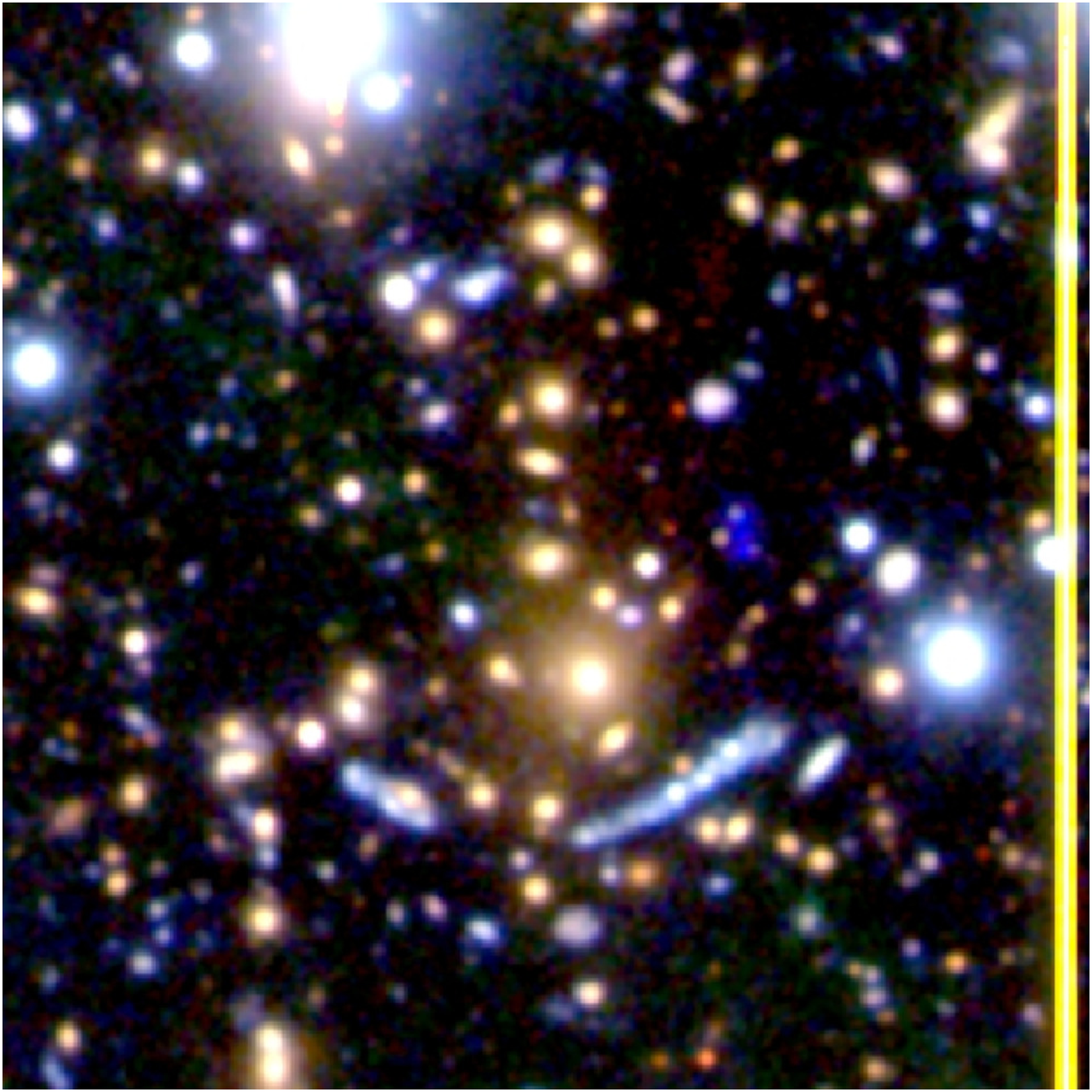} \\[3ex]
\includegraphics[width=87mm,angle=0, trim= 0mm 0mm 0mm 0mm,clip]{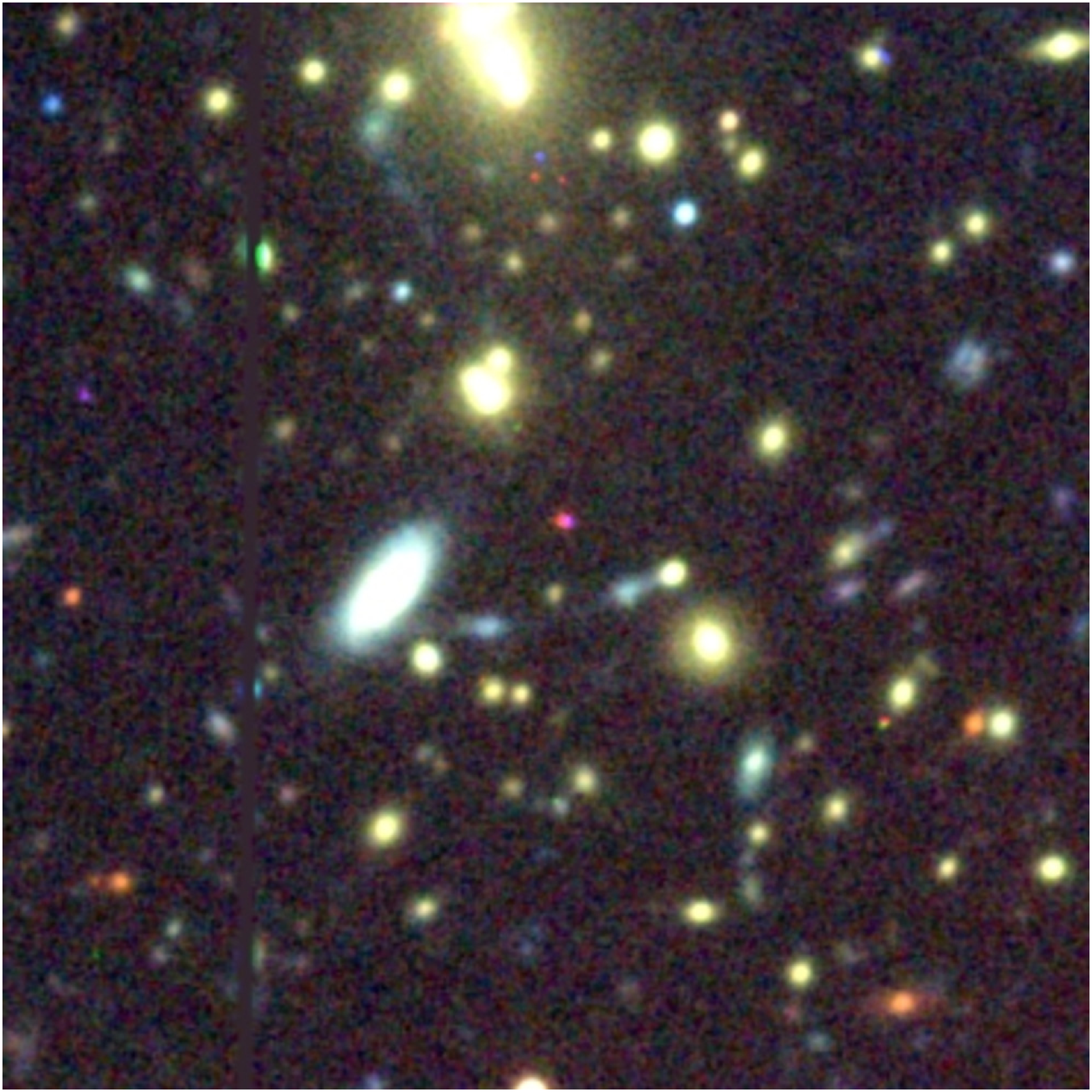} &
\includegraphics[width=87mm,angle=0, trim= 0mm 0mm 0mm 0mm,clip]{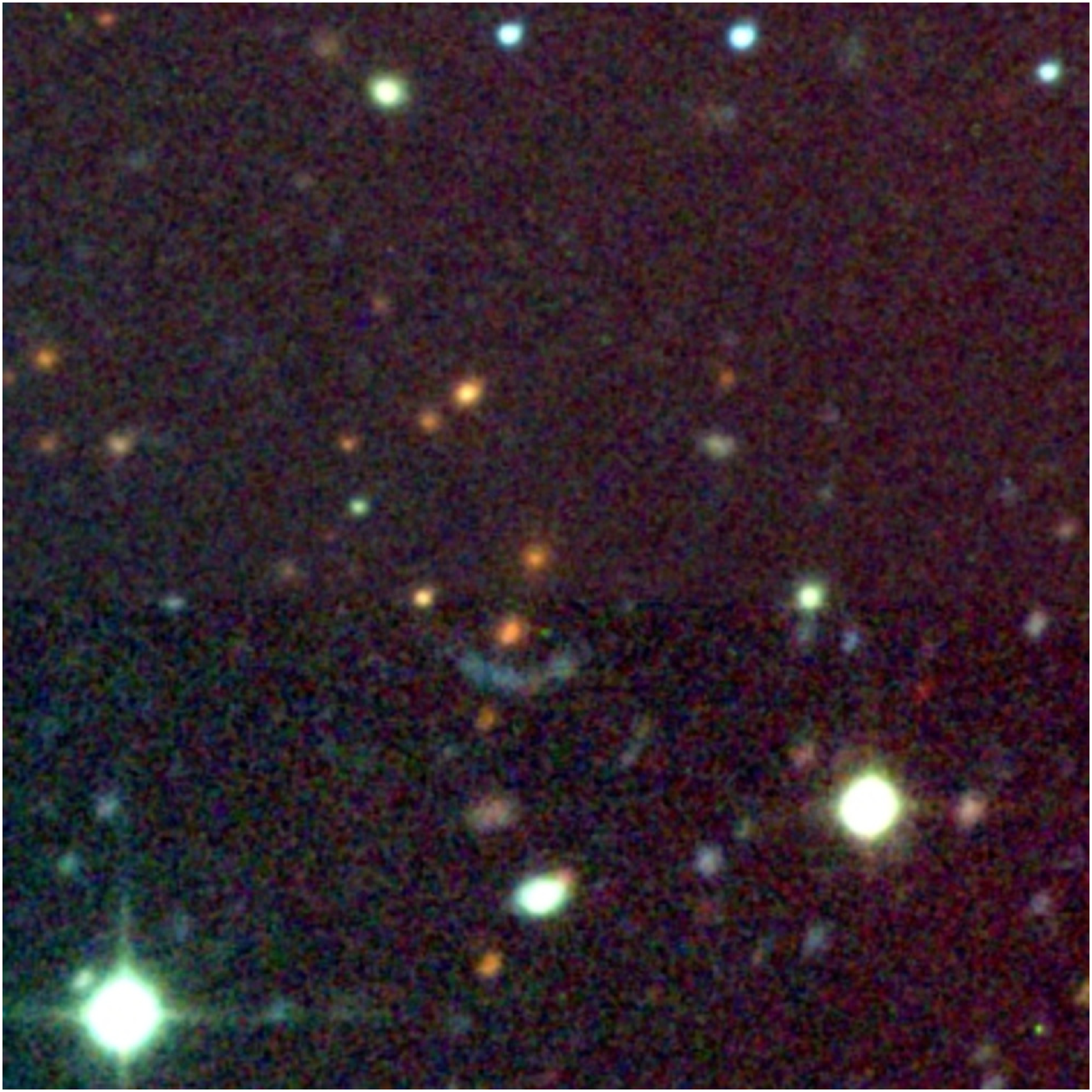} \\
\end{tabular}
\put(-435, 240){\bf \large \color{white} RCS2 J1526+0432}
    \put(-330, 175){\bf \color{yellow} S1.1}
    \put(-450, 115){\bf \color{yellow} S2.1}
\put(-175, 240){\bf \large \color{white} SDSS J2111$-$0114}
    \put(-101, 65){\bf \color{yellow} S1.1}
    \put(-172, 62){\bf \color{yellow} S1.2}
    \put(-180, 88){\bf \color{yellow} S1.3}
\put(-435, -20){\bf \large \color{white} SDSS J2135$-$0102}
    \put(-408,-157){\bf \color{yellow} S1.1}
    \put(-370,-151){\bf \color{yellow} S1.2}
    \put(-283,-88){\bf \color{yellow} S1.3}
    \put(-308,-149){\bf \color{yellow} S2.1}
\put(-175, -20){\bf \large \color{white} RCS2 J2147$-$0102}
    \put(-149,-169){\bf \color{yellow} S1.1}
\caption{ \label{fig:SLclusters_fov_06}
From the top-left to bottom-right panels we show the SL-selected galaxy clusters
RCS2 J1526+0432, SDSS J2111$-$0114, SDSS J2135$-$0102, and RCS2 J2147$-$0102. 
Lensed galaxies are labeled in the same manner as in Figure \ref{fig:SLclusters_fov_01}.
All images cover a field of view of 75\arcsec$\times$75\arcsec.
Galaxy cluster SDSS J2111$-$0114 is previously described in \cite{Bayliss11b}; its color composite 
image is made from \textit{gri} imaging obtained with Subaru/SuprimeCam \citep{Oguri09}.} 
\end{center}
\end{figure*}

\begin{figure*}[h!]
\begin{center}
\begin{tabular}{c c}
\includegraphics[width=87mm,angle=0, trim= 0mm 0mm 0mm 0mm,clip]{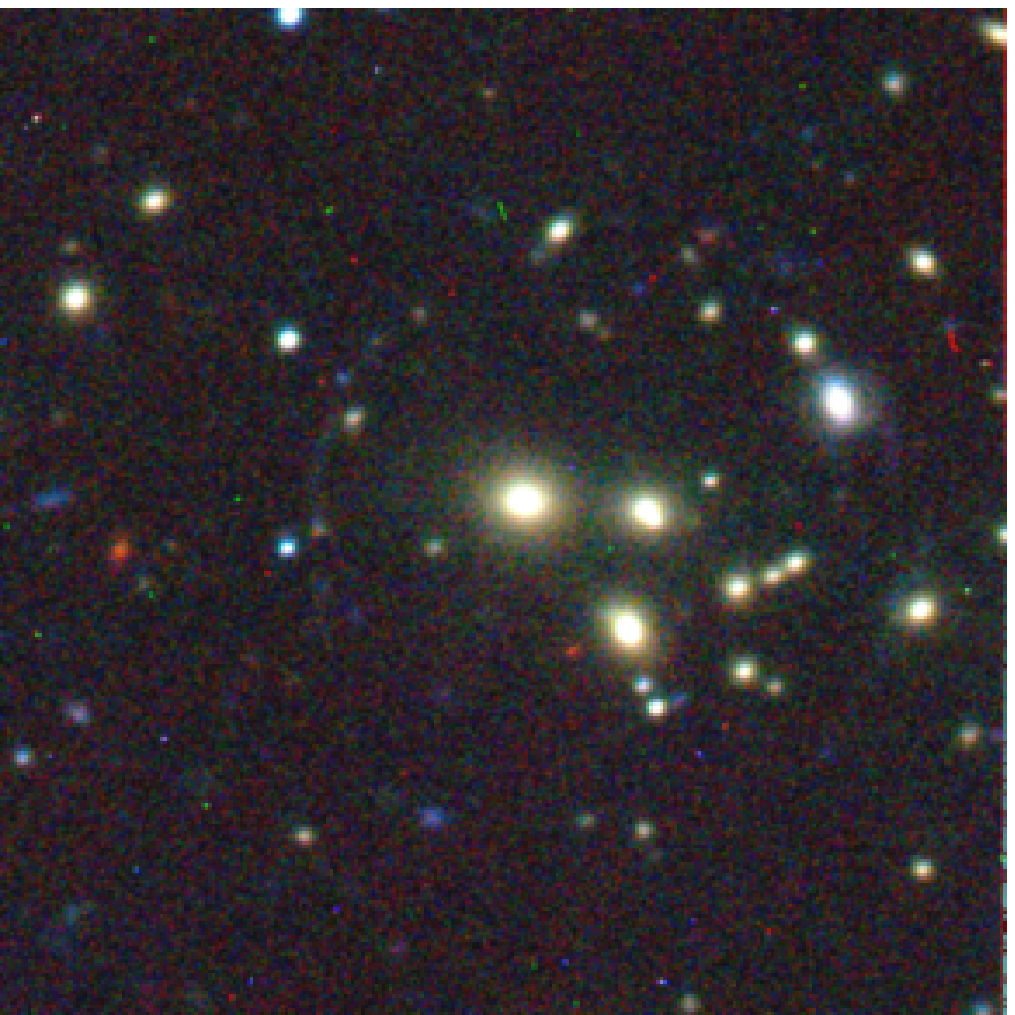} &
\includegraphics[width=87mm,angle=0, trim= 0mm 0mm 0mm 0mm,clip]{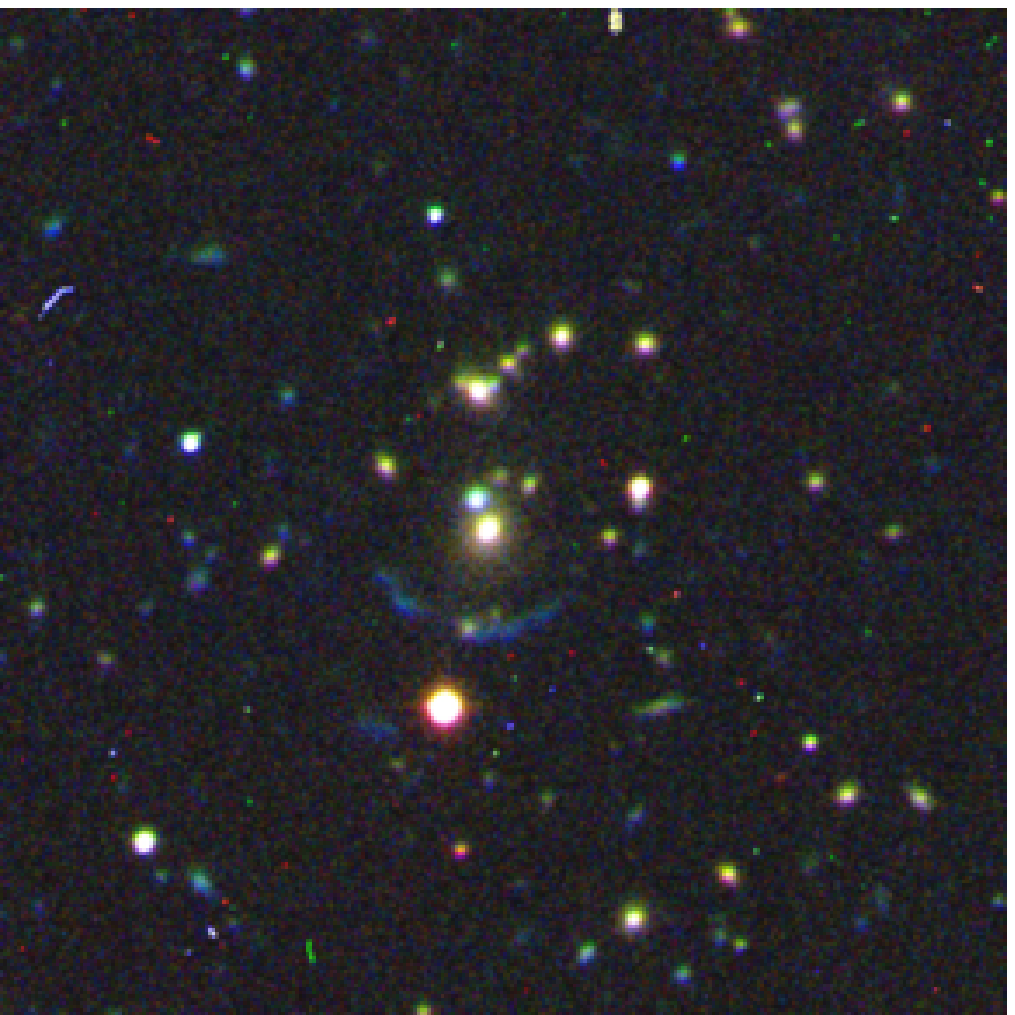} \\[3ex]
\includegraphics[width=87mm,angle=0, trim= 0mm 0mm 0mm 0mm,clip]{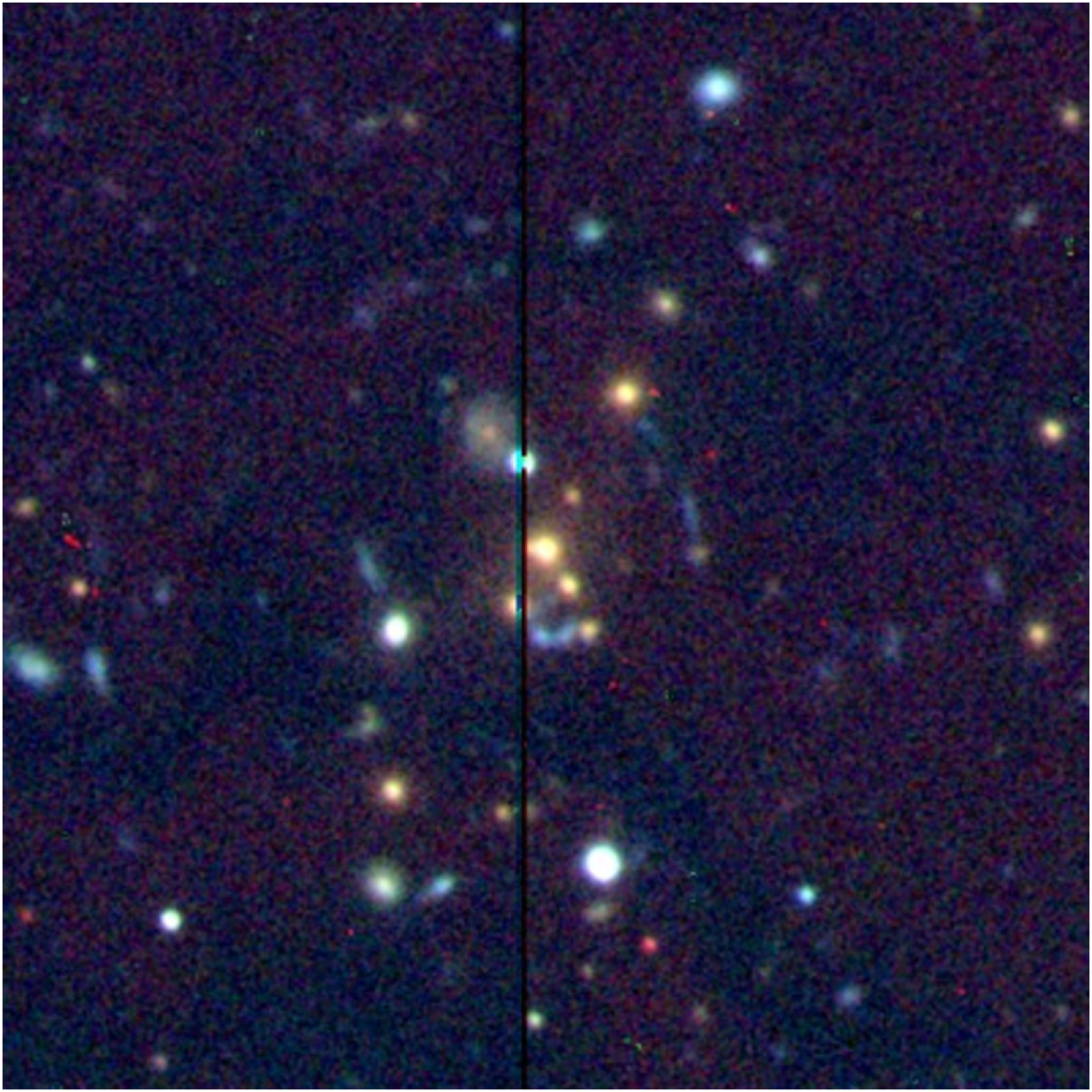} & 
\includegraphics[width=87mm,angle=0, trim= 0mm 0mm 0mm 0mm,clip]{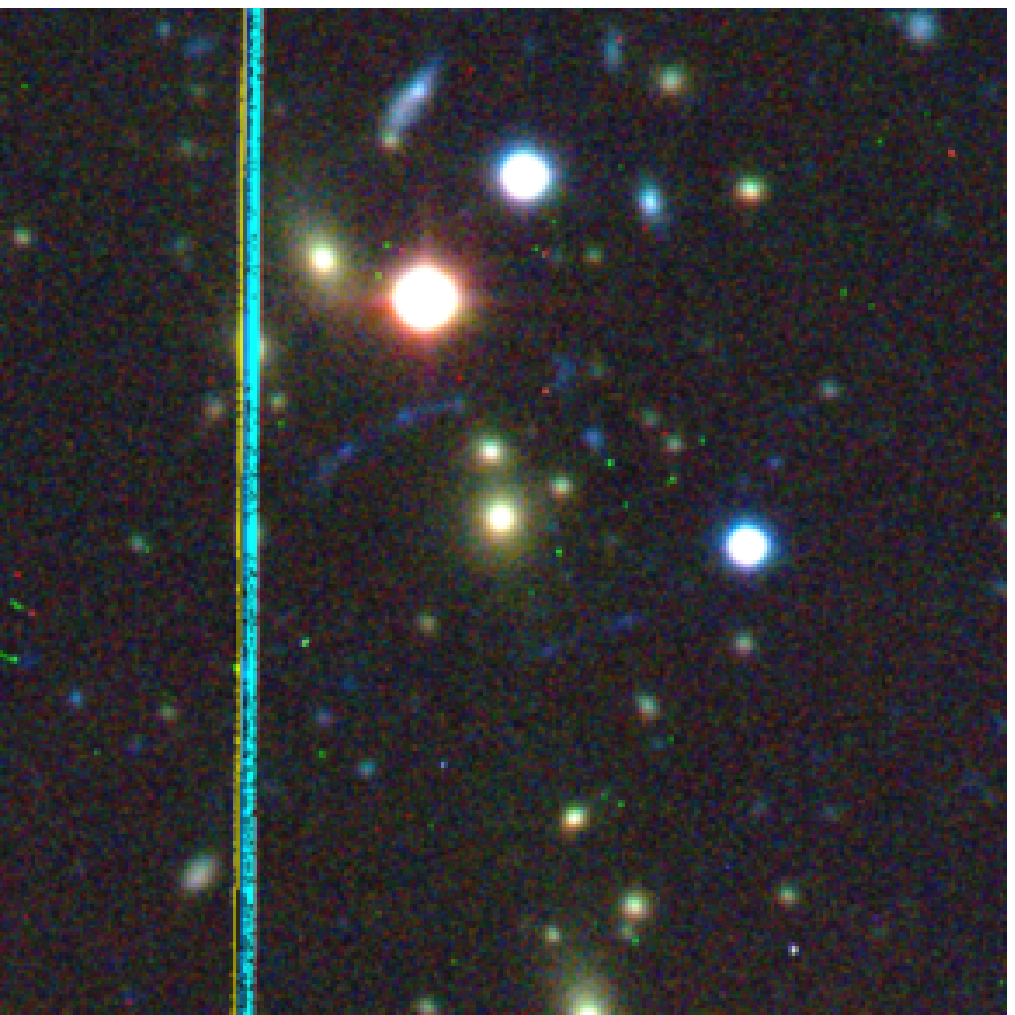} \\
\end{tabular}
\put(-435, 240){\bf \large \color{white} RCS2 J2151$-$0138}
    \put(-436, 172){\bf \color{yellow} S1.1}
\put(-175, 240){\bf \large \color{white} SDSS J2313$-$0104}
    \put(-134, 92){\bf \color{yellow} S1.1}
\put(-435, -20){\bf \large \color{white} RCS2 J2329$-$0120}
    \put(-348,-120){\bf \color{yellow} S1.1}
    \put(-410,-153){\bf \color{yellow} S2.2}
    \put(-378,-156){\bf \color{yellow} S2.1}
\put(-175, -20){\bf \large \color{white} RCS2 J2329$-$1317}
    \put(-184,-105){\bf \color{yellow} S1.1}
    \put(-157,-95){\bf \color{yellow} S1.2}
\caption{ \label{fig:SLclusters_fov_07}
From the top-left to bottom-right panels we show the SL-selected galaxy clusters
RCS2 J2151$-$0138, SDSS J2313$-$0104, RCS2 J2329$-$0120, and RCS2 J2329$-$1317. 
Lensed galaxies are labeled in the same manner as in Figure \ref{fig:SLclusters_fov_01}.
All images cover a field of view of 75\arcsec$\times$75\arcsec.} 
\end{center}
\end{figure*}

\begin{figure*}[h!]
\begin{center}
\begin{tabular}{c}
\includegraphics[width=87mm,angle=0, trim= 0mm 0mm 0mm 0mm,clip]{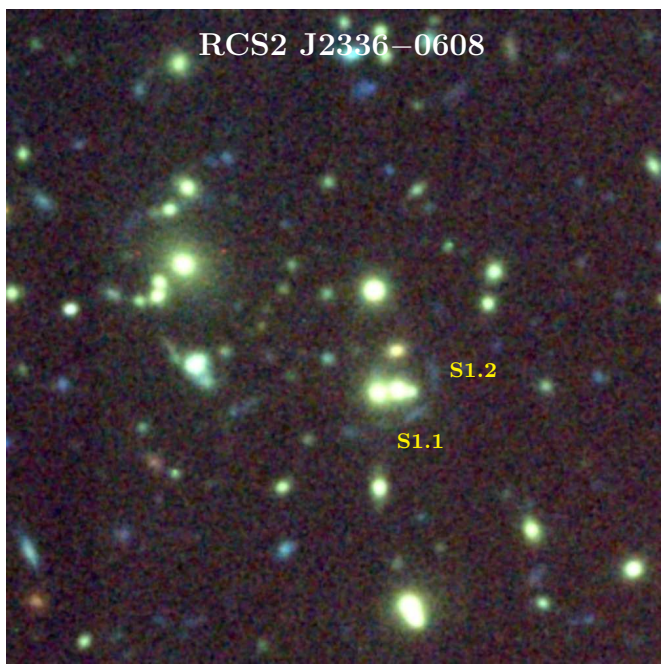}

\end{tabular}
\put(-180, 110){\bf \large \color{white} RCS2 J2336$-$0608}
    \put(-85, -12){\bf \color{yellow} S1.2}
    \put(-105, -39){\bf \color{yellow} S1.1}
\caption{ \label{fig:SLclusters_fov_08}
The SL-selected galaxy cluster RCS2 J2336$-$0608 is shown. 
Lensed galaxies are labeled in the same manner as in Figure \ref{fig:SLclusters_fov_01}.
As in previous panels, the image covers a field of view of 75\arcsec$\times$75\arcsec.} 
\end{center}
\end{figure*}

\end{document}